\documentclass[12pt,a4paper]{article}
\pdfoutput=1

\usepackage{amsmath,amssymb}
\usepackage[bookmarks=true]{hyperref}
\usepackage[nosort]{cite}
\usepackage[bulletsep]{collect}
\def\gfxon{\usepackage[final]{graphicx}}

\gfxon

\ifx\hypersetup\sadfkjashdfkxja\else
\hypersetup{pdftitle={Quantum Deformations of the One-Dimensional Hubbard Model}}
\hypersetup{pdfauthor={Niklas Beisert, Peter Koroteev}}
\hypersetup{pdfsubject={Mathematical Physics}}
\hypersetup{pdfkeywords={Spin Chain, R-Matrix, Quantum Deformation, Hubbard Model}}
\hypersetup{plainpages=false}
\hypersetup{bookmarksnumbered=true}
\hypersetup{pdfstartview=FitH}
\hypersetup{pdfpagemode=None}
\hypersetup{colorlinks=false}
\hypersetup{citebordercolor={.5 1 .5}}
\hypersetup{urlbordercolor={.5 1 1}}
\hypersetup{linkbordercolor={1 .7 .7}}
\fi

\makeatletter
\let\old@startsection=\@startsection
\renewcommand{\@startsection}[6]{\old@startsection{#1}{#2}{#3}{#4}{#5}{#6\mathversion{bold}}}
\makeatother

\makeatletter \@addtoreset{equation}{section} \makeatother

\allowdisplaybreaks[3]

\makeatletter
\let\old@makecaption=\@makecaption
\def\@makecaption{\small\old@makecaption}
\makeatother

\let\oldPhi=\Phi
\let\oldPsi=\Psi
\let\oldGamma=\Gamma
\let\oldDelta=\Delta
\let\oldSigma=\Sigma
\let\oldLambda=\Lambda
\let\oldTheta=\Theta
\let\oldPi=\Pi
\renewcommand{\Phi}{\mathnormal{\oldPhi}}
\renewcommand{\Psi}{\mathnormal{\oldPsi}}
\renewcommand{\Gamma}{\mathnormal{\oldGamma}}
\renewcommand{\Sigma}{\mathnormal{\oldSigma}}
\renewcommand{\Delta}{\mathnormal{\oldDelta}}
\renewcommand{\Theta}{\mathnormal{\oldTheta}}
\renewcommand{\Lambda}{\mathnormal{\oldLambda}}
\renewcommand{\Pi}{\mathnormal{\oldPi}}

\newcommand{\Twist}{\mathcal{T}}
\newcommand{\ham}{\mathcal{H}}
\newcommand{\gen}[1]{\mathfrak{#1}}
\newcommand{\smat}{\mathcal{S}}
\newcommand{\rmat}{\mathcal{R}}
\newcommand{\conj}{\mathcal{C}}

\newcommand{\perm}{\mathcal{P}}

\newcommand{\superN}{\mathcal{N}}
\newcommand{\copro}{\oldDelta}
\newcommand{\coproop}{\oldDelta\indup{op}}
\newcommand{\unit}{\eta}
\newcommand{\counit}{\varepsilon}
\newcommand{\antipode}{S}
\newcommand{\antimap}{s}
\newcommand{\specmap}{u}
\newcommand{\transpose}{\scriptscriptstyle\mathsf{ST}}

\newcommand{\Integers}{\mathbb{Z}}
\newcommand{\Reals}{\mathbb{R}}
\newcommand{\Complex}{\mathbb{C}}

\newcommand{\MMM}[2]{{\arraycolsep0pt\begin{array}[b]{c}\makebox[0cm]{$\atopfrac{#2}{\downarrow}$}\\#1\end{array}}}

\makeatletter
\newcommand{\ellSN}{\mathop{\operator@font sn}\nolimits}
\newcommand{\ellCN}{\mathop{\operator@font cn}\nolimits}
\newcommand{\ellDN}{\mathop{\operator@font dn}\nolimits}
\newcommand{\ellAM}{\mathop{\operator@font am}\nolimits}
\newcommand{\ellK}{\mathop{\smash{\operator@font K}\vphantom{a}}\nolimits}
\newcommand{\ellE}{\mathop{\smash{\operator@font E}\vphantom{a}}\nolimits}
\makeatother

\ifx\genfrac\sdflkaj
\newcommand{\atopfrac}[2]{{{#1}\above0pt{#2}}}
\else
\newcommand{\atopfrac}[2]{\genfrac{}{}{0pt}{}{#1}{#2}}
\fi
\newcommand{\sfrac}[2]{{\textstyle\frac{#1}{#2}}}
\newcommand{\half}{\sfrac{1}{2}}
\newcommand{\ihalf}{\sfrac{i}{2}}
\newcommand{\quarter}{\sfrac{1}{4}}

\newcommand{\indup}[1]{_{\mathrm{#1}}}

\newcommand{\rep}[1]{{\mathbf{#1}}}
\newcommand{\matr}[2]{\left(\begin{array}{#1}#2\end{array}\right)}
\newcommand{\lvl}[1]{^{\mathrm{#1}}}

\newcommand{\lrbrk}[1]{\left(#1\right)}
\newcommand{\bigbrk}[1]{\bigl(#1\bigr)}

\newcommand{\bigcomm}[2]{\big[#1,#2\big]}
\newcommand{\comm}[2]{[#1,#2]}

\newcommand{\acomm}[2]{\{#1,#2\}}
\newcommand{\bigacomm}[2]{\big\{#1,#2\big\}}

\newcommand{\lreval}[1]{\left.#1\right|}

\newcommand{\set}[1]{\{#1\}}
\newcommand{\state}[1]{\mathopen{|}#1\mathclose{\rangle}}
\newcommand{\bra}[1]{\mathopen{\langle}#1\mathclose{|}}
\newcommand{\ket}[1]{\mathopen{|}#1\mathclose{\rangle}}
\newcommand{\braket}[2]{\mathopen{\langle}#1\mathpunct{|}#2\mathclose{\rangle}}

\newcommand{\srep}[1]{\langle #1\rangle}
\newcommand{\lrep}[1]{\{#1\}}
\newcommand{\qnum}[1]{[#1]_q}

\newcommand{\xp}[1]{x^{+}_{#1}}
\newcommand{\xm}[1]{x^{-}_{#1}}
\newcommand{\xpm}[1]{x^{\pm}_{#1}}

\newcommand{\xbp}[1]{\bar{x}^{+}_{#1}}
\newcommand{\xbm}[1]{\bar{x}^{-}_{#1}}

\newcommand{\alg}[1]{\mathfrak{#1}}
\newcommand{\grp}[1]{\mathrm{#1}}
\newcommand{\Qgrp}{\mathrm{U}_q}

\newcommand{\nn}{\nonumber}
\newcommand{\nln}{\nonumber\\}
\newcommand{\nl}[1][0pt]{\nonumber\\[#1]&\hspace{-4\arraycolsep}&\mathord{}}
\newcommand{\nlnum}{\\&\hspace{-4\arraycolsep}&\mathord{}}
\newcommand{\earel}[1]{\mathrel{}&\hspace{-2\arraycolsep}#1\hspace{-2\arraycolsep}&\mathrel{}}
\newcommand{\eq}{\earel{=}}
\newcommand{\beq}{\begin{equation}}
\newcommand{\eeq}{\end{equation}}

\def\[{\begin{equation}}
\def\]{\end{equation}}
\def\<{\begin{eqnarray}}
\def\>{\end{eqnarray}}

\makeatletter
\def\mr@ignsp#1 {\ifx\:#1\@empty\else #1\expandafter\mr@ignsp\fi}%
\newcommand{\multiref}[1]{\begingroup
\xdef\mr@no@sparg{\expandafter\mr@ignsp#1 \: }%
\def\mr@comma{}%
\@for\mr@refs:=\mr@no@sparg\do{\mr@comma\def\mr@comma{,}\ref{\mr@refs}}%
\endgroup}
\makeatother

\newcommand{\hypref}[2]{\ifx\href\asklfhas #2\else\href{#1}{#2}\fi}

\newcommand{\secref}[1]{Sec.~\multiref{#1}}

\newcommand{\appref}[1]{App.~\multiref{#1}}

\newcommand{\tabref}[1]{Tab.~\multiref{#1}}

\newcommand{\figref}[1]{Fig.~\multiref{#1}}
\renewcommand{\eqref}[1]{(\multiref{#1})}

\ifx\href\asklfhas\newcommand{\href}[2]{#2}\fi

\newcommand{\arxivlink}[1]{\href{http://arxiv.org/abs/#1}{arxiv:#1}}

\begin{document}

\setcounter{page}{0}
\thispagestyle{empty}
\begin{flushright}\footnotesize
\texttt{\arxivlink{0802.0777}}\\
\texttt{AEI-2008-003}\\
\texttt{ITEP-TH-06/08}%
\vspace{0.5cm}
\end{flushright}
\vspace{0.5cm}

\renewcommand{\thefootnote}{\arabic{footnote}}
\setcounter{footnote}{0}
\begin{center}%
{\Large\textbf{\mathversion{bold}
Quantum Deformations of the\\
One-Dimensional Hubbard Model}%
\par}
\vspace{1cm}%

\textsc{Niklas Beisert$^{\natural}$
and Peter Koroteev$^{\sharp\natural\spadesuit}$}
\vspace{5mm}%

\textit{$^{\natural}$ Max-Planck-Institut f\"ur Gravitationsphysik\\%
Albert-Einstein-Institut\\%
Am M\"uhlenberg 1, 14476 Potsdam, Germany}%
\vspace{3mm}

\textit{$^{\sharp}$ Institute for Theoretical and Experimental Physics\\%
B.\ Cheremushkinskaya 25, Moscow 117259, Russia}%
\vspace{3mm}

\textit{$^{\spadesuit}$ Institute for Nuclear Research\\%
Prospekt 60-letiya Oktyabrya 7a, Moscow 117312, Russia}%
\vspace{3mm}

\thispagestyle{empty}

\texttt{nbeisert,koroteev@aei.mpg.de}
\par\vspace{1cm}

\vfill

\textbf{Abstract}\vspace{5mm}

\begin{minipage}{12.7cm}
The centrally extended superalgebra
$\alg{psu}(2|2)\ltimes\mathbb{R}^3$
was shown to play an important role for
the integrable structures of
the one-dimensional Hubbard model
and of the planar AdS/CFT correspondence.
Here we consider its quantum deformation
$\Qgrp(\alg{psu}(2|2)\ltimes\mathbb{R}^3)$
and derive the fundamental R-matrix.
From the latter we deduce an
integrable spin chain Hamiltonian with three independent parameters
and the corresponding Bethe equations
to describe the spectrum on periodic chains.
We relate our Hamiltonian to a two-parametric Hamiltonian
proposed by Alcaraz and Bariev
which can be considered a quantum deformation of the
one-dimensional Hubbard model.
\end{minipage}

\vspace*{\fill}

\end{center}

\newpage

\section{Introduction and Overview}\label{sec:Intro}

Finding the spectrum of a quantum mechanical model
is an intricate problem.
Indeed, for generic models there is no complete analytic solution
to the spectrum essentially because non-linear interaction terms in the
Hamiltonian easily make the problem chaotic and intractable.
Only very few models, such as the harmonic oscillator,
are \emph{solvable} exactly.
Somewhere in between these two extremes live the \emph{integrable} models.
They may contain highly non-trivial interactions, but they can
nevertheless be solved completely by the right ansatz for the wave functions.
Such a wave function will depend on a couple of parameters,
and quantization conditions will impose a system of equations on them.
There need not be a general analytic solution to these equations ---
after all the spectrum of integrable models is usually highly non-trivial ---
nevertheless the reduction to a small number of parameters is sufficient
to make the spectral problem much more tractable than for
generic quantum mechanical models.

Typical integrable models are formulated in 1+1 or 2 dimensions.
They include field theories, non-linear sigma models,
particle models, vertex models and spin chains
(we shall focus on the latter in this article).
A plethora of integrable models with all kinds of features
is known to date, and it appears near impossible to
make a complete census.
A central insight towards this goal was made in the 1980's
by the Leningrad/St.~Petersburg--School
who related integrability to the existence of large hidden symmetries.
Through the enumeration of suitable symmetry algebras
one can hope to classify the integrable models.

For example, a very large class of integrable spin chains
can be derived and investigated using
Yangians and quantum affine algebras $\Qgrp(\hat{\alg{g}})$
\cite{Drinfeld:1986in,Jimbo:1985ua}.
In particular, the Heisenberg XXX (algebraic) spin chain
and its relatives with different symmetry algebra $\alg{g}$
and/or different representation are all based
on the Yangian double of $\alg{g}$.
Likewise, the quantum-deformed XXZ-like (trigonometric) spin chain
cousins are related to the quantum affine algebra $\Qgrp(\hat{\alg{g}})$.
XYZ-like (elliptic) spin chains also have a similar but much more elaborate
underlying symmetry algebra.

A famous integrable spin chain model that has escaped this classification
for a long time is the one-dimensional Hubbard model,
see \cite{Essler:2005aa}.
It is a model of electrons propagating on a chain of nuclei.
Each nucleus site can either be unoccupied, singly occupied
with electron spin pointing up/down or doubly occupied
with opposing electron spins.
In total there are four states per site
($c^\dagger_\alpha$ is a fermionic electron creation operator)
\[
\state{\circ},\quad
\state{\mathord{\uparrow}}\sim c^\dagger_1\state{\circ},\quad
\state{\mathord{\downarrow}}\sim c^\dagger_2\state{\circ},\quad
\state{\mathord{\updownarrow}}\sim c^\dagger_1c^\dagger_2\state{\circ}.
\]
The middle two states are considered fermionic
while the outer two states are overall bosonic.
This model is exciting because it shows some characteristics of superconductivity,
and therefore it is very desirable to understand its foundations well.
Integrability was established by Lieb and Wu who also
solved the spectrum by means of the Bethe ansatz \cite{Lieb:1968aa}.
An R-matrix which encodes the integrable structure
was later found by Shastry \cite{Shastry:1986bb}.
On the one hand, the R-matrix is the foundation
for much of the integrable machinery,
such as the algebraic Bethe ansatz \cite{Ramos:1996us,Martins:1997aa}.
On the other hand, this particular R-matrix is rather exceptional
because unlike most other known R-matrices
it cannot be written as a function
of the difference of two spectral parameters.
Altogether, the algebraic origin of the R-matrix remained mysterious.
It is well-known that it is symmetric under two undeformed $\alg{su}(2)$ algebras:
spin and (twisted) eta-pairing \cite{Lieb:1989aa,Yang:1989aa} symmetry.
Therefore one may expect the underlying algebra
to be of Yangian (algebraic)
rather than of quantum affine (trigonometric) type.
Indeed, two $\alg{su}(2)$ Yangians algebras were identified \cite{Uglov:1993jy},
but they are not sufficient to explain the R-matrix.
A fusion procedure of two XX models was used to
derive the R-matrix and explain its features,
but it seems very specialized to the model at hand
and it hardly illuminates the symmetries.

New insight into the algebraic structure came from a very different
and unexpected direction:
the field of gauge theory, string theory
and the so-called AdS/CFT correspondence,
which relates certain pairs of gauge and string theories.
In that context it was observed that
$\superN=4$ superconformal Yang--Mills theory in the 't~Hooft limit
and its dual, IIB string theory on the $AdS_5\times S^5$,
both display signs of integrability \cite{Minahan:2002ve,Beisert:2003tq,Bena:2003wd,Beisert:2003yb},
see \cite{Beisert:2004ry,Plefka:2005bk} for reviews and further references.
The asymptotic coordinate Bethe ansatz \cite{Staudacher:2004tk}
for the gauge theory spin chain leads to (two copies of)
an interesting scattering matrix \cite{Beisert:2005tm} which is not of difference form.
A construction for strings in light-cone gauge \cite{Frolov:2006cc}
leads to an equivalent S-matrix \cite{Arutyunov:2006yd}.
The scattering particles have four flavors $\state{\phi^1},\state{\phi^2},\state{\psi^1},\state{\psi^2}$,
the former two being bosonic and the latter two being fermionic.
The set of particle flavors is equivalent to the states of a site in the Hubbard model
\[
\state{\circ}\sim\state{\phi^1},\quad
\state{\mathord{\uparrow}}\sim\state{\psi^1},\quad
\state{\mathord{\downarrow}}\sim\state{\psi^2},\quad
\state{\mathord{\updownarrow}}\sim\state{\phi^2},
\]
and it was observed that the S-matrix has a structure reminiscent of
the R-matrix for the Hubbard model \cite{Staudacher:2005aa}.
Indeed, the two matrices can be mapped
into each other exactly \cite{Beisert:2006qh}
which leads to a very curious connection between
string theory and the integrable structure of the Hubbard model.
This link is also reflected
in the asymptotic Bethe equations for planar AdS/CFT \cite{Beisert:2005fw}
which contain (two copies) of the Lieb--Wu equations in disguise.

The large amount of supersymmetry present in the string/gauge theory system,
the superalgebra $\alg{psu}(2,2|4)$,
thus made its way into the integrable structure of the one-dimensional Hubbard model:
What remains of this symmetry in the above scattering picture is
(two copies of) $\alg{su}(2|2)$ \cite{Beisert:2004ry}.
A crucial point for the further understanding
was that the symmetry is centrally extended
by gauge transformations inherent to the gauge theory \cite{Beisert:2005tm}
or by residual transformations in the light cone gauge for string theory \cite{Arutyunov:2006ak}.
The symmetry of the S-matrix turns out to be an (exceptional)
threefold central extension $\alg{h}$
of the $\alg{psu}(2|2)$ superalgebra%
\footnote{For simplicity we shall consider the algebra to be complex
and do not distinguish between $\alg{psu}(2|2)$,
$\alg{psl}(2|2)$ or $\alg{psl}(2|2,\Complex)$.
Reality conditions refer to the real version $\alg{psu}(2|2)$ of the algebra.}
\[
\alg{h}:=\alg{psu}(2|2)\ltimes \Reals^3=\alg{su}(2|2)\ltimes \Reals^2.
\]
This algebra contains the two well-known bosonic $\alg{su}(2)$ symmetries
of the Hubbard model which relate the two bosonic
and two fermionic states, respectively.
However, the additional fermionic generators of the algebra
also relate the bosons to the fermions and vice versa.
In fact the algebra is strong enough
to fully constrain the form of the R-matrix \cite{Beisert:2005tm}.

The proper framework for the symmetries of the R-matrix
and thus for the integrable structure of the one-dimensional Hubbard model
is expected to be a quasi-triangular Hopf algebra \cite{Gomez:2006va,Plefka:2006ze}.
The goal is then to find the universal R-matrix
of which the above R-matrix is the fundamental representation.
However this requires to first identify the complete symmetry algebra
of the R-matrix.
Generically one may expect the algebra to be a Yangian double:
a deformation of the universal enveloping algebra of
the loop algebra of the underlying symmetry $\alg{h}$.
Indeed many of the Yangian generators
have been identified \cite{Beisert:2007ds,Matsumoto:2007rh}.
An investigation of the classical limit \cite{Torrielli:2007mc}
of the R-matrix has then revealed the complete classical structure
in terms of a quasi-triangular bialgebra \cite{Beisert:2007ty}
based on a curious deformation of the loop algebra $\alg{u}(2|2)[u,u^{-1}]$.
This result shows that in addition to the central charges,
there must also be inner automorphisms.

At least two important steps remain to be taken:
First, the classical bialgebra needs to be quantized to a Yangian double.
Second, the universal R-matrix for the Yangian needs to be established
which makes the Yangian into a quasi-triangular Hopf algebra.
However, it is not easy to deal with Yangian doubles
and their algebraic structure because proper quantization
of the higher levels is somewhat unintuitive
and specialized to the algebra $\alg{g}$.
Instead one usually considers
the corresponding quantum affine algebra $\Qgrp(\hat{\alg{g}})$
of which the Yangian is a contraction limit for $q\to 1$.
Here one pays the price that the $\alg{g}$ symmetry is
quantum deformed and not as manifest as in the Yangian.
Instead one gains a uniform treatment for the quantum deformation
of the Kac--Moody structure of the whole of the affine algebra $\hat{\alg{g}}$.
It is the aim of the present paper to lay the foundations
for the quantum deformation of the integrable structure
of the one-dimensional Hubbard model.
Here we shall start with the quantum deformation
$\Qgrp(\alg{h})$ of the algebra $\alg{h}$ and leave the full quantum
affine algebra $\Qgrp(\hat{\alg{h}})$
(or rather the deformed $\Qgrp(\alg{u}(2|2)[u,u^{-1}])$)
and its quasi-triangular structure for future work;
the corresponding Yangian double would follow as the limit $q\to 1$.
We will then derive the fundamental R-matrix which should be understood
as the quantum deformation of Shastry's R-matrix for the Hubbard model.
We apply it to derive the Bethe equations for periodic wave functions
and a three-parameter family of Hamiltonians with
$\Qgrp(\alg{su}(2)\times\alg{su}(2))$ symmetry
which includes the Hubbard Hamiltonian as a special case.

In fact, many attempts have been made to modify and generalize the Hubbard Hamiltonian
due to the exceptional properties of the Hubbard model.
Widely discussed modifications are the EKS model \cite{Essler:1992py},
the supersymmetric U-model \cite{Bracken:1995aa,Bedurftig:1995aa,Ramos:1996my,Pfannmuller:1996vp}
and its quantum deformation \cite{Maassarani:1994ac,Bariev:1995aa,Gould:2005aa}
as well as the $\alg{su}(n)$ Hubbard models
\cite{Maassarani:1997aa,Martins:1997bb,Maassarani:1997bb}
These can all be explained with the available integrability toolkit:
The EKS model is a model based on the fundamental
representation of $\alg{u}(2|2)$ and the supersymmetric U-model
is based on the four-dimensional representation of $\alg{su}(2|1)$.
These models are somewhat similar to the Hubbard model,
but they do not include it as a special case.
The $\alg{su}(n)$ Hubbard models employ an
external coloring of states which preserves integrability.
This coloring can be applied to any integrable model with conserved charges
\cite{Maassarani:1998aa} and it does not alter the underlying symmetry.
Further similar models have been discussed in
\cite{Gould:1996aa,Martins:1997ex,Montorsi:1998aa,Foerster:2000aa,Bracken:2001aa,Gohmann:2001wh}.
An important class of models which is also discussed
in this context consists of the
supersymmetric t-J model \cite{Schlottmann:1987aa,Lai:1974aa,Sutherland:1975vr,Essler:1992nk}
and some of its deformations such as the Bariev model
\cite{Bariev:1991bb,Bariev:1991aa} and others
\cite{Bariev:1994bb,Bariev:1994aa,Bariev:1995cc,Alcaraz:1998aa}.
The main difference is that these models use a three-dimensional representation
on each site and thus the Hilbert space is very different from one of the Hubbard model.
The only known true deformation of the Hubbard model
appears to be a Hamiltonian composed by Alcaraz and Bariev \cite{Alcaraz:1999aa}.
The Hamiltonian contains substantially more terms and
so far it has not been investigated further in the literature.
The Bethe equations for this model were given in \cite{Alcaraz:1999aa},
and the bear some resemblance with those for the XXZ model.
Thus it is conceivable that the Alcaraz--Bariev model is
a quantum-deformation of the Hubbard model.
We shall address the question whether we can recover
this Hamiltonian at the end of our work.
\bigskip

The present paper is organized as follows:
We start with a technical part concerning the
algebra, R-matrix and Bethe ansatz in \secref{sec:hopfalgebra,sec:rmat,sec:diagonalisation},
respectively. In the second part consisting of \secref{sec:hubbard}
we apply the obtained results to a concrete spin chain model.
It is not necessary to read the earlier sections (in full detail)
to understand the later sections.

First the $\Qgrp(\alg{h})$ symmetry
is introduced in \secref{sec:hopfalgebra}.
We also sketch finite representations of the algebra with particular focus
on the fundamental representation needed for the derivation of
the fundamental R-matrix in \secref{sec:rmat}.
We then perform the nested Bethe ansatz for this R-matrix in \secref{sec:diagonalisation}
to obtain the Bethe equations for a periodic chain.
Finally, in \secref{sec:hubbard} we derive
a class of integrable Hamiltonians associated to the R-matrix.
These constitute quantum deformations of the Hubbard Hamiltonian.
In particular, we recover one of the models proposed by Alcaraz and Bariev.
We conclude and give an outlook of
open problems in \secref{sec:conclusions}.

\section{The Hopf Algebra $\Qgrp(\alg{su}(2|2)\ltimes \Reals^2)$}\label{sec:hopfalgebra}

We start with the the construction of the symmetry
algebra underlying quantum deformations of the one-dimensional Hubbard model.
This part is rather technical in nature and can be skipped or be used as a
reference for the following sections.

\subsection{From $\alg{su}(2|2)$ to $\grp{U}(\alg{su}(2|2)\ltimes \Reals^2)$}

We start by introducing
the universal enveloping algebra $\grp{U}(\alg{su}(2|2)\ltimes \Reals^2)$
step by step starting from the Lie superalgebra $\alg{su}(2|2)$.
This will help us to understand and derive the quantum deformation
$\Qgrp(\alg{su}(2|2)\ltimes \Reals^2)$.

\paragraph{Lie Superalgebra.}

The Lie superalgebra $\alg{su}(2|2)$ is generated by the
$\alg{su}(2)\times\alg{su}(2)$ generators $\gen{R}^{a}{}_{b}$,
$\gen{L}^{\alpha}{}_{\beta}$, the supercharges
$\gen{Q}^{\alpha}{}_{b}$, $\gen{S}^{a}{}_{\beta}$ and the central
charge $\gen{C}$.%
\footnote{To obtain the simple Lie algebra $\alg{psu}(2|2)$ from $\alg{su}(2|2)$
we would have to project out this central element.}
The Lie brackets
of the $\alg{su}(2)$ generators
take the standard form
\[
\begin{array}[b]{rclcrcl}
\comm{\gen{R}^a{}_b}{\gen{R}^c{}_d}\eq
\delta^c_b\gen{R}^a{}_d
-\delta^a_d\gen{R}^c{}_b,
&&
\comm{\gen{L}^\alpha{}_\beta}{\gen{L}^\gamma{}_\delta}\eq
\delta^\gamma_\beta\gen{L}^\alpha{}_\delta
-\delta^\alpha_\delta\gen{L}^\gamma{}_\beta,
\\[4pt]
\comm{\gen{R}^a{}_b}{\gen{Q}^\gamma{}_d}\eq
-\delta^a_d\gen{Q}^\gamma{}_b
+\half \delta^a_b\gen{Q}^\gamma{}_d,
&&
\comm{\gen{L}^\alpha{}_\beta}{\gen{Q}^\gamma{}_d}\eq
\delta^\gamma_\beta\gen{Q}^\alpha{}_d
-\half \delta^\alpha_\beta\gen{Q}^\gamma{}_d,
\\[4pt]
\comm{\gen{R}^a{}_b}{\gen{S}^c{}_\delta}\eq
\delta^c_b\gen{S}^a{}_\delta
-\half \delta^a_b\gen{S}^c{}_\delta,
&&
\comm{\gen{L}^\alpha{}_\beta}{\gen{S}^c{}_\delta}\eq
-\delta^\alpha_\delta\gen{S}^c{}_\beta
+\half \delta^\alpha_\beta\gen{S}^c{}_\delta.
\end{array}
\]
The Lie brackets of two supercharges yield
\[
\acomm{\gen{Q}^\alpha{}_b}{\gen{S}^c{}_\delta} =
\delta^c_b\gen{L}^\alpha{}_\delta +\delta^\alpha_\delta\gen{R}^c{}_b
+\delta^c_b\delta^\alpha_\delta\gen{C}.
\]
The remaining Lie brackets vanish.

\paragraph{Central Extension.}

This algebra
has two further possible central extensions $\gen{P}$, $\gen{K}$.
They are generated by Lie brackets of alike supercharges
\[
\acomm{\gen{Q}^{\alpha}{}_{b}}{\gen{Q}^{\gamma}{}_{d}}
=\varepsilon^{\alpha\gamma}\varepsilon_{bd}\gen{P}, \qquad
\acomm{\gen{S}^{a}{}_{\beta}}{\gen{S}^{c}{}_{\delta}}
=\varepsilon^{ac}\varepsilon_{\beta\delta}\gen{K}.
\]
The centrally extended algebra with these
charges shall be denoted by
\[
\alg{h}:=\alg{su}(2|2)\ltimes \Reals^2
=\alg{psu}(2|2)\ltimes\Reals^3.
\]
%

\paragraph{Universal Enveloping Algebra.}

The universal enveloping algebra
$\grp{U}(\alg{h})$
of $\alg{h}$ is generated by
polynomials of the Lie algebra generators.
The Lie brackets are represented as commutators or
anti-commutators (depending on the statistics of generators)
\[
\comm{X}{Y}\to XY-YX,\qquad
\acomm{X}{Y}\to XY+YX.
\]
They respect the Lie algebra relations by
identification of certain polynomials, e.g.
\[\label{eq:PinEnv}
\gen{Q}^{\alpha}{}_{b}
\gen{Q}^{\gamma}{}_{d}
+
\gen{Q}^{\gamma}{}_{d}
\gen{Q}^{\alpha}{}_{b}
=\varepsilon^{\alpha\gamma}
\varepsilon_{bd}
\gen{P}.
\]
%

\paragraph{Chevalley Basis.}

\begin{figure}\centering
\includegraphics[width=6cm]{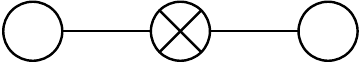}
\caption{Distinguished Dynkin diagram of $\alg{su}(2|2)$.}
\label{fig:DynkinSU22}
\end{figure}

Within the universal enveloping algebra
it is not necessary to keep all generators of the Lie algebra explicitly.
For example, the central charge $\gen{P}$ can be represented through
a quadratic combination of supercharges, see \eqref{eq:PinEnv}.
A minimal set of generators for this rank-three algebra is given by
three Cartan generators $\gen{H}_j$, three simple positive
roots $\gen{E}_j$ and three simple negative roots $\gen{F}_j$,
$j=1,2,3$.
One may identify them with the Lie generators as follows
\[\label{eq:chevalleyident}
\begin{array}[b]{rclcrclcrcl}
\gen{H}_1\eq\gen{R}^2{}_2-\gen{R}^1{}_1 = 2\gen{R}^2{}_2 ,&&
\gen{E}_1\eq\gen{R}^2{}_1,&&
\gen{F}_1\eq\gen{R}^1{}_2,
\\
\gen{H}_2\eq-\gen{C}-\half\gen{H}_1-\half\gen{H}_3,&&
\gen{E}_2\eq\gen{Q}^2{}_2,&&
\gen{F}_2\eq\gen{S}^2{}_2,
\\
\gen{H}_3\eq\gen{L}^2{}_2-\gen{L}^1{}_1 = 2\gen{L}^2{}_2 ,&&
\gen{E}_3\eq\gen{L}^1{}_2,&&
\gen{F}_3\eq\gen{L}^2{}_1.
\end{array}\]
This basis corresponds to the distinguished Dynkin diagram
of $\alg{su}(2|2)$ in \figref{fig:DynkinSU22}.
The symmetric Cartan matrix in this basis reads
\[
A_{jk}=\matr{rrr}{+2&-1&0\\-1&0&+1\\0&+1&-2}.
\]
Note that the algebra $\alg{su}(2|2)$ has a degenerate Cartan matrix
and therefore has a null vector
which reads $v_j=(1,2,1)$.
It can be used to express the
central charge of $\alg{su}(2|2)$
\[\label{eq:CChevalley}
\gen{C}=-\frac{1}{2} \sum_{j=1}^3v_j\gen{H}_j=-\half\gen{H}_1-\gen{H}_2-\half\gen{H}_3.
\]
The other two central charges in the Chevalley basis take the form
\[\label{eq:PKChevalley}
\gen{P}=
\bigacomm{\comm{\gen{E}_1}{\gen{E}_2}}{\comm{\gen{E}_3}{\gen{E}_2}},
\qquad
\gen{K}=
\bigacomm{\comm{\gen{F}_1}{\gen{F}_2}}{\comm{\gen{F}_3}{\gen{F}_2}}.
\]
%

\paragraph{Commutation Relations.}

Let us now state the commutation relations of the universal enveloping algebra.
Commutators with the Cartan generators $\gen{H}_j$ are given by
($j,k=1,2,3$)
\[
\comm{\gen{H}_j}{\gen{H}_k}=0,
\qquad
\comm{\gen{H}_j}{\gen{E}_k}=+A_{jk}\gen{E}_k,
\qquad
\comm{\gen{H}_j}{\gen{F}_k}=-A_{jk}\gen{F}_k.
\]
The non-trivial commutators of positive and negative simple roots
read
\[
\comm{\gen{E}_1}{\gen{F}_1}= +\gen{H}_1,
\qquad
\acomm{\gen{E}_2}{\gen{F}_2}= -\gen{H}_2,
\qquad
\comm{\gen{E}_3}{\gen{F}_3}= -\gen{H}_3.
\]
Note that we have normalized the generators in a way such
that all relations can be expressed using the symmetric
Cartan matrix $A_{jk}$. For superalgebras this leads to a
negative sign in $\comm{\gen{E}_j}{\gen{F}_j}$
for one of the two bosonic subalgebras.
The remaining commutators between positive and negative simple roots
vanish in this basis
\[\comm{\gen{E}_j}{\gen{F}_k}=0 \quad\mbox{for }j\neq k.\]
Finally, we need to impose the Serre relations between
positive and between negative simple roots.
\<\label{eq:ClassSerre}
0\eq
\comm{\gen{E}_1}{\gen{E}_3}
=\gen{E}_2\gen{E}_2
=\bigcomm{\gen{E}_1}{\comm{\gen{E}_1}{\gen{E}_2}}
=\bigcomm{\gen{E}_3}{\comm{\gen{E}_3}{\gen{E}_2}}
\nln\eq\comm{\gen{F}_1}{\gen{F}_3}
=\gen{F}_2\gen{F}_2
=\bigcomm{\gen{F}_1}{\comm{\gen{F}_1}{\gen{F}_2}}
=\bigcomm{\gen{F}_3}{\comm{\gen{F}_3}{\gen{F}_2}}.
\>

Note that for superalgebras these standard Serre relations are not sufficient.
For the algebra $\grp{U}(\alg{su}(2|2))$ we need
two additional Serre relations which effectively read
\[
0=\gen{P}=\gen{K}.
\]
It is however consistent to drop them altogether which leads to
the centrally extended algebra $\grp{U}(\alg{h})$.

\subsection{Quantum Deformation}

The quantum algebra $\Qgrp(\alg{g})$ is a deformation
of the universal enveloping algebra $\grp{U}(\alg{g})$
of some Lie algebra $\alg{g}$. It is obtained by inserting
various factors and exponents of $q$ in various relations.
A convenient combination to use
in this context is the so-called quantum number
\[
\qnum{x}:=\frac{q^{x}-q^{-x}}{q-q^{-1}}\,.
\]
This relation is defined in the same way if $x$ is a generator.
There are two ways to achieve this:
Either one defines $q^x=1+x\log q+\half x^2\log^2 q+\ldots$
as a formal power series.
For practical purposes one would assume that $q\approx 1$ and thus
$\log q\approx 0$.
Alternatively one can define $q^x$ as an element of the Hopf algebra
and $q^{-x}$ as its inverse.
This is particularly useful if $x$
appears only with integer multiplicity in exponents, $q^{nx}=(q^x)^n$,
as will be the case here.

In the previous section we have seen that the
two central charges can be obtained by dropping
two Serre relations.
It is straightforward to apply the same central extension for
the quantum deformed algebra $\Qgrp(\alg{su}(2|2))$
to obtain $\Qgrp(\alg{h})$.

\paragraph{Commutation Relations.}

Let us begin with the deformation of the
commutation relations of the algebra $\Qgrp(\alg{su}(2|2))$.
The commutators with Cartan elements do not receive
deformations
\[
\comm{\gen{H}_j}{\gen{H}_k}=0,
\qquad
\comm{\gen{H}_j}{\gen{E}_k}=+A_{jk}\gen{E}_k,
\qquad
\comm{\gen{H}_j}{\gen{F}_k}=-A_{jk}\gen{F}_k.
\]
The Cartan elements usually appear in exponents,
and it is convenient to note the exponentiated form
of these relations
\[
q^{\gen{H}_j}\gen{E}_k=q^{+A_{jk}}\gen{E}_k q^{\gen{H}_j},
\qquad
q^{\gen{H}_j}\gen{F}_k=q^{-A_{jk}}\gen{F}_k q^{\gen{H}_j},
\]
The non-trivial commutators of simple roots read
in the deformed algebra
\[
\comm{\gen{E}_1}{\gen{F}_1}=\qnum{\gen{H}_1}\,,
\qquad
\acomm{\gen{E}_2}{\gen{F}_2}=-\qnum{\gen{H}_2}\,,
\qquad
\comm{\gen{E}_3}{\gen{F}_3}=-\qnum{\gen{H}_3}\,,
\]
and the remaining mixed commutators vanish
\[
\comm{\gen{E}_j}{\gen{F}_k}=0
\quad\mbox{for }j\neq k.
\]
The Serre relations have the same form as in the undeformed algebra
\eqref{eq:ClassSerre}, but with some additional factors
of $q$ due to the quantum-adjoint action.
Spelled out they yield
\< 0\eq
\comm{\gen{E}_1}{\gen{E}_3}
=\comm{\gen{F}_1}{\gen{F}_3} =\gen{E}_2\gen{E}_2
=\gen{F}_2\gen{F}_2
\\
\eq\gen{E}_1\gen{E}_1\gen{E}_2
-(q+q^{-1})\gen{E}_1\gen{E}_2\gen{E}_1
+\gen{E}_2\gen{E}_1\gen{E}_1
=\gen{E}_3\gen{E}_3\gen{E}_2
-(q+q^{-1})\gen{E}_3\gen{E}_2\gen{E}_3
+\gen{E}_2\gen{E}_3\gen{E}_3
\nln\nonumber
\eq\gen{F}_1\gen{F}_1\gen{F}_2
-(q+q^{-1})\gen{F}_1\gen{F}_2\gen{F}_1
+\gen{F}_2\gen{F}_1\gen{F}_1
=\gen{F}_3\gen{F}_3\gen{F}_2
-(q+q^{-1})\gen{F}_3\gen{F}_2\gen{F}_3
+\gen{F}_2\gen{F}_3\gen{F}_3
.
\>
%

\paragraph{Central Elements.}

The standard central element from the Cartan subalgebra remains
undeformed
\[\label{eq:quantumC}
\gen{C}=-\half\gen{H}_1-\gen{H}_2-\half\gen{H}_3.
\]
As before, the ordinary Serre relations obtained from the
Cartan matrix are not sufficient to
define $\Qgrp(\alg{su}(2|2))$, but we need the constraint
$\gen{P}=\gen{K}=0$ with the
quantum deformation of \eqref{eq:PKChevalley}
\<
\label{eq:quantumPK}
\gen{P}\eq \gen{E}_1\gen{E}_2\gen{E}_3\gen{E}_2
+\gen{E}_2\gen{E}_3\gen{E}_2\gen{E}_1
+\gen{E}_3\gen{E}_2\gen{E}_1\gen{E}_2
+\gen{E}_2\gen{E}_1\gen{E}_2\gen{E}_3
-(q+q^{-1})\gen{E}_2\gen{E}_1\gen{E}_3\gen{E}_2,
\nln
\gen{K}\eq \gen{F}_1\gen{F}_2\gen{F}_3\gen{F}_2
+\gen{F}_2\gen{F}_3\gen{F}_2\gen{F}_1
+\gen{F}_3\gen{F}_2\gen{F}_1\gen{F}_2
+\gen{F}_2\gen{F}_1\gen{F}_2\gen{F}_3
-(q+q^{-1})\gen{F}_2\gen{F}_1\gen{F}_3\gen{F}_2.
\>
In the centrally extended algebra $\Qgrp(\alg{h})$ we
will not impose the constraint and thus obtain
two non-trivial central elements $\gen{P},\gen{K}$.
It is straightforward, if tedious, to confirm that these polynomials
are indeed central elements of the quantum-deformed algebra.

\subsection{Representations}

It is commonly the case for a finite-dimensional simple Lie algebra
$\alg{g}$ and for generic values of $q$
that the representation theory of
the quantum deformed algebra $\Qgrp(\alg{g})$
is analogous to the one of $\alg{g}$.%
\footnote{In quantum algebras
one singles out the case when $q$ is a root of unity because the
representation theory is very special at these points.}
We have studied some of the simplest finite-dimensional representations of
$\Qgrp(\alg{h})$ and found agreement with this rule.
Here we would like to give an overview over some basic representations of
$\Qgrp(\alg{h})$ assuming that there is no qualitative
difference to the undeformed case.

\paragraph{Outer Automorphism.}

Finite representations of $\alg{h}$
were studied in \cite{Beisert:2006qh}.
The starting point was the representation theory
of $\alg{su}(2|2)$ with fixed eigenvalue $C_0$
of the central charge $\gen{C}$,
studied in, e.g.\ \cite{Kamupingene:1989wj,Palev:1990wm,Zhang:2004qx}.
The $\grp{SL}(2)$ outer automorphism of $\gen{h}$ then allows
to rotate the triplet of eigenvalues $(C_0,0,0)$
of the central charges $(\gen{C},\gen{P},\gen{K})$
to any desired triplet $(C,P,K)$ with
$C_0^2=C^2-PK$.%
\footnote{For simplicity we shall consider the algebra $\gen{h}$ to be complex,
and thus there is no distinction between positive and negative values
of $C^2-PK$.}
The representation of the $\alg{psu}(2|2)$ generators $\gen{Q},\gen{S}$
is obtained by conjugation with the $\grp{SL}(2)$ element.
Note that the combination $\vec{\gen{C}}^2=\gen{C}^2-\gen{P}\gen{K}$ is
invariant under the automorphism.

It appears that for the
quantum-deformed algebra $\Qgrp(\alg{h})$
there also exists a similar outer automorphism.
We could use it to relate representations
of $\Qgrp(\alg{h})$ to those of $\Qgrp(\alg{su}(2|2))$,
see \cite{Ky:1994we,Ky:1994cr},
which in turn are analogous to those of $\alg{su}(2|2)$.
We however do not yet understand the automorphism explicitly
and therefore the existence of the below representations
is an educated guess. The above combination $\vec{\gen{C}}^2$
should be quantum-deformed to some $\qnum{\vec{\gen{C}}^2}$.
The classification of representations
would then use the eigenvalues of this operator.
We find that the smallest representations are indeed characterized
by the eigenvalues of the operator
\[
\label{eq:InvariantOperator}
\qnum{\vec{\gen{C}}^2}:=\qnum{\gen{C}}^2-\gen{P}\gen{K}.
\]
Presumably this combination is invariant under the tentative automorphism.

Let us consider typical (long)
and atypical (short) representations of $\Qgrp(\alg{h})$
in analogy to the $\alg{su}(2|2)$ representations studied in \cite{Kamupingene:1989wj,Palev:1990wm}.

\paragraph{Long Multiplets.}

The standard finite-dimensional type of representation
shall be denoted by
\[
\lrep{m,n;C,P,K}.
\]
It corresponds to the typical highest-weight representations of $\alg{su}(2|2)$
with Dynkin labels $[m;r;n]$ and $r=\pm (\qnum{\vec{C}^2})^{1/2}-\half n + \half m$.
The non-negative integers $n,m$ represent the Dynkin labels of
the $\alg{su}(2)\times\alg{su}(2)$ subalgebra.

We can decompose a $\Qgrp(\alg{h})$ multiplet into irreducible multiplets
of the subalgebra $\Qgrp(\alg{su}(2)\times\alg{su}(2))$.
Let the symbol $[k]$ represent the $\Qgrp(\alg{su}(2))$
representation with spin $k/2$;
$[-1]$ is the empty (zero-dimensional) representation.
Then the long multiplet decomposes as follows
\<\label{eq:DecoLong}
\lrep{m,n}\earel{\to}
([m]\otimes[1]\otimes[1],[n])
\oplus
([m],[n]\otimes[1]\otimes[1])
\nl
\oplus
([m]\otimes[1],[n]\otimes[1])
\oplus
([m]\otimes[1],[n]\otimes[1]).
\>
Note the well-known tensor product of $\Qgrp(\alg{su}(2))$  representations
\[
[m]\otimes[n]=\bigoplus_{k=0}^{\min(m,n)}[m+n-2k].
\]
The dimension of a long multiplet is thus $16{(m+1)}{(n+1)}$.

\paragraph{Short Multiplets.}

A short multiplet shall be labelled by
\[
\srep{m,n;C,P,K}.
\]
It can only exist when
the constraint
\[\label{eq:short}
\qnum{\vec{C}^2}=\qnum{C}^2-PK=\qnum{\half(m+n+1)}^2
\]
holds.
The decomposition into irreducible representations of
$\Qgrp(\alg{su}(2)\times\alg{su}(2))$ takes the form
\<\label{eq:DecoShort}
\srep{m,n}\earel{\to}
([m]\otimes [1],[n])
\oplus
([m-1],[n-1]\otimes [1])
\nl
\oplus
([m],[n]\otimes [1])
\oplus
([m-1]\otimes [1],[n-1]).
\>
This multiplet has dimension $4(m+1)(n+1)+4mn$. For example,
the four-dimensional fundamental representation is the special case $m=n=0$
which decomposes into $\srep{0,0}\to ([1],[0])\oplus ([0],[1])$.

\paragraph{Multiplet Splitting.}

\begin{figure}\centering
\includegraphics[width=3.5cm]{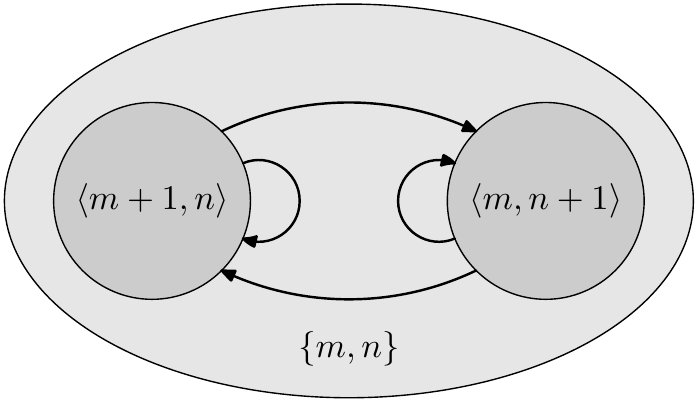}\quad
\includegraphics[width=3.5cm]{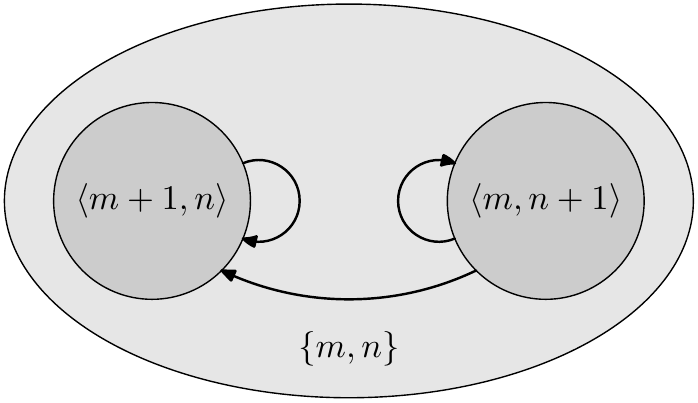}\quad
\includegraphics[width=3.5cm]{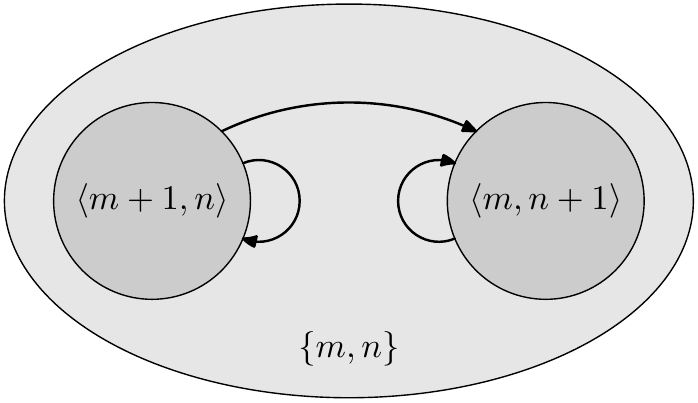}\quad
\includegraphics[width=3.5cm]{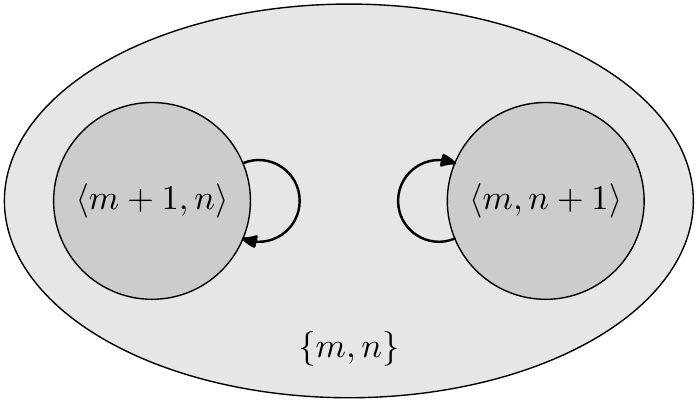}
\caption{Multiplet Splitting.
The long multiplet $\lrep{m,n}$ consists
of two short multiplets $\srep{m+1,n}$ and $\srep{m,n+1}$
and a representation of $\Qgrp(\alg{h})$
maps between the short multiplets (arrows).
Generically, the short multiplets are connected in all possible ways (left).
When the shortening condition holds one (middle) or both (right)
arrows between the short multiplets are broken.
In the second figure from the left, $\srep{m+1,n}$ is a subrepresentation
while $\srep{m,n+1}$ is a factor representation,
and the long multiplet is indecomposable.
The long multiplet in the right figure is fully decomposable.}
\label{fig:Splitting}
\end{figure}

The vector space of a long multiplet $\lrep{m,n;C,P,K}$
can be decomposed into the vector spaces of two
short multiplets
\[\label{eq:split}
\lrep{m,n;\vec{C}}\to
\srep{m+1,n;\vec{C}}\oplus
\srep{m,n+1;\vec{C}}.
\]
However, the typical representation of $\Qgrp(\alg{h})$
is in general irreducible.
A special case is when the labels and central charges
obey the shortening condition
\[\label{eq:shortening}
\qnum{\vec{C}^2}
=\qnum{\half(m+n+2)}^2,
\]
which is equivalent to the
constraint \eqref{eq:short}
for the short multiplets in \eqref{eq:split}.
The long representation may then be reduced into the
above two short representations.
However, in general one cannot expect the long representation to be decomposable,
it merely has one short subrepresentation
which closes on one of the smaller vector spaces.
When projecting out this small vector space
one obtains a short factor representation on the other small
vector space.
This fact is perhaps best illustrated by a figure, see \figref{fig:Splitting}.

\paragraph{Tensor Products.}

Let us state some formulas for the
decomposition of tensor products
of long and short representations.
It is convenient to introduce some short hand notation
for sums of similar representations
\[
\lrep{[k]\oplus[l],n}:=\lrep{k,n}\oplus\lrep{l,n},\quad \mbox{etc.}.
\]
The tensor product of two long representations of $\Qgrp(\alg{h})$
decomposes as follows
\<\label{eq:longlong}
\lrep{m,n}\otimes\lrep{k,l}\eq
\lrep{[m]\otimes[k]\otimes[1]\otimes[1],[n]\otimes[l]}
\nl \oplus
\lrep{[m]\otimes[k]\otimes[1],[n]\otimes[l]\otimes[1]}
\nl \oplus
\lrep{[m]\otimes[k]\otimes[1],[n]\otimes[l]\otimes[1]}
\nl \oplus
\lrep{[m]\otimes[k],[n]\otimes[l]\otimes[1]\otimes[1]}.
\>
For a tensor product of a long and a short representation the following
decomposition applies
\<\label{eq:longshort}
\lrep{m,n}\otimes\srep{k,l}\eq
\lrep{[m]\otimes[k]\otimes[1],[n]\otimes[l]}
\nl \oplus
\lrep{[m]\otimes[k-1]\otimes[1],[n]\otimes[l-1]}
\nl \oplus
\lrep{[m]\otimes[k],[n]\otimes[l]\otimes[1]}
\nl \oplus
\lrep{[m]\otimes[k-1],[n]\otimes[l-1]\otimes[1]}.
\>
Finally, the tensor product of two short representations reads
\<\label{eq:shortshort}
\srep{m,n}\otimes\srep{k,l}\eq
\lrep{[m]\otimes[k],[n]\otimes[l]}
\nl \oplus
\lrep{[m]\otimes[k-1],[n]\otimes[l-1]}
\nl \oplus
\lrep{[m-1]\otimes[k],[n-1]\otimes[l]}
\nl \oplus
\lrep{[m-1]\otimes[k-1],[n-1]\otimes[l-1]}.
\>
These formulas agree with the decompositions
\eqref{eq:DecoLong,eq:DecoShort}.

\subsection{Hopf Algebra}

Here we will complete the Hopf algebra structure of
our quantum deformed algebra. We first define
the coproduct and then state the
unit, counit and antipode.

\paragraph{Coproduct.}

We shall use the standard coproduct for the generators of $\Qgrp(\alg{su}(2|2))$
\<\label{eq:qcoproduct}
\copro(1)\eq 1\otimes 1,\nln
\copro(\gen{H}_j)\eq \gen{H}_j\otimes 1+1\otimes \gen{H}_j, \nln
\copro(\gen{E}_j)\eq \gen{E}_j\otimes 1+q^{-\gen{H}_j}\otimes \gen{E}_j, \nln
\copro(\gen{F}_j)\eq \gen{F}_j\otimes q^{\gen{H}_j}+1\otimes \gen{F}_j,\>
where $1$ denotes the unit element of the Hopf algebra.
Note that the relations are fully compatible with
the Serre relations and commutators.
The coproducts for the central charges follow
by substituting their expressions in terms of
the above generators
\<
\copro(\gen{C})\eq \gen{C}\otimes 1+1\otimes \gen{C}, \nln
\copro(\gen{P})\eq \gen{P}\otimes 1+q^{2\gen{C}}\otimes \gen{P}, \nln
\copro(\gen{K})\eq \gen{K}\otimes q^{-2\gen{C}}+1\otimes \gen{K}.
\>
Note that the coproduct for the generators $\gen{P},\gen{K}$
turns out to be proportional to the generators themselves as expected.
This fact is related to the the role of
$\gen{P}=\gen{K}=0$ as Serre relations
of $\Qgrp(\alg{su}(2|2))$.
The coproduct of such consistent identifications has to be
proportional to these or other identifications.

\paragraph{Unit, Counit and Antipode.}

The unit element of the Hopf algebra is denoted by $\unit(1)=1$.
The counit $\counit:\Qgrp(\alg{h})\to\Complex$ takes the usual form
\[
\counit(1)=1,\qquad
\counit(\gen{H}_j)=\counit(\gen{E}_j)=\counit(\gen{F}_j)=0.
\]
Finally, the antipode $\antipode:\Qgrp(\alg{h})\to\Qgrp(\alg{h})$
is uniquely fixed by the compatibility condition
\[\label{eq:HopfAntipode}
\mu\circ(\antipode\otimes1)\circ\copro(\gen{J})
=\mu\circ(1\otimes \antipode)\circ\copro(\gen{J})
=\unit\circ\counit(\gen{J}),
\]
for all $\gen{J}\in\Qgrp(\alg{h})$,
where $\unit:\Complex\to\Qgrp(\alg{h})$ denotes the unit
and $\mu:\Qgrp(\alg{h})\otimes \Qgrp(\alg{h})\to \Qgrp(\alg{h})$
denotes the product.
One gets
\[\label{eq:Antipodes}
\antipode(1)=1,\qquad
\antipode(\gen{H}_j)=-\gen{H}_j, \qquad
\antipode(\gen{E}_j)=-q^{\gen{H}_j}\gen{E}_j, \qquad
\antipode(\gen{F}_j)=-\gen{F}_jq^{-\gen{H}_j},
\]
and similarly for the central charges
\[
\antipode(\gen{C})=-\gen{C},\qquad
\antipode(\gen{P})=-q^{-2\gen{C}}\gen{P},\qquad
\antipode(\gen{K})=-q^{2\gen{C}}\gen{K}.
\]
%

\paragraph{Hermitian Conjugation.}

So far we have for simplicity assumed a complex algebra.
To restrict to a real algebra we must identify the generators
with their conjugates.
The proper hermiticity relations compatible with the coproduct are
\[\label{eq:hermi}
\gen{H}_j^\dagger=\gen{H}_j,\qquad
\gen{E}_j^\dagger=q^{-\gen{H}_j}\gen{F}_j.
\]
This implies the following relations for the central charges
\[
\gen{C}^\dagger=\gen{C},
\qquad
\gen{P}^\dagger=q^{2\gen{C}}\gen{K}.
\]
To have a hermitian coproduct we furthermore have to constrain $q$ to be real.

\paragraph{Braiding.}

Just as for the undeformed algebra $\grp{U}(\alg{h})$ we shall
deform the coproduct slightly and according to a $\Integers$-grading
of the algebra. This grading associates the charges
$+2,+1,-1,-2$ to the generators $\gen{P},\gen{E}_2,\gen{F}_2,\gen{K}$,
respectively; the other generators are uncharged.
This braiding will later lead to a very non-trivial R-matrix.

For the braiding we introduce a new abelian generator $\gen{U}$
and deform the coproduct \eqref{eq:qcoproduct} as follows
\<\label{eq:bradedcopr}
\copro(\gen{E}_2)\eq
\gen{E}_2\otimes 1+q^{-\gen{H}_2}\gen{U}\otimes \gen{E}_2,
\nln
\copro(\gen{F}_2)\eq
\gen{F}_2\otimes q^{\gen{H}_2}+\gen{U}^{-1}\otimes \gen{F}_2,
\nln
\copro(\gen{P})\eq\gen{P}\otimes 1+q^{2\gen{C}}\gen{U}^2\otimes \gen{P},
\nln
\copro(\gen{K})\eq\gen{K}\otimes q^{-2\gen{C}}+\gen{U}^{-2}\otimes \gen{K},
\nln
\copro(\gen{U})\eq\gen{U}\otimes\gen{U}.
\>
The coproduct of the other elements remains undeformed.
This deformation of the coproduct is consistent with the algebra
because the exponents of $\gen{U}$ follow the $\Integers$-grading of the algebra.
The counit for $\gen{U}$ should be defined as $\counit(\gen{U})=1$.
The antipode must then obey the relation \eqref{eq:HopfAntipode}
which leads to the following modifications of \eqref{eq:Antipodes}
\[
\antipode(\gen{E}_2)=-q^{\gen{H}_2}\gen{U}^{-1}\gen{E}_2, \qquad
\antipode(\gen{F}_2)=-q^{-\gen{H}_2}\gen{U}\gen{F}_2, \qquad
\antipode(\gen{U})=\gen{U}^{-1}
\]
and for the central charges
\[
\antipode(\gen{C})=-\gen{C},\qquad
\antipode(\gen{P})=-q^{-2\gen{C}}\gen{U}^{-2}\gen{P},\qquad
\antipode(\gen{K})=-q^{2\gen{C}}\gen{U}^2\gen{K}.
\]
The hermitian conjugation remains untouched by the braiding
and
\[
\gen{U}^\dagger=\gen{U}^{-1}.
\]
%

\paragraph{Cocommutativity of the Center.}

A Hopf algebra is called quasi-cocommutative
if the coproduct $\copro(\gen{J})$ and the opposite
coproduct $\coproop(\gen{J}):=\perm(\copro(\gen{J}))$
are related by conjugation.
Here $\perm$ defines the graded permutation operator
for the tensor product.
The conjugation is specified by an R-matrix
$\rmat\in \Qgrp(\alg{h})\otimes\Qgrp(\alg{h})$
satisfying the cocommutativity condition%
\[\label{eq:cocorel}
\coproop(\gen{J})\rmat=
\rmat\copro(\gen{J}).
\]

For generators $\gen{J}$ from the center of $\Qgrp(\alg{h})$
this relation implies that the coproduct must cocommute
$\coproop(\gen{J})=\copro(\gen{J})$.
With our coproduct this is obvious for the central charge $\gen{C}$
\[
\copro(\gen{C})=
\gen{C}\otimes 1+1\otimes\gen{C}
=\coproop(\gen{C}).
\]
The other two central charges $\gen{P}$ and $\gen{K}$ however do
not enjoy this property
\[\begin{array}{rcl}
\copro(\gen{P})\eq\gen{P}\otimes 1+q^{2\gen{C}}\gen{U}^2\otimes \gen{P},
\\[3pt]
\copro(\gen{K})\eq\gen{K}\otimes q^{-2\gen{C}}+\gen{U}^{-2}\otimes \gen{K},
\end{array}
\quad\mbox{but}\quad
\begin{array}{rcl}
\coproop(\gen{P})\eq\gen{P}\otimes \gen{U}^2q^{2\gen{C}}+1\otimes \gen{P},
\\[3pt]
\coproop(\gen{K})\eq\gen{K}\otimes \gen{U}^{-2}+q^{-2\gen{C}}\otimes \gen{K}.
\end{array}
\]
By taking the difference of the coproducts
one can see that the central charges $\gen{P},\gen{K}$
have to be identified with $\gen{C}$ and $\gen{U}$ as follows
\[\label{eq:PKident}
\gen{P}=g\alpha(1-q^{2\gen{C}}\gen{U}^2),\qquad
\gen{K}=g\alpha^{-1}(q^{-2\gen{C}}-\gen{U}^{-2}).
\]
Here $g$ and $\alpha$ are two global constants of the reduced algebra.

An alternative derivation of the braiding and identifications
based on a consistent factorised scattering picture
is presented in \appref{sec:scatterbraid}.

\subsection{Fundamental Representation}

The fundamental representation is the short multiplet
\[
\srep{C,P,K}=\srep{0,0;C,P,K} .
\]
The vector space is spanned by four states
\[
\state{\phi^1},\state{\phi^2},\state{\psi^1},\state{\psi^2},
\]
where $\state{\phi^a}$ are bosonic and $\state{\psi^\alpha}$ are fermionic.
The fundamental action of the Chevalley-Serre generators is given by
\[\begin{array}{rclrclrclrcl}
\gen{H}_1\state{\phi^1}\eq -\state{\phi^1},&
\gen{H}_2\state{\phi^1}\eq -(C-\half)\state{\phi^1},&
\gen{E}_1\state{\phi^1}\eq q^{+1/2}\state{\phi^2},&
\gen{F}_2\state{\phi^1}\eq c\state{\psi^1}, \\[3pt]
\gen{H}_1\state{\phi^2}\eq +\state{\phi^2},&
\gen{H}_2\state{\phi^2}\eq -(C+\half)\state{\phi^2},&
\gen{E}_2\state{\phi^2}\eq a\state{\psi^2},&
\gen{F}_1\state{\phi^2}\eq q^{-1/2}\state{\phi^1}, \\[3pt]
\gen{H}_3\state{\psi^2}\eq +\state{\psi^2},&
\gen{H}_2\state{\psi^2}\eq -(C+\half)\state{\psi^2},&
\gen{E}_3\state{\psi^2}\eq q^{-1/2}\state{\psi^1},&
\gen{F}_2\state{\psi^2}\eq d\state{\phi^2}, \\[3pt]
\gen{H}_3\state{\psi^1}\eq -\state{\psi^1},&
\gen{H}_2\state{\psi^1}\eq -(C-\half)\state{\psi^1},&
\gen{E}_2\state{\psi^1}\eq b\state{\phi^1},&
\gen{F}_3\state{\psi^1}\eq q^{+1/2}\state{\psi^2}.
\end{array}
\]
The braiding generator $\gen{U}$ will act
with the same eigenvalue $U$ on all four states.

\paragraph{Constraints.}

It is not too hard to see that the closure of the
algebra of supercharges leads to the constraints
\[\label{eq:constraint}
ad=\qnum{C+\half},\qquad
bc=\qnum{C-\half},\qquad
ab=P,\qquad
cd=K.
\]
This is in agreement with the constraint \eqref{eq:short} which is required
for the fundamental representation
\[
\qnum{\vec{C}^2}=\qnum{\half}^2.
\]
It furthermore follows that the constraint $ad-bc=1$ for $q=1$ is
deformed to
\[\label{eq:abcdq}
(ad-qbc)(ad-q^{-1}bc)=1.
\]

Note that there is a subtlety in taking the limit $q\to 1$
in the above constraint \eqref{eq:abcdq}:
It leads to two inequivalent constraints $ad-bc=\pm 1$
which correspond to different embeddings of the Lie algebra into
the Chevalley basis.
The plus sign yields the above constraint
with the identifications in \eqref{eq:chevalleyident} or
equivalently $\gen{E}_2=-\gen{Q}^2{}_2$, $\gen{F}_2=-\gen{S}^2{}_2$.
The minus sign corresponds to one of the following embeddings
\[
\gen{E}_2=\mp\gen{Q}^2{}_2,
\qquad
\gen{F}_2=\pm\gen{S}^2{}_2,
\qquad
\gen{H}_2=\gen{C}+\half\gen{H}_1+\half\gen{H}_3.
\]
%

\paragraph{$\xpm{}$ Parameters.}

For later convenience we shall introduce the new representation parameters
$\xpm{},\gamma$ and rewrite $a,b,c,d$ as follows
\<\label{eq:xpmparmeters}
a\eq\sqrt{g}\,\gamma,
\nln
b\eq\frac{\sqrt{g}\,\alpha}{\gamma}\,\frac{1}{\xm{}}\,\bigbrk{\xm{}-q^{2C-1}\xp{}},
\nln
c\eq \frac{i\sqrt{g}\gamma}{\alpha}\,\frac{q^{-C+1/2}}{\xp{}}\,,
\nln
d\eq \frac{i\sqrt{g}}{\gamma}\,q^{C+1/2}\bigbrk{\xm{} -q^{-2C-1}\xp{}}.
\>
In terms of these parameters the constraint \eqref{eq:constraint}
implies the following quadratic relation between $\xpm{}$
\[\label{eq:xpmrel}
\frac{\xp{}}{q}+\frac{q}{\xp{}}-q\xm{}-\frac{1}{q\xm{}}
+ig(q-q^{-1})\lrbrk{\frac{\xp{}}{q\xm{}}-\frac{q\xm{}}{\xp{}}}=\frac{i}{g}\,.
\]
The central charge $C$ cannot be written unambiguously using $\xpm{}$,
but the combination $q^{2C}$ is well-defined
\[\label{eq:q2c}
q^{2C}=q\,\frac{(q-q^{-1})/\xp{}-ig^{-1}}{(q-q^{-1})/\xm{}-ig^{-1}}
=q^{-1}\frac{(q-q^{-1})\xp{}+ig^{-1}}{(q-q^{-1})\xm{}+ig^{-1}}\,.\]
These two expressions are equivalent upon \eqref{eq:xpmrel}.
Finally, the central charges $P,K$ read
\[
P=g\alpha\lrbrk{1-q^{2C}\frac{\xp{}}{q\xm{}}},
\qquad
K=
g\alpha^{-1}
\lrbrk{q^{-2C}-\frac{q\xm{}}{\xp{}}}.
\]
With the above identification \eqref{eq:PKident}
of $\gen{P}$ and $\gen{K}$ with $\gen{U}$,
the squared eigenvalue of the latter on the fundamental representation reads
\[\label{eq:U2}
U^2=\frac{\xp{}}{q\xm{}}\,.
\]
%

\paragraph{Fermion Normalization.}

The parameter $\gamma$ adjusts the normalization
of fermions $\state{\psi^\alpha}$ with respect
to bosons $\state{\phi^a}$.
Furthermore the parameter $\alpha$ adjusts the
normalization of $\gen{E}_2,\gen{P}$ vs.\ $\gen{F}_2,\gen{K}$.
A particularly useful choice for $\gamma$ will turn out to be
\[\label{eq:gammafix}
\gamma=\frac{\sqrt{-i\alpha q^{C+1/2}U(\xp{}-\xm{})}}{\sqrt[\scriptstyle4]{{1-(q-q^{-1})^2g^2}}}\,.
\]
Apart from the features discussed below,
we expect it to have nice analytic properties
analogous to the case of the undeformed case
discussed in \cite{Arutyunov:2007tc}.
Despite some simplifications
we will largely keep $\alpha$ and $\gamma$ unspecified in this paper.

\paragraph{Uniformization.}

For a fixed parameter $g$ the constraint \eqref{eq:xpmrel}
defines a complex torus with modulus
\[\label{eq:TorusModulus}
k=4ig\sqrt{1-g^2(q-q^{-1})^2}.
\]
The quotient of the parameters $\xpm{}$ can be expressed conveniently through
the point $z$ on a doubly-periodic complex plane using
\[
\frac{\xp{}}{q\xm{}}=\frac{\ellCN(z)+ir\ellSN(z)}{\ellCN(z)-ir\ellSN(z)}\,,
\]
where the constant $r$ is given by
\[
r=\sqrt{\frac{1+16g^2-16g^4(q-q^{-1})^2}{1+4g^2(q^{1/2}+q^{-1/2})^2}}\,.
\]
The individual parameters $\xpm{}$ are then written as rational functions
of $\ellSN(z),\ellCN(z),\ellDN(z)$, but we shall refrain from
stating these functions explicitly here.

\paragraph{Transposition.}

The antipode map corresponds to
transposing the representation matrices
\[
\antipode(\gen{J}) \simeq \conj^{-1}\bar{\gen{J}}^{\transpose}\conj,
\]
where the supertransposition is
defined by $A_{jk}^{\transpose} := (-1)^{(|j|+1)|k|}A_{kj}$
and $\conj$ is the charge conjugation matrix
\[\begin{array}{rclcrcl}
\conj\state{\phi^1} \eq -q^{+1/2}\state{\phi^2},&\quad&
\conj\state{\psi^1} \eq -q^{+1/2}\state{\psi^2},\\[3pt]
\conj\state{\phi^2} \eq +q^{-1/2}\state{\phi^1},&\quad&
\conj\state{\psi^2} \eq +q^{-1/2}\state{\psi^1}.
\end{array}
\]
The conjugation matrix was constructed such that the equation is satisfied for
the bosonic Chevalley-Serre generators $\gen{E}_1,\gen{E}_3,\gen{F}_1,\gen{F}_3$.
By solving this equation for the fermionic generators $\gen{E}_2,\gen{F}_2$,
we obtain the antipode representation parameters
$\bar{a},\bar{b},\bar{c},\bar{d}$ in terms of the original
parameters $a,b,c,d$
(see also \cite{Arutyunov:2006yd})
\<
\bar{a} \eq -U^{-1}q^{-C+1/2}b,
\nln
\bar{b} \eq U^{-1}q^{-C-1/2} a,
\nln
\bar{c} \eq -Uq^{C + 1/2}d,
\nln
\bar{d} \eq Uq^{C - 1/2}c.
\>
We can solve these equations for the crossed spectral parameters and get%
\footnote{Note that $\bar{\bar\gamma}=-\gamma$ which is consistent
with the $\Integers_4$-periodicity of supertransposition.
Alternatively one could define $\bar U=-1/U$
leading to $\bar{\bar\gamma}=\gamma$.}
\<\label{eq:CrossingRel}
\xbp{}= \antimap(\xp{})
,\qquad
\xbm{}=\antimap(\xm{})
,\qquad
\bar\gamma =\frac{q^{1/2}\alpha}{\gamma} (q^{C}U - q^{-C}U^{-1})  .
\>
Note that $\bar\gamma$ is compatible
with the choice \eqref{eq:gammafix} of $\gamma$ as
a function of $\xpm{}$.
The antipode map $s(x)$ for the parameters $\xpm{}$ is defined as
\[\label{eq:antimap}
\antimap(x)=\frac{ig^{-1}+(q-q^{-1})x}{ig^{-1}x-(q-q^{-1})}\,.
\]
This map $\xpm{}\mapsto\antimap(\xpm{})$ acts on the braiding factor
and the central charge as follows
\[
\bar C= -C,\qquad \bar U= \frac{1}{U}\,.
\]
%

\paragraph{Hermiticity.}

Let us introduce the canonical scalar product
for our states
\[
\braket{\phi_a}{\phi^b}=\delta_a^b,
\qquad
\braket{\psi_\alpha}{\psi^\beta}=\delta_\alpha^\beta.
\]
The representation is hermitian if the conditions in \eqref{eq:hermi}
hold. They imply the relations $a^\ast=q^{C+1/2}d$
and $b^\ast=q^{C-1/2}c$.
Expressed in terms of $\xpm{}$-parameters they correspond to
\[\label{eq:xpmconj}
(\xp{})^\ast=\frac{1}{\antimap(\xm{})}\,,
\qquad
(\xm{})^\ast=\frac{1}{\antimap(\xp{})}\,.
\]
This is equivalent to the condition that the uniformized parameter $z$
on the complex torus is real.
Furthermore the moduli of the parameters $\gamma$ and $\alpha$ are
constrained
\[\label{eq:Unitaritygamma}
|\gamma|^2=-i\xp{}+iq^{2C+1}\xm{}\,,
\qquad
|\alpha|^2=1.
\]
Again, the first constraint is automatically satisfied
when $\gamma$ is given by \eqref{eq:gammafix}
as a function of $\xpm{}$.

\section{The Fundamental R-Matrix}\label{sec:rmat}

To determine the full universal R-matrix is a formidable task.
Here we will satisfy ourselves with the lesser task of finding
the representation of the R-matrix on two fundamentals.

\subsection{Matrix Structure}

We construct the fundamental R-matrix by demanding
that it satisfies the cocommutativity relation \eqref{eq:cocorel}
\[
\coproop(\gen{J})\rmat=
\rmat\copro(\gen{J}).
\]
The most general ansatz which satisfies the relation
for the $\Qgrp(\alg{su}(2)\times\alg{su}(2))$ subalgebra,
i.e.\ for $\gen{E}_1,\gen{E}_3,\gen{F}_1,\gen{F}_3$,
is given in \tabref{tab:Rmatrix}.
The ten functions $A,B,C,D,E,F,G,H,K,L$ are
a priori free. They are however fully constrained
up to one overall phase factor $R^0_{12}$
by cocommutativity w.r.t.\ the supersymmetry generators
$\gen{E}_2,\gen{F}_2$.
This finding is related to the irreducibility of the
tensor product of two fundamental representations
\eqref{eq:shortshort}
\[\label{eq:fundsqr}
\srep{0,0}\otimes\srep{0,0}=\lrep{0,0}.
\]
We present our findings for the ten functions in \tabref{tab:Scoeffs}.
This derivation parallels completely the case of the
undeformed algebra in \cite{Beisert:2006qh}
and for $q=1$ the results agree.
Note that $C_{12}=F_{12}$
and $H_{12}=K_{12}$ if
$\gamma$ is determined by \eqref{eq:gammafix}.
\begin{table}
\<
\rmat\state{\phi^1\phi^1}\eq
A_{12}\state{\phi^1\phi^1}
\nln
\rmat\state{\phi^1\phi^2}\eq
\frac{q A_{12}+q^{-1} B_{12}}{q+q^{-1}}\state{\phi^2\phi^1}
+\frac{A_{12}-B_{12}}{q+q^{-1}}\state{\phi^1\phi^2}
-\frac{q^{-1}C_{12}}{q+q^{-1}}\state{\psi^2\psi^1}
+\frac{C_{12}}{q+q^{-1}}\state{\psi^1\psi^2}
\nln
\rmat\state{\phi^2\phi^1}\eq
\frac{A_{12}-B_{12}}{q+q^{-1}}\state{\phi^2\phi^1}
+\frac{q^{-1}A_{12}+q B_{12}}{q+q^{-1}}\state{\phi^1\phi^2}
+\frac{C_{12}}{q+q^{-1}}\state{\psi^2\psi^1}
-\frac{q C_{12}}{q+q^{-1}}\state{\psi^1\psi^2}
\nln
\rmat\state{\phi^2\phi^2}\eq
A_{12}\state{\phi^2\phi^2}
\nn\\[10pt]
\rmat\state{\psi^1\psi^1}\eq
- D_{12}\state{\psi^1\psi^1}
\nln
\rmat\state{\psi^1\psi^2}\eq
-\frac{qD_{12}+q^{-1} E_{12}}{q+q^{-1}}\state{\psi^2\psi^1}
-\frac{D_{12}-E_{12}}{q+q^{-1}}\state{\psi^1\psi^2}
+\frac{q^{-1}F_{12}}{q+q^{-1}}\state{\phi^2\phi^1}
\!-\!\frac{F_{12}}{q+q^{-1}}\state{\phi^1\phi^2}
\nln
\rmat\state{\psi^2\psi^1}\eq
-\frac{D_{12}-E_{12}}{q+q^{-1}}\state{\psi^2\psi^1}
-\frac{q^{-1}D_{12}+q E_{12}}{q+q^{-1}}\state{\psi^1\psi^2}
-\frac{F_{12}}{q+q^{-1}}\state{\phi^2\phi^1}
\!+\!\frac{q F_{12}}{q+q^{-1}}\state{\phi^1\phi^2}
\nln
\rmat\state{\psi^2\psi^2}\eq
- D_{12}\state{\psi^2\psi^2}
\nn\\[10pt]
\rmat\state{\phi^a\psi^\beta}\eq
G_{12}\state{\phi^a\psi^\beta}
+H_{12}\state{\psi^\beta\phi^a}
\nln
\rmat\state{\psi^\alpha\phi^b}\eq
K_{12}\state{\phi^b\psi^\alpha}
+L_{12}\state{\psi^\alpha\phi^b}\nonumber
\>
\caption{The fundamental R-matrix of $\Qgrp(\alg{h})$. }
\label{tab:Rmatrix}
\end{table}
\begin{table}
\<
A_{12}\eq
R^0_{12}\,
\frac{q^{C_1}U_1}{q^{C_2}U_2}\,
\frac{\xp{2}-\xm{1}}{\xm{2}-\xp{1}}
\nln
B_{12}\eq
R^0_{12}\,
\frac{q^{C_1}U_1}{q^{C_2}U_2}\,
\frac{\xp{2}-\xm{1}}{\xm{2}-\xp{1}}
\lrbrk{1-(q+q^{-1})q^{-1}
\frac{\xp{2}-\xp{1}}{\xp{2}-\xm{1}}\,
\frac{\xm{2}-\antimap(\xp{1})}{\xm{2}-\antimap(\xm{1})}
}
\nln
C_{12}\eq
R^0_{12}\,
(q+q^{-1})\,
\frac{ig\alpha^{-1}\gamma_2\gamma_1q^{C_1}U_1}{q^{2C_2+3/2}U^2_2}\,
\frac{ig^{-1}\xp{2}-(q-q^{-1})}{\xm{2}-\antimap(\xm{1})}\,
\frac{\antimap(\xp{2})-\antimap(\xp{1})}{\xm{2}-\xp{1}}
\nln
D_{12}\eq
-R^0_{12}\nln
E_{12}\eq
-R^0_{12}
\lrbrk{1-(q+q^{-1})q^{-2C_2-1}U^{-2}_2
\frac{\xp{2}-\xp{1}}{\xm{2}-\xp{1}}\,
\frac{\xp{2}-\antimap(\xm{1})}{\xm{2}-\antimap(\xm{1})}
}
\nln
F_{12}\eq
-R^0_{12}\,
(q+q^{-1})\,
\frac{ig\alpha^{-1}\gamma_2\gamma_1 q^{C_1}U_1}{q^{2C_2+3/2}U^2_2}\,
\frac{ig^{-1}\xp{2}-(q-q^{-1})}{\xm{2}-\antimap(\xm{1})}\,
\frac{\antimap(\xp{2})-\antimap(\xp{1})}{\xm{2}-\xp{1}}\,
\nl\qquad\qquad\cdot
\frac{\alpha^2}{1-g^2(q-q^{-1})^2}\,
\frac{U_2q^{C_2+1/2}(\xp{2}-\xm{2})}{\gamma^2_2}\,
\frac{U_1q^{C_1+1/2}(\xp{1}-\xm{1})}{\gamma^2_1}
\nln
G_{12}\eq
R^0_{12}\,
\frac{1}{q^{C_2+1/2}U_2}\,
\frac{\xp{2}-\xp{1}}{\xm{2}-\xp{1}}
\nln
H_{12}\eq R^0_{12}\,
\frac{\gamma_1}{\gamma_2}\,
\frac{\xp{2}-\xm{2}}{\xm{2}-\xp{1}}
\nln
K_{12}\eq
R^0_{12}\,
\frac{q^{C_1}U_1}{q^{C_2}U_2}\,
\frac{\gamma_2}{\gamma_1}\,
\frac{\xp{1}-\xm{1}}{\xm{2}-\xp{1}}
\nln
L_{12}\eq
R^0_{12}\,
q^{C_1+1/2}U_1
\,\frac{\xm{2}-\xm{1}}{\xm{2}-\xp{1}}
\nonumber
\>

\caption{The coefficients for the fundamental R-matrix of $\Qgrp(\alg{h})$.}
\label{tab:Scoeffs}
\end{table}
%

\subsection{Discrete Symmetries of the R-matrix}

\paragraph{Braiding Unitarity.}

An R-matrix is expected to obey the so-called unitarity condition
\[
\rmat_{12}\rmat_{21}=1\otimes 1.
\]
This condition says that the operation of flipping the order of two sites
along a chain is an involution.
Unitarity implies the following ten relations for the
coefficients of the operator in \tabref{tab:Rmatrix}
\[
A_{12}A_{21}=B_{12}B_{21}+C_{12}F_{21}=G_{12}L_{21}+H_{12}H_{21}=1
\]
as well as
\[\label{eq:UniQuo}
\frac{A_{12}}{D_{21}}
=
\frac{D_{12}}{A_{21}}\,,
\qquad
\frac{B_{12}}{E_{21}}
=
\frac{E_{12}}{B_{21}}
=
-\frac{C_{12}}{C_{21}}
=
-\frac{F_{12}}{F_{21}}\,,
\qquad
\frac{H_{12}}{K_{21}}
=
\frac{K_{12}}{H_{21}}
=
-\frac{G_{12}}{G_{21}}
=
-\frac{L_{12}}{L_{21}}\,.
\]

Given the R-matrix coefficients in \tabref{tab:Scoeffs}
it is easy to convince ourselves that this property indeed holds
if the undetermined factor $R^0_{12}$ obeys the equation
$R^0_{12}R^0_{21}=1$.
Curiously, the coefficients in \tabref{tab:Scoeffs} even satisfy the
stronger relation that all three terms in \eqref{eq:UniQuo}
are actually equal.
This amounts to
\[\label{eq:ABCRel}
A_{12}D_{12}=B_{12}E_{12}-C_{12}F_{12}=H_{12}K_{12}-G_{12}L_{12}.
\]

\paragraph{Yang--Baxter Equation.}

Furthermore we have considered the Yang--Baxter equation
\[\label{eq:YBE}
\rmat_{12}\rmat_{13}\rmat_{23}=\rmat_{23}\rmat_{13}\rmat_{12}.
\]
It amounts to around hundred cubic equations among the coefficients
$A,\ldots,L$. We have confirmed in \texttt{Mathematica}
that all relations hold subject to the constraint \eqref{eq:xpmrel}.
An alternative argument \cite{Beisert:2006qh}
for the validity of \eqref{eq:YBE}
uses the decomposition
of the threefold tensor product of fundamental multiplets,
see \eqref{eq:longshort,eq:shortshort}
\[
\srep{0,0}\otimes\srep{0,0}\otimes\srep{0,0}=\lrep{1,0}\oplus \lrep{0,1}.
\]
Effectively we have to prove the YBE only for one component
in each of the resulting multiplets.
Representative states for the two multiplets
are given by $\state{\phi^1\phi^1\phi^1}$ and
$\state{\psi^1\psi^1\psi^1}$.
The YBE is trivially satisfied for both and thus
it is valid in general.

\paragraph{Matrix Unitarity.}

The R-matrix is also a unitary matrix
\[
(\rmat_{12})^\dagger\rmat_{12}=1\otimes 1.
\]
Using the above unitarity we can rewrite the
condition as
$(\rmat_{12})^\dagger=\rmat_{21}$
which makes it straightforward to read off
unitarity conditions for the coefficients
\[\begin{array}[b]{rclrclrclrclrcl}
(A_{12})^\ast\eq A_{21},&
(B_{12})^\ast\eq B_{21},&
(C_{12})^\ast\eq F_{21},&
(G_{12})^\ast\eq L_{21},&
(H_{12})^\ast\eq H_{21},\\
(D_{12})^\ast\eq D_{21},&
(E_{12})^\ast\eq E_{21},&
(F_{12})^\ast\eq C_{21},&
(L_{12})^\ast\eq G_{21},&
(K_{12})^\ast\eq K_{21}.
\end{array}
\]
These conditions are satisfied when
the conjugate parameters are given by \eqref{eq:xpmconj}
and when the phase factor
is a pure phase
$(R^0_{12})^\ast=R^0_{21}=(R^0_{12})^{-1}$.

\paragraph{Crossing Symmetry.}

Finally, the R-matrix may in principle have crossing symmetry.
The crossing equation for the fundamental R-matrix
takes the standard form
\[
 (\conj^{-1}\otimes 1)\rmat_{\bar{1}2}^{\transpose\otimes 1}(\conj\otimes 1)\rmat_{12} = 1\otimes 1.
\]
Using the same trick as above for matrix unitarity, we can write the
crossing relation in terms of the R-matrix coefficients as follows
\[\begin{array}[b]{rclrclrclrclrcl}
A_{\bar 12}\eq \displaystyle\frac{A_{21}-B_{21}}{q+q^{-1}}\,,&
\displaystyle\frac{A_{\bar 12}-B_{\bar 12}}{q+q^{-1}}\eq A_{21},&
G_{\bar 12}\eq L_{21},&
\displaystyle\frac{C_{\bar 12}}{q+q^{-1}}\eq H_{21},&
H_{\bar 12}\eq \displaystyle\frac{F_{21}}{q+q^{-1}}\,,
\\[12pt]
D_{\bar 12}\eq \displaystyle\frac{D_{21}-E_{21}}{q+q^{-1}}\,,&
\displaystyle\frac{D_{\bar 12}-E_{\bar 12}}{q+q^{-1}}\eq D_{21},&
L_{\bar 12}\eq G_{21},&
\displaystyle\frac{F_{\bar 12}}{q+q^{-1}}\eq K_{21},&
K_{\bar 12}\eq \displaystyle\frac{C_{21}}{q+q^{-1}}\,.
\end{array}
\]
All these relations hold simultaneously if
the phase factor satisfies the crossing relation
\<\label{eq:crossingfactor}
R^0_{12}R^0_{\bar 12}
\eq
q\, \frac{\xm{1}-\xm{2}}{\xm{1}-\xp{2}}\,
\frac{\xp{1}-\antimap(\xm{2})}{\xp{1}-\antimap(\xp{2})}
=
q^{-1}
\frac{\xp{1}-\xp{2}}{\xm{1}-\xp{2}}\,
\frac{\antimap(\xp{1})-\xm{2}}{\antimap(\xm{1})-\xm{2}}
\nln
\eq
q\, \frac{\antimap(\xm{1})-\antimap(\xm{2})}{\antimap(\xm{1})-\antimap(\xp{2})}\,
\frac{\antimap(\xp{1})-\xm{2}}{\antimap(\xp{1})-\xp{2}}
=
q^{-1}
\frac{\antimap(\xp{1})-\antimap(\xp{2})}{\antimap(\xm{1})-\antimap(\xp{2})}\,
\frac{\xp{1}-\antimap(\xm{2})}{\xm{1}-\antimap(\xm{2})}\,,
\>
which can be written in many (more) alternative ways.
This is the quantum-deformed analog of the crossing relation
obtained in \cite{Janik:2006dc}.
It would be interesting to find a solution of \eqref{eq:crossingfactor}.

\subsection{Special Points}

Let us investigate the behavior of the R-matrix
at special points of the parameters $\xpm{1,2}$.
The analysis is analogous to the undeformed case in
\cite{Beisert:2006qh,Chen:2006gp},
here we merely give a brief summary.

\paragraph{Permutation.}

As usual when the representation parameters
coincide, $\xpm{1}=\xpm{2}$,
the R-matrix becomes a graded permutation operator times $-1$.

\paragraph{Standard Poles and Zeros.}

\begin{figure}\centering
\includegraphics[scale=1]{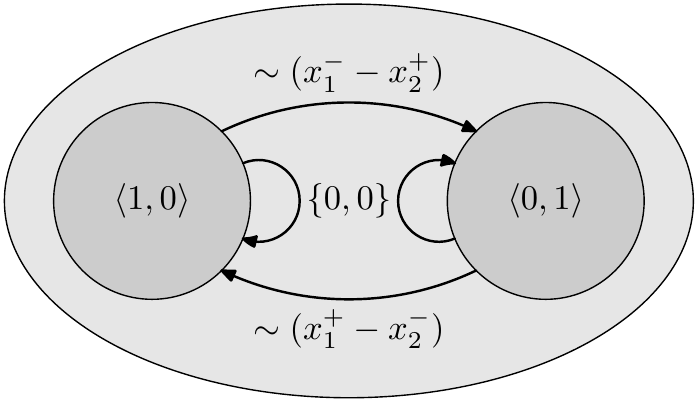}
\caption{Tensor product structure of two fundamental representations
along a chain.
$\srep{1,0}$ contains the state $\state{\phi^1\phi^1}$
while
$\srep{0,1}$ contains the state $\state{\psi^1\psi^1}$.
The arrows indicate the action of the long representation.
The long arrows break when $\xp{2}=\xm{1}$ or
$\xm{2}=\xp{1}$, respectively.}
\label{fig:ChainSplitting}
\end{figure}

The R-matrix has a pole at $\xp{2}=\xm{1}$
and a zero at $\xm{2}=\xp{1}$.
At these points the shortening condition
\eqref{eq:shortening} holds,
and thus the long multiplet in \eqref{eq:fundsqr}
splits up as follows \eqref{eq:split}
\[
\lrep{0,0}\to\srep{1,0}\oplus\srep{0,1}.
\]
Note that the action of $\Qgrp(\alg{h})$
on two fundamental multiplets
is defined via the coproduct,
and typically the resulting long representation
is not decomposable, see \figref{fig:ChainSplitting}.
In particular, at $\xp{2}=\xm{1}$ the
representation closes on the short multiplet $\srep{1,0}$
which contains the state $\state{\phi^1\phi^1}$.
The residue of the corresponding pole of the R-matrix projects
to this submultiplet.
Conversely at $\xm{2}=\xp{1}$ the
representation closes on the short multiplet $\srep{0,1}$
which contains the state $\state{\psi^1\psi^1}$.
The R-matrix projects to this submultiplet at this point.
This behavior is in fact standard for Yangian and quantum affine
algebras, but here it appears already at the level of the
finite-dimensional algebra $\Qgrp(\gen{h})$.

\paragraph{Singlet Pole.}

Furthermore, there is a pole at $\xpm{2}=\antimap(\xpm{1})$
whose residues project to the singlet state $\state{\rep{1}}$ discussed below.
It corresponds to the decomposition
\[
\lrep{0,0}\to\rep{\bullet}\oplus\rep{adj}_{\alg{psu}(2|2)}\oplus\rep{\bullet},
\]
where $\bullet$ represents a singlet.

\subsection{Singlet State and Quasi-Triangularity}

A singlet state is annihilated by all generators of the Hopf algebra $\Qgrp(\alg{h})$.
Consequently, a universal R-matrix in $\Qgrp(\alg{h})\otimes\Qgrp(\alg{h})$
would have to act trivially on this state. In a quasi-triangular
Hopf algebra this also applies to singlet states
which are composed from non-singlets (fusion).
We can use this property to obtain an alternative derivation
of the crossing relation \eqref{eq:crossingfactor}
for the phase factor $R^0_{12}$ as in \cite{Beisert:2005tm}.

\paragraph{Singlet State.}

The singlet component $\state{\rep{1}}$
in a two-spin state is defined by $\copro(\gen{J})\state{\rep{1}}=0$
which should hold for all generators $\gen{J}$.
Without specifying the spin orientations in $\state{\rep{1}}$,
the constraints $\copro(\gen{J})\simeq 0$
for the central charges $\gen{C},\gen{P},\gen{K}$
lead to a relation between the representation labels
\[\label{eq:adjsingl}
\xp{1}=\antimap(\xp{2}),\qquad
\xm{1}=\antimap(\xm{2}).
\]
We make an ansatz which satisfies the constraint
for the bosonic generators $\gen{E}_1,\gen{E}_3,\gen{F}_1,\gen{F}_3$
\[
\state{\rep{1}}=
q^{-1/2}\state{\phi^1\phi^2}
-q^{+1/2}\state{\phi^2\phi^1}
+\kappa q^{-1/2}\state{\psi^1\psi^2}
-\kappa q^{+1/2}\state{\psi^2\psi^1}.
\]
The coefficient $\kappa$ is determined by the constraint for
the fermionic generators $\gen{E}_2,\gen{F}_2$
\[
\kappa=\frac{\gamma_1\gamma_2q^{-1/2}}{\alpha(q^{C_1}U_1-q^{C_2}U_2)}\,.
\]
By means of the crossing relation \eqref{eq:CrossingRel} for $\gamma_2=\bar \gamma_1$
it simply equals $\kappa=1$.

\paragraph{Representation of the R-Matrix.}

The singlet state obeys the curious identity
\[
\rmat_{13}\rmat_{23}\state{\rep{1}_{12}X_3}=\lambda \state{\rep{1}_{12}X_3}
\]
with a common eigenvalue $\lambda$ for all $X\in\set{\phi^a,\psi^\alpha}$.
The eigenvalue takes the form
\[
\lambda
=\frac{A_{13}(A_{23}-B_{23})}{q+q^{-1}}
=\frac{D_{13}(D_{23}-E_{23})}{q+q^{-1}}
=R^0_{13}R^0_{23}
q^{-1}\,
\frac{q^{C_1}U_1}{q^{C_3}U_3}\,
\frac{q^{C_2}U_2}{q^{C_3}U_3}\,
\frac{\xp{3}-\xm{1}}{\xm{3}-\xm{1}}\,\frac{\xp{3}-\xp{2}}{\xm{3}-\xp{2}}
\,.
\]
This observation is in agreement with
quasi-triangularity of the Hopf algebra $\Qgrp(\alg{h})$:
If we view the two-particle state as
transforming under one representation (the singlet)
then the product $\rmat_{13}\rmat_{23}$ must
equal the representation $\rmat_{\rep{1}3}$
of the R-matrix on a singlet and a fundamental (fusion).
However, singlet representations of the R-matrix
are trivial and thus the eigenvalue must be $\lambda=1$.
This condition leads to a constraint on the overall phase factor
\[
R^0_{12}R^0_{\bar 12}=
q\,
\frac{\xm{1}-\xm{2}}{\xm{1}-\xp{2}}\,
\frac{\xp{1}-\antimap(\xm{2})}{\xp{1}-\antimap(\xp{2})}
\,.
\]
It is the same as the crossing relation obtained earlier in
\eqref{eq:crossingfactor}.
Note that we have mapped the representations
according to $(1,2,3)\to(1,\bar 1,2)$.
The map $(1,2,3)\to(\bar 1,1,2)$
leads to a contradictory result.
A consistent result is achieved by
$(1,2,3)\to(\bar 1,\bar{\bar 1},2)$
for which we obtain
\[
R^0_{\bar 12}R^0_{\bar{\bar1}2}=
q\,
\frac{\antimap(\xm{1})-\xm{2}}{\antimap(\xm{1})-\xp{2}}\,
\frac{\antimap(\xp{1})-\antimap(\xm{2})}{\antimap(\xp{1})-\antimap(\xp{2})}
\,.
\]
Note that $R^0_{\bar{\bar1}2}\neq R^0_{12}$, more precisely
there is a double crossing relation
which is the quantum-deformed analog of the relation solved
in \cite{Beisert:2006zy,Beisert:2006ib}
\[
\frac{R^0_{\bar{\bar1}2}}{R^0_{12}}=
\frac{\xm{1}-\xp{2}}{\xm{1}-\xm{2}}\,
\frac{\antimap(\xm{1})-\xm{2}}{\antimap(\xm{1})-\xp{2}}\,
\frac{\xp{1}-\antimap(\xp{2})}{\xp{1}-\antimap(\xm{2})}\,
\frac{\antimap(\xp{1})-\antimap(\xm{2})}{\antimap(\xp{1})-\antimap(\xp{2})}\,.
\]

\section{Diagonalizing the R-matrix}\label{sec:diagonalisation}

The R-matrix can be used to determine the eigenstates
for certain spin chains.
In particular, it is required to write down
quantization conditions for eigenstates of
finite closed or open spin chains.
However, as the R-matrix is a matrix,
the resulting equations would be matrix
equations and rather hard to handle.
Instead of dealing with matrix equations we can first
``diagonalize'' the R-matrix by introducing a suitable
vacuum state and excitations.
This is done by means of the nested coordinate Bethe ansatz \cite{Yang:1967bm},
and the procedure follows along the lines of \cite{Beisert:2006qh}
for the undeformed setup.
Alternatively, one may perform a nested algebraic Bethe ansatz
similar to the one in \cite{Ramos:1996us,Martins:2007hb}
which should lead to the same set of Bethe equations.

\subsection{Vacuum and Propagation}
\label{eq:BAVacProp}

We would like to find suitable eigenstates
of a chain of $K$ fundamental representations with
labels $\xpm{k}$, $k=1,\ldots,K$.
We define the level-II vacuum state $\state{0}\lvl{II}$ to consist
only of $\phi^1$'s. The top-level R-matrix $\rmat\lvl{I}_{\pi}$
is a product of pairwise R-matrices $\rmat\lvl{I}_{k,j}=\rmat_{k,j}$
representing a permutation $\pi\in S_K$ of the $K$ sites.
It multiplies the vacuum state by a phase factor
\[
\rmat\lvl{I}_\pi \state{0}\lvl{II}=R\lvl{I}_\pi\state{0}\lvl{II}.
\]
The total phase factor is a product over pairwise phase factors
\[ R\lvl{I}_\pi=\prod\limits_{(k,j)\in\pi}R\lvl{I,I}(x_k, x_j),\quad
R\lvl{I,I}(x_k,x_j)=A(x_k,x_j)=
R^0(x_k,x_j)
\frac{q^{C_k}U_k}{q^{C_j}U_j}\,
\frac{\xp{j}-\xm{k}}{\xm{j}-\xp{k}}\,.
\]

The level-II vacuum has two flavors of elementary excitations.
An excitation replaces one of the $\phi^1$ spins by a $\psi^\alpha$ spin
\[
\state{\psi^\alpha}\lvl{II}_\pi=\sum\limits_{k=1}^{K}\Psi_{\pi,k}(y)
\state{\phi^1 \dots\MMM{\psi^\alpha}{\pi(k)}\dots\phi^1}\lvl{I}.
\]
The wave function $\Psi_{\pi,k}(y)$
is parametrized by a rapidity $y$ and it
depends locally on the parameters $\xpm{k}$ of the spin representations
\[
\Psi_{\pi,k}(y)=f(y,x_{\pi(k)})\prod\limits_{j=1}^{k-1}
R\lvl{II,I}(y,x_{\pi(j)}) .
\]
The permutation $\pi$ defines the ordering of the top-level
spins along the chain.

In the nested Bethe ansatz the wave function must be compatible
with the action of the R-matrix
\[
\label{eq:compatibility}
\rmat_\pi\lvl{I}\state{\psi^\alpha}\lvl{II}=
R\lvl{I}_\pi\state{\psi^\alpha}\lvl{II}_\pi.
\]
It suffices to consider a chain with two sites and
the permutation $\pi$ interchanges them.
The solution to the compatibility condition reads
\[
R\lvl{II,I}(y,x_k)
=q^{C_k+1/2}U_k\frac{y-\xm{k}}{y-\xp{k}}\,,\qquad
f(y,x_k)=\frac{y\gamma_k}{y-\xp{k}}\,.
\]
Clearly the rapidity $y$ could be replaced by any function of $y$;
we chose it such that the functions $R\lvl{II,I}(y,x_k)$
and $f(y,x_k)$ are rational.
Furthermore $f(y,x_k)$ could be multiplied by any function of $y$.

\subsection{Scattering}
\label{sec:scatstate}

We now have to solve
the analog of the compatibility condition \eqref{eq:compatibility}
for a two-excitation state.
We make the following ansatz
\[
\state{\psi^\alpha\psi^\beta}\lvl{II}_\pi=\sum\limits_{k<j=1}^{K}
\Psi_{\pi,k}(y_1)\Psi_{\pi,j}(y_2)
\state{\phi^1\dots\MMM{\psi^\alpha}{\pi(k)}\dots\MMM{\psi^\beta}{\pi(j)}\dots\phi^1}\lvl{I}
\]
which solves the compatibility condition except when the two excitations are neighbors.
This state can mix with a state where one spin $\phi^1$ is replaced by
$\phi^2$
\[
\state{\phi^2}\lvl{II}_\pi=
\sum\limits_{k=1}^K
\Psi_{\pi,k}(y_1)\Psi_{\pi,k}(y_2)f(y_1,y_2,x_{\pi(k)})
\state{\phi^1\dots \MMM{\phi^2}{\pi(k)}\dots\phi^1}\lvl{I}.
\]
The spin $\phi^2$ should be interpreted as the overlap
of two fermionic excitations.
Thus a generic two-excitation eigenstate reads
\[
\state{\Psi}\lvl{II}_\pi=
\state{\psi^\alpha\psi^\beta}\lvl{II}_\pi
+C^{\alpha\beta}\state{\phi^2}\lvl{II}_\pi
+\perm\lvl{II}_{12}\rmat\lvl{II}\state{\psi^\alpha\psi^\beta}\lvl{II}_\pi.
\]
The matrix $C^{\alpha\beta}$ has the non-zero
elements $C^{12}=q^{-1/2}$ and $C^{21}=-q^{+1/2}$.
The operator $\perm\lvl{II}_{12}$ is a (graded) permutation
which interchanges the excitations along
with their rapidities $y_1,y_2$.
Finally the level-II R-matrix $\rmat\lvl{II}$
must be invariant under $\Qgrp(\alg{su}(2))$ which restricts its form to
\<
\rmat\lvl{II}\state{\psi^1\psi^1}\lvl{II}\eq
M_{12}\state{\psi^1\psi^1}\lvl{II},
\nln
\rmat\lvl{II}\state{\psi^1\psi^2}\lvl{II}\eq
\frac{M_{12}-N_{12}}{q+q^{-1}}\state{\psi^1\psi^2}\lvl{II}
+\frac{qM_{12}+q^{-1}N_{12}}{q+q^{-1}}\state{\psi^2\psi^1}\lvl{II},
\nln
\rmat\lvl{II}\state{\psi^2\psi^1}\lvl{II}\eq
\frac{q^{-1}M_{12}+qN_{12}}{q+q^{-1}}\state{\psi^1\psi^2}\lvl{II}
+\frac{M_{12}-N_{12}}{q+q^{-1}}\state{\psi^2\psi^1}\lvl{II},
\nln
\rmat\lvl{II}\state{\psi^2\psi^2}\lvl{II}\eq
M_{12}\state{\psi^2\psi^2}\lvl{II}.
\>

Imposing the analog of the compatibility condition \eqref{eq:compatibility}
yields a set of equations from which the unknown functions
$M_{12}(y_1,y_2),N_{12}(y_1,y_2),f(y_1,y_2,x_k)$ can be extracted.
The level-II R-matrix $\rmat\lvl{II,II}$ can easily be
determined by considering any of the triplet of states
$\state{\psi^1\psi^1}\lvl{II}$,
$q^{1/2}\state{\psi^1\psi^2}\lvl{II}+q^{-1/2}\state{\psi^2\psi^1}\lvl{II}$
or
$\state{\psi^2\psi^2}\lvl{II}$.
It follows that
\[
M(y_1,y_2)=1.
\]

To determine the other functions we pick the
singlet state $q^{-1/2}\state{\psi^1\psi^2}\lvl{II}-q^{1/2}\state{\psi^2\psi^1}\lvl{II}$.
A lengthy calculation shows that
\<
N(y_1,y_2)\eq\frac{q\specmap(y_2)-q^{-1}\specmap(y_1)+ig^{-1}}
{q^{-1}\specmap(y_2)-q\specmap(y_1)-ig^{-1}}\,,
\nln
f(y_1,y_2,x_k)\eq
\frac{i\alpha(q+q^{-1})q^{1/2}}{2g\gamma_k^2}\,
\frac{\xp{k}(\xp{k}-\xm{k})}{ig^{-1}+(q-q^{-1})\xm{k}}\,
\frac{y_1-y_2}{q^{-1}\specmap(y_2)-q\specmap(y_1)-ig^{-1}}
\nlnum\nn\qquad\cdot
\lrbrk{\antimap(y_1)\antimap(y_2)-
  \frac{y_1y_2}{\xp{k}\xm{k}}\,
  \frac{ig^{-1}+(q-q^{-1})\xm{k}}{ig^{-1}y_1-(q-q^{-1})}\,
  \frac{ig^{-1}+(q-q^{-1})\xp{k}}{ig^{-1}y_2-(q-q^{-1})}
},
\>
is the solution to the compatibility condition if $\specmap(y)$
is the function, see \eqref{eq:antimap},
\[
\label{eq:vy}
\specmap(y) =y + \antimap(y),
\qquad
\antimap(y)=\frac{ig^{-1}+(q-q^{-1})y}{ig^{-1}y-(q-q^{-1})}\,.
\]
Note that the level-II R-matrix indeed solves the YBE
and $\specmap(y)$ is the rapidity variable
for this $\Qgrp(\alg{su}(2))$ R-matrix.
The function $\specmap(y)$ is invariant under taking the antipode
\[
\specmap(y)=\specmap(\antimap(y)),
\]
because the map $\antimap(y)$ is an involution.

\subsection{Final Level}

The final-level R-matrix has the standard form for $\Qgrp(\alg{su}(2))$ symmetry.
The diagonalization by means of the nested Bethe ansatz
leads to the following phase factors
\<
R\lvl{I,I}(x_j,x_k)\eq A(x_j,x_k)=
R^0(x_j,x_k)
\frac{q^{C_j}U_j}{q^{C_k}U_k}\,
\frac{\xp{k}-\xm{j}}{\xm{k}-\xp{j}}\,,
\nln
R\lvl{II,I}(y_j,x_k)
\eq q^{C_k+1/2}U_k\,\frac{\xm{k}-y_j}{\xp{k}-y_j}\,,
\nln
R\lvl{II,II}(y_j,y_k)\eq M_{12}(y_j,y_k)=1,
\nln
R\lvl{III,II}(w_j,y_k)\eq
q^{-1} \frac{q\specmap(y_k)-w_j+\ihalf g^{-1}}{q^{-1}\specmap(y_k)-w_j-\ihalf g^{-1}}\,,
\nln
R\lvl{III,III}(w_j,w_k)\eq\frac{q^{-1}w_k-q w_j-\ihalf (q+q^{-1}) g^{-1}}{q w_k-q^{-1}w_j+\ihalf (q+q^{-1})g^{-1}}\,.
\>
Here $w_k$ are the level-III rapidities.
For completeness we have included the previously derived
phase factors from lower levels.

\subsection{Bethe Equations}

Let us now consider a closed spin chain with $K$ sites.
The wave function for excitations must be periodic
in order to define suitable eigenstates.
Periodicity is imposed by means of the Bethe equations
which use the elements of the diagonalized R-matrix.
Generically for a nested Bethe ansatz with two levels
they can be written as follows
\<\label{eq:BetheEquationsGeneral}
1\eq
\prod_{j=1}^{K} R\lvl{I,II}(x_j,y_k)
\mathop{\prod_{j=1}}_{j\neq k}^{N} R\lvl{II,II}(y_j,y_k)
\prod_{j=1}^{M} R\lvl{III,II}(w_j,y_k),
\nln
1\eq
\prod_{j=1}^{K} R\lvl{I,III}(x_j,w_k)
\prod_{j=1}^{N} R\lvl{II,III}(y_j,w_k)
\mathop{\prod_{j=1}}_{j\neq k}^{M} R\lvl{III,III}(w_j,y_k),
\>
Thus for the system in question they take the form
\<
\label{eq:BetheEquations}
1\eq\prod_{j=1}^{K}q^{-C_j-1/2}U_j^{-1} \frac{y_k-\xp{j}}{y_k-\xm{j}}
\prod\limits_{j=1}^{M}
q^{-1} \frac{q\specmap(y_k)-w_j+\ihalf g^{-1}}{q^{-1}\specmap(y_k)-w_j-\ihalf g^{-1}}\,,
\nln
1\eq\prod_{j=1}^{N}
q\,\frac{w_k-q^{-1}\specmap(y_j)+\ihalf g^{-1}}{w_k-q\specmap(y_j)-\ihalf g^{-1}}
\mathop{\prod_{j=1}}_{j\neq k}^{M}
\frac{q^{-1}w_k-qw_j-\ihalf(q+q^{-1})g^{-1}}{qw_k-q^{-1}w_j+\ihalf(q+q^{-1})g^{-1}}\,.
\>
%

\subsection{Dualization}
\label{sec:Dual}

At the very beginning of the nested Bethe ansatz one can in fact choose between
two different vacua, composed from only bosons $\phi^1$'s as above
(alternatively $\phi^2$'s) or composed from only fermions
$\psi^1$'s (alternatively $\psi^2$'s).
In both cases the NBA will proceed in a very
similar fashion, but lead to two different
but equivalent sets of Bethe equations.

Instead of performing the alternative NBA,
we shall derive the alternative Bethe equations
by means of dualization.
The dualization of our Bethe equations for the undeformed case
was performed in \cite{Beisert:2005fw}
(which is equivalent to the dualization
of the Lieb--Wu equations \cite{Lieb:1968aa} in \cite{Woynarovich:1983aa}).
Here the procedure is qualitatively the same, but requires
special care due to various insertions of $q$.
Let us outline it in the following:

The first Bethe equation in \eqref{eq:BetheEquations} can be viewed as
an algebraic equation $P(y_k)=0$ with the polynomial
\<\label{eq:dualityPol}
P(y)\eq
\prod_{j=1}^{K}q^{C_j+1/2}U_j\lrbrk{y-\xm{j}}
\prod_{j=1}^{M}q\lrbrk{ig^{-1}y-(q-q^{-1})} \lrbrk{q^{-1}\specmap(y)-w_j-\ihalf g^{-1}}
\nl
-\prod_{j=1}^{K}\lrbrk{y-\xp{j}}
\prod_{j=1}^{M}\lrbrk{ig^{-1}y-(q-q^{-1})} \lrbrk{q\specmap(y)-w_j+\ihalf g^{-1}}.
\>
Note that the common factor $(ig^{-1}y-(q-q^{-1}))$ was introduced
in order to cancel the denominator of $\antimap(y)$ in $\specmap(y)$.
This polynomial has degree $K+2M$, i.e.\ $N$ of its roots
are the $y_k$ and the remaining $\tilde N=K+2M-N$ roots
will be denoted by $\tilde y_k$. We can construct a
constant function $F(y)$ by dividing by
all root monomials
\[
F(y)=P(y)
\prod_{j=1}^{N}\frac{1}{y-y_j}
\prod_{j=1}^{\tilde N}\frac{1}{y-\tilde y_j}
=
(ig^{-1})^M\Bigg(
\prod_{j=1}^{K}q^{C_j+1/2}U_j-
q^{M}
\Bigg).
\]

Next let us define $\xpm{w}$ as a function of $w_k$ implicitly through the relation
\[
\label{eq:xpmw}
w_k=q^{-1}\specmap(\xp{w})-\ihalf g^{-1}= q\specmap(\xm{w})+\ihalf g^{-1}.
\]
The $\xpm{w}$ obey the $\xpm{}$-constraint \eqref{eq:xpmrel}
which can be written in a particularly convenient form using the
map $\specmap(y)$ \eqref{eq:vy}
\[
\label{eq:xpmrel2}
q^{-1}\specmap(\xp{})-q\specmap(\xm{})=\frac{i}{g}\,.
\]
Now any ratio of the constant function $F(y)$ evaluated for two different
values of the parameter equals $1$.
There are four useful points
$y=\xpm{w},\antimap(\xpm{w})$ where one of the
two terms in the polynomial $P(y)$ drops out by means of \eqref{eq:xpmw}.
At these points we obtain the following identity
which holds by virtue of the first Bethe equation in \eqref{eq:BetheEquations}
\<
1=
\frac{F(\xp{w})F(\antimap(\xp{w}))}
     {F(\xm{w})F(\antimap(\xm{w}))}
\eq
\prod_{j=1}^{K}q^{-2C_j-1}U_j^{-2}\frac{\xp{w}-\xp{j}}{\xm{w}-\xm{j}}\,
\frac{\antimap(\xp{w})-\xp{j}}{\antimap(\xm{w})-\xm{j}}
\nl\cdot
\prod_{j=1}^{N}\frac{\xm{w}-y_j}{\xp{w}-y_j}\,\frac{\antimap(\xm{w})-y_j}{\antimap(\xp{w})-y_j}
\prod_{j=1}^{\tilde N}\frac{\xm{w}-\tilde y_j}{\xp{w}-\tilde y_j}\,
\frac{\antimap(\xm{w})-\tilde y_j}{\antimap(\xp{w})-\tilde y_j}
\nl\cdot
\lrbrk{
q^{-2}
\frac{ig^{-1}\xp{w}-(q-q^{-1})}{ig^{-1}\xm{w}-(q-q^{-1})}\,
\frac{ig^{-1}\antimap(\xp{w})-(q-q^{-1})}{ig^{-1}\antimap(\xm{w})-(q-q^{-1})}
}^M
\nl\cdot
\prod_{j=1}^{M}\lrbrk{\frac{q\specmap(\xp{w})-w_j+\ihalf g^{-1}}{q^{-1}\specmap(\xm{w})-w_j-\ihalf g^{-1}}}^2.
\>
For the first line we use the identity
\[
\frac{\xp{k}-\xp{j}}{\xm{k}-\xm{j}}\,
\frac{\antimap(\xp{k})-\xp{j}}{\antimap(\xm{k})-\xm{j}}
=q^{2C_j+2}U_j^{2}
\]
and obtain simply $q^K$. The second line is simplified by means of the identity
\[
\frac{\xp{}-y}{\xm{}-y}\,\frac{\antimap(\xp{})-y}{\antimap(\xm{})-y}
=
\frac{\specmap(\xp{})-\specmap(y)}{\specmap(\xm{})-\specmap(y)}
\]
and for the third line the identity
\[\label{eq:q2CU}
\frac{ig^{-1}\xp{}-(q-q^{-1})}{ig^{-1}\xm{}-(q-q^{-1})}
=q^{2C}U^2
\]
leads to $q^{-2M}$. In the last line we substitute
the definition \eqref{eq:xpmw} of $\xpm{w}$.
Altogether this yields the identity
\<
1
\eq
\prod_{j=1}^{N}
q^{-1}\frac{w_k-q\specmap(y_j)-\ihalf g^{-1}}{w_k-q^{-1}\specmap(y_j)+\ihalf g^{-1}}
\prod_{j=1}^{\tilde N}
q^{-1}\frac{w_k-q\specmap(\tilde y_j)-\ihalf g^{-1}}{w_k-q^{-1}\specmap(\tilde y_j)+\ihalf g^{-1}}
\nl\cdot
\mathop{\prod_{j=1}}_{j\neq k}^{M}\lrbrk{\frac{qw_k-q^{-1}w_j+\ihalf (q+q^{-1}) g^{-1}}{q^{-1}w_k-qw_j-\ihalf (q+q^{-1})g^{-1}}}^2.
\>
Two of these terms coincide with terms in the second Bethe in \eqref{eq:BetheEquations}.
We multiply the Bethe equation and the identity and obtain the
dual Bethe equations for $\tilde y_k$ and $w_k$
\<\label{eq:BetheEquationsDual}
1\eq\prod_{j=1}^{K}q^{C_j+1/2}U_j\,\frac{\tilde y_k-\xm{j}}{\tilde y_k-\xp{j}}
\prod_{j=1}^{M}q\,\frac{q^{-1}\specmap(\tilde y_k)-w_j-\ihalf g^{-1}}{q\specmap(\tilde y_k)-w_j+\ihalf g^{-1}}\,,
\nln
1\eq
\prod_{j=1}^{\tilde N}
q^{-1}\frac{w_k-q\specmap(\tilde y_j)-\ihalf g^{-1}}{w_k-q^{-1}\specmap(\tilde y_j)+\ihalf g^{-1}}
\mathop{\prod_{j=1}}_{j\neq k}^{M}
\frac{qw_k-q^{-1}w_j+\ihalf (q+q^{-1}) g^{-1}}{q^{-1}w_k-qw_j-\ihalf (q+q^{-1})g^{-1}}\,.
\>

These Bethe equations do not contain the scattering term $R\lvl{I,I}(x_j,x_k)$
which should be dualized as well for completeness.
In fact it can easily be obtained from the scattering
in the dual level-II vacuum
composed from $\psi^1$'s: it equals $-D(x_j,x_k)$.
An alternative procedure is to consider the $\xpm{k}$ as dynamical degrees of freedom
as in \cite{Beisert:2005fw}.
In other words they obey some Bethe equation
of the form
\<\label{eq:BetheExtra}
\ldots\eq
\mathop{\prod_{j=1}}_{j\neq k}^{K} R\lvl{I,I}(x_j,x_k)
\prod_{j=1}^{N} R\lvl{II,I}(y_j,x_k)
\prod_{j=1}^{M} R\lvl{III,I}(w_j,x_k)
\nln\eq
\mathop{\prod_{j=1}}_{j\neq k}^{K}
R^0(x_j,x_k)
\frac{q^{C_j}U_j}{q^{C_k}U_k}\,
\frac{\xp{k}-\xm{j}}{\xm{k}-\xp{j}}
\prod_{j=1}^{N} q^{C_k+1/2}U_k\,\frac{\xm{k}-y_j}{\xp{k}-y_j}\,.
\>
We can now derive another identity from the
constant function $F(y)$
\<
1=
\frac{F(\xm{k})}{F(\xp{k})}
\eq
-
\prod_{j=1}^{N}\frac{\xp{k}-y_j}{\xm{k}-y_j}
\prod_{j=1}^{\tilde N}\frac{\xp{k}-\tilde y_j}{\xm{k}-\tilde y_j}
\prod_{j=1}^{K}q^{-C_j-1/2}U_j^{-1}\frac{\xm{k}-\xp{j}}{\xp{k}-\xm{j}}
\nl\cdot
\lrbrk{q\,\frac{ig^{-1}\xp{k}-(q-q^{-1})}{ig^{-1}\xm{k}-(q-q^{-1})}}^{-M}
\prod_{j=1}^{M}\frac{q\specmap(\xm{k})-w_j+\ihalf g^{-1}}{q^{-1}\specmap(\xp{k})-w_j-\ihalf g^{-1}}\,.
\>
Using the identities \eqref{eq:xpmrel2,eq:q2CU} we find
\[\label{eq:AllMomConstr}
1=
\prod_{j=1}^{N}q^{-C_k-1/2}U_k^{-1}\frac{\xp{k}-y_j}{\xm{k}-y_j}
\prod_{j=1}^{\tilde N}q^{-C_k-1/2}U_k^{-1}\frac{\xp{k}-\tilde y_j}{\xm{k}-\tilde y_j}
\mathop{\prod_{j=1}}_{j\neq k}^{K} \frac{q^{C_k}U_k}{q^{C_j}U_j}\,\frac{\xm{k}-\xp{j}}{\xp{k}-\xm{j}}\,,
\]
which yields when multiplied to \eqref{eq:BetheExtra} the dual Bethe equation
\[
\ldots=
\mathop{\prod_{j=1}}_{j\neq k}^{K} R^0(x_j,x_k)
\prod_{j=1}^{\tilde N}q^{-C_k-1/2}U_k^{-1}\frac{\xp{k}-\tilde y_j}{\xm{k}-\tilde y_j}\,.
\]

In conclusion the dual phase factors read
\<
\tilde R\lvl{I,I}(x_j,x_k)\eq -D(x_j,x_k)=
R^0(x_j,x_k),
\nln
\tilde R\lvl{II,I}(\tilde y_j,x_k)
\eq q^{-C_k-1/2}U_k^{-1}\frac{\xp{k}-\tilde y_j}{\xm{k}-\tilde y_j}\,,
\nln
\tilde R\lvl{II,II}(\tilde y_j,\tilde y_k)\eq 1,
\nln
\tilde R\lvl{III,II}(w_j,\tilde y_k)\eq
q\, \frac{q^{-1}\specmap(\tilde y_k)-w_j-\ihalf g^{-1}}{q\specmap(\tilde y_k)-w_j+\ihalf g^{-1}}\,,
\nln
\tilde R\lvl{III,III}(w_j,w_k)\eq\frac{q w_k-q^{-1}w_j+\ihalf (q+q^{-1})g^{-1}}{q^{-1}w_k-q w_j-\ihalf (q+q^{-1}) g^{-1}}\,.
\>
%

\section{Quantum Deformation of the Hubbard Model}\label{sec:hubbard}

In this section we derive an integrable homogeneous nearest-neighbor Hamiltonian
from the $\Qgrp(\alg{h})$ R-matrix in \tabref{tab:Rmatrix}.
Its Hilbert space is the same as for the one-dimensional Hubbard model
and we will show that our Hamiltonian is a three-parameter deformation
of the Hubbard Hamiltonian.
We then discuss the relation to the integrable two-parameter deformations
of the Hubbard model proposed by Alcaraz and Bariev \cite{Alcaraz:1999aa}
and introduce additional sets of two-parameter deformations.

\subsection{Hamiltonian}

In the following we derive a nearest-neighbor spin chain Hamiltonian
from the $\Qgrp(\alg{h})$ R-matrix in \tabref{tab:Rmatrix}.

\paragraph{Integrable Hamiltonian.}

For an integrable spin chain based on an R-matrix
there is a standard procedure to obtain a
homogeneous Hamiltonian
\[\label{eq:PerHam}
\ham=\sum_{k=1}^L \ham_{k,k+1}.
\]
The pairwise interaction $\ham_{12}$
is the following logarithmic derivative of the R-matrix
\[\label{eq:1loopHamq}
\mathcal{H}_{12}=
-i\frac{\bigbrk{\xp{}-\antimap(\xp{})}\bigbrk{\xm{}-\antimap(\xm{})}}{q^{-1}\xp{}\antimap(\xp{})}
\lrbrk{\frac{du^\ast}{du}}^{-1/2}
\lreval{\rmat^{-1}_{12}\frac{d}{du_1}\rmat_{12}}_{x^{\pm}_{12}=x^{\pm}}.
\]
The spectral parameters $u_k$ are defined as functions of the parameters $\xpm{k}$, see \eqref{eq:vy},
\[
u_k=q^{-1}\specmap(\xp{k})-\frac{i}{2g}=q\specmap(\xm{k})+\frac{i}{2g}\,.
\]
Note that $u_k$ is not real
\[
u_k^\ast=\frac{(q+q^{-1})u_k+\ihalf g^{-1}(q-q^{-1})}{(q+q^{-1})-2ig(q-q^{-1})u_k}\,,
\qquad
\frac{du_k^\ast}{du_k}=\bigbrk{q^{-1}\xp{k}\antimap(\xp{k})}^{-2}
=\bigbrk{q\xm{k}\antimap(\xm{k})}^{-2},
\]
and therefore we have included a compensating prefactor
$(du^\ast/du)^{-1/2}=q^{-1}\xp{k}\antimap(\xp{k})$
to make the Hamiltonian hermitian.
The additional real prefactor is meant to simplify the resulting expressions.
\begin{table}
\<
\ham_{12}\state{\phi^1\phi^1}\eq
 A\state{\phi^1\phi^1}
\nln
\ham_{12}\state{\phi^1\phi^2}\eq
\frac{q A+q^{-1}B}{q+q^{-1}}\state{\phi^1\phi^2}
+\frac{A-B}{q+q^{-1}}\state{\phi^2\phi^1}
+\frac{q^{-1}C}{q+q^{-1}}\state{\psi^1\psi^2}
-\frac{C}{q+q^{-1}}\state{\psi^2\psi^1}
\nln
\ham_{12}\state{\phi^2\phi^1}\eq
\frac{A-B}{q+q^{-1}}\state{\phi^1\phi^2}
+\frac{q^{-1}A+q B}{q+q^{-1}}\state{\phi^2\phi^1}
-\frac{C}{q+q^{-1}}\state{\psi^1\psi^2}
+\frac{qC}{q+q^{-1}}\state{\psi^2\psi^1}
\nln
\ham_{12}\state{\phi^2\phi^2}\eq A\state{\phi^2\phi^2}
\nln[10pt]
\ham_{12}\state{\psi^1\psi^1}\eq D\state{\psi^1\psi^1}
\nln
\ham_{12}\state{\psi^1\psi^2}\eq
\frac{qD+q^{-1}E}{q+q^{-1}}\state{\psi^1\psi^2}
+\frac{D-E}{q+q^{-1}}\state{\psi^2\psi^1}
+\frac{q^{-1}F}{q+q^{-1}}\state{\phi^1\phi^2}
-\frac{F}{q+q^{-1}}\state{\phi^2\phi^1}
\nln
\ham_{12}\state{\psi^2\psi^1}\eq
\frac{D-E}{q+q^{-1}}\state{\psi^1\psi^2}
+\frac{q^{-1}D+q E}{q+q^{-1}}\state{\psi^2\psi^1}
-\frac{F}{q+q^{-1}}\state{\phi^1\phi^2}
+\frac{q F}{q+q^{-1}}\state{\phi^2\phi^1}
\nln
\ham_{12}\state{\psi^2\psi^2}\eq D\state{\psi^2\psi^2}
\nln[10pt]
\ham_{12}\state{\phi^a\psi^\beta}\eq
G\state{\psi^\beta\phi^a}+ H\state{\phi^a\psi^\beta}
\nln
\ham_{12}\state{\psi^\alpha\phi^b}\eq
K\state{\psi^\alpha\phi^b}+L\state{\phi^b\psi^\alpha}
\nonumber
\>
\caption{General form of the nearest-neighbor
$\Qgrp(\alg{su}(2)\times\alg{su}(2))$ spin chain
Hamiltonian}
\label{tab:HamiltonianGeneral}
\end{table}

The resulting Hamiltonian clearly has $\Qgrp(\alg{su}(2)\times\alg{su}(2))$
symmetry%
\footnote{It does not necessarily have
$\Qgrp(\alg{h})$ symmetry due to the
$u$-dependence of the coefficients $a,b,c,d$
in the representation of the fermionic generators.
In fact there is no
$\Qgrp(\alg{h})$-invariant operator apart
from the identity for generic $q$.}
 and the general form with such symmetry
is listed in \tabref{tab:HamiltonianGeneral}.
Just like the R-matrix in \tabref{tab:Rmatrix} is has
ten independent coefficients $X=A,B,C,D,E,F,G,H,K,L$.
We have arranged them such that they are directly related
to the corresponding coefficients $X_{12}$ of the R-matrix in
\tabref{tab:Scoeffs} according to
\<
X
\eq
i\bigbrk{\xp{}-\antimap(\xp{})}\bigbrk{\xm{}-\antimap(\xm{})}
\lreval{\frac{d X_{12}}{d u_1}}_{\xpm{1,2}=\xpm{}}
\nln\eq
i\bigbrk{\xp{}-\antimap(\xp{})}\bigbrk{\xm{}-\antimap(\xm{})}
\lreval{
\lrbrk{
\frac{d \xp{}}{du}\,
\frac{\partial X_{12}}{\partial\xp{1}}
+
\frac{d\xm{}}{du}\,
\frac{\partial X_{12}}{\partial\xm{1}}
}
}_{\xpm{1,2}=\xpm{}}
.
\>
Note the following two useful identities in evaluating the derivatives
\<\label{eq:deru}
\frac{d\xpm{}}{du}\eq
\frac{\xp{}-\xm{}}{\xpm{}-\antimap(\xpm{})}\,
\frac{q^{\pm C\pm 1}U^{\pm 1}}{q^CU-q^{-C}U^{-1}}
\,,
\nln
\frac{d}{du}\log (Uq^{C})\eq
\frac{1}{2}\,
\lrbrk{\frac{q}{\xp{}-\antimap(\xp{})}
-\frac{q^{-1}}{\xm{}-\antimap(\xm{})}}.
\>

The coefficients $X$ still depend on the parameters $\alpha$, $\gamma_k$ and
the phase factor $R^0_{12}$.
The latter two can furthermore depend non-trivially on $\xpm{k}$
which would influence the Hamiltonian explicitly
or implicitly through the derivatives.
For definiteness we set the phase factor to
\[
R^0_{12}=\sqrt{\frac{q^{C_2}U_2}{q^{C_1}U_1}\,\frac{\xm{2}-\xp{1}}{\xp{2}-\xm{1}}}\,;
\]
a different phase factor would induce an overall
shift of the spectrum which we shall incorporate explicitly later.
The most suitable expression for $\gamma_k$ is given in \eqref{eq:gammafix}
and we shall set the global parameter $\alpha$ to unity, $\alpha=1$.

Now we are ready to obtain an explicit expression
for the nearest-neighbor Hamiltonian.
Taking into account that for $\xpm{1}=\xpm{2}$ the R-matrix becomes
minus the graded permutation operator one can show that
the Hamiltonian has the coefficients
listed in \tabref{tab:HamCoeff}.%
\begin{table}
\<
A=-D\eq
\frac{1}{4g}\,
\frac{(q^{C}U+q^{-C}U^{-1})(q^{C}U^{-1}+q^{-C}U)}{(q^{C}U-q^{-C}U^{-1})(q^{C}U^{-1}-q^{-C}U)}
\nln
A-B=E-D\eq
\frac{q+q^{-1}}{g}\,
\frac{1}{(q^{C}U-q^{-C}U^{-1})(q^{C}U^{-1}-q^{-C}U)}
\nln
C=F\eq
(q+q^{-1})
\sqrt{1-(q-q^{-1})^2g^2}\,
\nln
G\eq
\frac{q^{C+1/2}U^{-1}-q^{-C-1/2}U^{-1}-q^{C-1/2}U+q^{-C+1/2}U}
     {g(q-q^{-1})(q^{C}U-q^{-C}U^{-1})(q^{C}U^{-1}-q^{-C}U)}
\nln
L\eq
\frac{q^{C+1/2}U-q^{-C-1/2}U-q^{C-1/2}U^{-1}+q^{-C+1/2}U^{-1}}
     {g(q-q^{-1})(q^{C}U-q^{-C}U^{-1})(q^{C}U^{-1}-q^{-C}U)}
\nln
H=K \eq
0
\nn
\>
\caption{The coefficients for the nearest-neighbor Hamiltonian.}
\label{tab:HamCoeff}
\end{table}
%

\paragraph{Integrability Constraints.}

The coefficients obey certain relations:
There are two linear relations
and two quadratic relations
\<\label{eq:IntConstr}
A+D\eq B+E=H+K,
\nln
(A-B)^2+CF\eq (q+q^{-1})^2GL,
\nln
2(A-B)(A-D)\eq (q+q^{-1})\bigbrk{G^2+(q+q^{-1})GL+L^2}.
\>
Note that the two linear relations can be derived from the
identity \eqref{eq:ABCRel};
the origin of the two quadratic relations remains unclear.
These equations are invariant under four trivial transformations:
(\emph{i}) A rescaling of all coefficients by a common factor.
(\emph{ii}) A shift of the two-site Hamiltonian by the
two-site identity operator $\copro(1)$,
taking the form in
\tabref{tab:HamiltonianGeneral}
with
\[
\label{eq:IdentityHam}
A=B=D=E=H=K=1,\qquad C=F=G=L=0.
\]
(\emph{iii}) A reciprocal rescaling of $C$ and $F$
corresponding to a
different rescaling of fermions $\state{\psi^\alpha}$
w.r.t.\ bosons $\state{\phi^a}$.
(\emph{iv}) An opposite shift of $H$ and $K$
which has no impact on the spectrum.
Altogether the ten coefficients together with the
parameter $q$ subject to the four
constraints \eqref{eq:IntConstr} and
four identifications (\emph{i}--\emph{iv})
define a three-parameter family of models.
This is the same number of degrees of freedom
as for the fundamental R-matrix given in terms of
$q,g,\xpm{}$ subject to the one constraint \eqref{eq:xpmrel}.
Therefore the constraints \eqref{eq:IntConstr}
are expected to be sufficient to ensure integrability
of the Hamiltonian in \tabref{tab:HamiltonianGeneral}.

\paragraph{Hermiticity.}

There are ten independent complex coefficients in
the Hamiltonian in \tabref{tab:HamiltonianGeneral}.
The hermiticity condition $\ham_{12}^\dagger = \ham_{12}$
imposes certain reality conditions on the coefficients
\[
\label{eq:HermiticityConstraints}
C^\ast = F,\quad G^\ast = L,\qquad A,B,D,E,H,K\in\Reals.
\]
Furthermore we have to require that%
\footnote{Usually in quantum-deformed spin chains
the case of $q$ on the unit circle also leads
to hermitian Hamiltonians. Here we have not
been managed to establish hermiticity in this case.}
\[
q\in \Reals.
\]
Due to our choice \eqref{eq:gammafix} for $\gamma_k$,
we have that $C=F$ and thus $C,F$ must both be real.

\paragraph{Generalizations.}

Before performing an explicit comparison of the models in question
we would like to introduce certain integrable generalizations
of the Hamiltonian which change the spectrum in a controllable fashion.
Thus given a Hamiltonian $\ham_{12}$ as in \tabref{tab:HamiltonianGeneral}
one can transform it as follows
\<\label{eq:allowedtransf}
\ham'_{12}\eq
a_0 \,\Twist\ham_{12}\Twist^{-1}
+\half a_{1} \copro(\gen{H}_1)
+a_{2} \copro(1)
+\half a_{3} \copro(\gen{H}_3)
\nlnum\nonumber
+\half b_1 (\gen{H}_1\otimes 1-1\otimes \gen{H}_1)
+b_2 (\gen{H}_1\gen{H}_1\otimes 1-1\otimes \gen{H}_1\gen{H}_1)
+\half b_3 (\gen{H}_3\otimes 1-1\otimes \gen{H}_3).
\>
Here $a_k,b_k$ are arbitrary constants.
Indeed $a_0$ is an overall multiplier of the operator,
$a_2$ is an overall shift
and $a_1,a_3$
correspond to a shift of the energy eigenvalues proportional
to the $\gen{H}_1,\gen{H}_3$ Cartan generator eigenvalues.
The terms multiplied by $b_k$ vanish
after summation over the whole spin chain
with periodic boundary conditions \eqref{eq:PerHam}.

The similarity transformation induced by $\Twist$ is
the following Reshetikhin twist
\cite{Reshetikhin:1990ep}
of the integrable structure
\[\label{eq:twisting}
\copro'(\gen{J})=\Twist\copro(\gen{J})\Twist^{-1},
\qquad
\rmat'=\perm(\Twist)\rmat\Twist^{-1}.
\]
It is not hard to verify that the twist preserves
the cocommutativity property.
We shall consider a twist $\Twist$ constructed from
the identity and Cartan generators
\[\label{eq:twistT}
\Twist =
\exp \bigbrk{
if_1(1\otimes \gen{H}_1)
+\ihalf f_2(\gen{H}_1\otimes\gen{H}_3-\gen{H}_3\otimes\gen{H}_1)
+if_3(1\otimes \gen{H}_3)
},
\]
which can be generalized consistently to arbitrarily many sites
\[
\Twist =
\exp \lrbrk{
if_1
\sum_{j=1}^K (j-1)\gen{H}_{1,j}
+\ihalf f_2 \sum_{j<k=1}^K (\gen{H}_{1,j}\gen{H}_{3,k}-\gen{H}_{3,j}\gen{H}_{1,k})
+if_3 \sum_{j=1}^K (j-1)\gen{H}_{3,k}
}.
\]
The coefficients $f_k$ are arbitrary parameters.
Hermiticity requires them to be real.

\subsection{Bethe Equations and Spectrum}
\label{sec:BetheSpec}

The spectrum of the above Hamiltonian on a closed homogeneous spin chain
is determined by Bethe equations.
Here we specify the Bethe equations and energy relations.

\paragraph{Energies.}

We use the Bethe ansatz based on a ferromagnetic vacuum
consisting of $K$ spins $\phi^1$ as in \eqref{eq:BAVacProp}.
We assume there are $N$ main excitations (magnons)
with momenta $p_k$ which turn a $\phi^1$ into a $\psi^1$.
Finally, there are $M$ auxiliary excitations
with rapidities $w_k$ which turn a $\psi^1$ into a $\psi^2$.
We have shown in \secref{sec:diagonalisation}
that the Bethe equations for this system are given by \eqref{eq:BetheEquations}
where we set all representation parameters to be equal $\xpm{k}=\xpm{}$
for a homogeneous chain
\<\label{eq:BetheEquationsHomo}
1\eq \lrbrk{q^{-C-1/2}U^{-1} \frac{y_k-\xp{}}{y_k-\xm{}}}^K
\prod\limits_{j=1}^{M}
q^{-1} \frac{q\specmap(y_k)-w_j+\ihalf g^{-1}}{q^{-1}\specmap(y_k)-w_j-\ihalf g^{-1}}\,,
\nln
1\eq\prod_{j=1}^{N}
q\,\frac{w_k-q^{-1}\specmap(y_j)+\ihalf g^{-1}}{w_k-q\specmap(y_j)-\ihalf g^{-1}}
\mathop{\prod_{j=1}}_{j\neq k}^{M}
\frac{q^{-1}w_k-qw_j-\ihalf(q+q^{-1})g^{-1}}{qw_k-q^{-1}w_j+\ihalf(q+q^{-1})g^{-1}}\,.
\>

First of all, a relation
between the magnon momenta $p_k$ and the magnon rapidities $y_k$
has to be established. In the Bethe equations
\eqref{eq:BetheEquationsGeneral} the term $R\lvl{I,II}(x,y_k)$
serves the purpose of $e^{-ip_k}$,
i.e.~the rapidity relation is
\[\label{eq:momspec}
e^{ip_k}=R\lvl{II,I}(y_k,x)=q^{C+1/2}U\,\frac{y_k-\xm{}}{y_k-\xp{}}
\,,\qquad
e^{iP}=\prod_{j=1}^N e^{ip_k}\,.
\]
Here the total spin chain momentum is given by $P$.
The energy for a solution of the Bethe equations is given
as the sum of vacuum energy and magnon energies
(here $K$ refers to the length of the chain)
\[
E=E_0 K+\sum_{j=1}^N E(y_j).
\]
The vacuum energy density $E_0$ and magnon dispersion relation $E(y_k)$ follow
readily from the expression for the Hamiltonian
in \tabref{tab:HamiltonianGeneral}
(here $K$ refers to the coefficient $K$ of the Hamiltonian)
\[\label{eq:engABC}
E_0=A,\qquad
E(y_k)=H+K-2A+G e^{ip_k}+L e^{-ip_k},
\]
where $e^{ip_k}$ is defined in \eqref{eq:momspec}
as a function of $y_k$.
The energy relation can also be obtained formally
from the diagonalized elements of the R-matrix
in analogy to \eqref{eq:momspec}
\<\label{eq:engspec}
E_0\eq i(\xp{}-\antimap(\xp{}))(\xm{}-\antimap(\xm{}))\lreval{\frac{d}{du}\log R\lvl{I,I}(x_0,x)}_{\xpm{0}=\xpm{}}\,,
\nln
E(y_k)\eq i(\xp{}-\antimap(\xp{}))(\xm{}-\antimap(\xm{}))\, \frac{d}{du}\log R\lvl{II,I}(y_k,x).
\>
This leads to the same as the above expressions.

\paragraph{Transformation.}

The above generalization of the Hamiltonian
\eqref{eq:allowedtransf}
requires certain modifications of the just derived expressions.
Most importantly, the Bethe equations receive further phase factors
due to the twist induced by the $f_k$
\footnote{Clearly the twist can also be applied to
the inhomogeneous equations where $\xpm{k}\neq\xpm{j}$.}
\<
\label{eq:BetheEquationsPhase}
1\eq
\lrbrk{
e^{i(f_3-f_1-f_2)}
q^{-C-1/2}U^{-1}
\frac{y_k-\xp{}}{y_k-\xm{}}
}^K
\prod\limits_{j=1}^{M}
e^{2if_2}
q^{-1}\frac{q\specmap(y_k)-w_j+\ihalf g^{-1}}{q^{-1}\specmap(y_k)-w_j-\ihalf g^{-1}}\,,
\\\nn
1\eq
e^{2i(f_2-f_3) K}
\prod_{j=1}^{N}
e^{-2if_2}
q\,
\frac{w_k-q^{-1}\specmap(y_j)+\ihalf g^{-1}}{w_k-q\specmap(y_j)-\ihalf g^{-1}}
\mathop{\prod_{j=1}}_{j\neq k}^{M}
\frac{q^{-1}w_k-qw_j-\ihalf(q+q^{-1})g^{-1}}{qw_k-q^{-1}w_j+\ihalf(q+q^{-1})g^{-1}} \,.
\>
These phase factors therefore enter the momentum relation
$e^{ip_k}=R\lvl{II,I}(y_k,x)$ as follows
\[
e^{ip_k}=e^{i(f_1+f_2-f_3)} q^{C+1/2}U\,\frac{y_k-\xm{}}{y_k-\xp{}}
\,.
\]
Due to the logarithmic derivatives in
\eqref{eq:engspec} the expression for the energy is unaffected by the
phase factors.
It however receives contributions from the rescaling
and shifts in \eqref{eq:allowedtransf}
\[\label{eq:EnergyTrans}
E'=(a_0 E_0-a_1+a_2) K+2a_3M+\sum_{j=1}^{N} \bigbrk{a_0E(y_j)+a_1-a_3}.
\]

\paragraph{Dual Picture.}

The procedure to obtain the spectrum in the dual picture
discussed in \secref{sec:Dual} is the same.
Here the ferromagnetic vacuum is based on $\psi^1$'s,
the magnons flip them to $\phi^1$'s and
auxiliary excitations flip the latter to $\phi^2$'s.
The twisted version of the dual Bethe equations in \eqref{eq:BetheEquationsDual}
have the following insertions of phase factors
\<
\label{eq:BetheEquationsDualPhase}
1\eq
\lrbrk{
e^{i(f_1+f_2-f_3)}
q^{C+1/2}U\,\frac{\tilde y_k-\xm{}}{\tilde y_k-\xp{j}}}^K
\prod_{j=1}^{M}e^{-2if_2}q\,\frac{q^{-1}\specmap(\tilde y_k)-w_j-\ihalf g^{-1}}{q\specmap(\tilde y_k)-w_j+\ihalf g^{-1}}\,,
\\\nn
1\eq
e^{2i(-f_1-f_2)K}\prod_{j=1}^{\tilde N}
e^{2if_2}
q^{-1}\frac{w_k-q\specmap(\tilde y_j)-\ihalf g^{-1}}{w_k-q^{-1}\specmap(\tilde y_j)+\ihalf g^{-1}}
\mathop{\prod_{j=1}}_{j\neq k}^{M}
\frac{qw_k-q^{-1}w_j+\ihalf (q+q^{-1}) g^{-1}}{q^{-1}w_k-qw_j-\ihalf (q+q^{-1})g^{-1}}\,.
\>
To obtain the correct momentum relation we have to
remember that the vacuum as well as the magnons are fermionic excitations.
Therefore there is an implicit factor of $-1$ in the momentum
relation $e^{i\tilde p_k}=-\tilde R\lvl{II,I}(\tilde y_k,x)$
\[
e^{i\tilde p_k}=-e^{i(f_3-f_1-f_2)} q^{-C-1/2}U^{-1}\,\frac{\tilde y_k-\xp{}}{\tilde y_k-\xm{}}
\,,
\qquad
e^{iP}=(-1)^{K+\tilde N-1}
\prod_{j=1}^{\tilde N} e^{i\tilde p_j}.
\]
Note that the definition of the total momentum $P$
via the cyclic shift operator requires taking
into account the statistics of the vacuum sites and excitations.
The energy relation uses the above dispersion relation $E(y)$ up to a sign
\[
E'=(-a_0E_0-a_3+a_2) K+2a_1M+\sum_{j=1}^{\tilde N} \bigbrk{-a_0E(\tilde y_j)-a_1+a_3}.
\]
The equality of the total momentum $P$ and total energy $E'$
in both pictures makes use of the identity
\eqref{eq:AllMomConstr}
which guarantees for all eigenstates that
\[
\prod_{j=1}^N e^{ip_j}
\prod_{j=1}^{\tilde N} e^{-i\tilde p_j}
=(-1)^{K+\tilde N-1},
\qquad
\sum_{j=1}^N E(y_j)
+\sum_{j=1}^{\tilde N} E(\tilde y_j)
=-2E_0 K.
\]

\subsection{Electronic Oscillator Notation}
\label{eq:Electronic}

We have established the Hamiltonian in terms of a matrix acting
on a spin chain based on four-dimensional vector spaces.
Let us rewrite the four-dimensional vector space
in terms of fermionic ``electron'' creation and annihilation operators $c^\dagger_\alpha$ and $c_\alpha$
as it is usually done in the Hubbard model literature.

The map between vector states and electronic states reads%
\footnote{More generally, we may introduce different
normalization factors for the four states.
By rescaling the state $\state{\circ}$ and the generators
$c^\dagger_{1},c^\dagger_{2}$ these can all
be absorbed into a single variable $\kappa$.}
\[
\state{\phi^1}=\state{\circ},\qquad
\state{\phi^2}=\kappa c^\dagger_{1}c^\dagger_{2}\state{\circ}, \qquad
\state{\psi^1}=c^\dagger_{1}\state{\circ},\qquad
\state{\psi^2}=c^\dagger_{2}\state{\circ}.
\]
where creation and annihilation operators satisfy the following
algebra
\[
\acomm{c_{\alpha}^{}}{c^{\dagger}_{\beta}}=\delta_{\alpha\beta},
\qquad
\acomm{c_{\alpha}^{}}{c_{\beta}^{}} = \acomm{c^{\dagger}_{\alpha}}{c^{\dagger}_{\beta}} = 0.
\]
We define the number operator by
\[\label{eq:DensityOperator}
n_{\alpha}=c_{\alpha}^\dagger c^{}_{\alpha}.
\]
Note that $n_{\alpha}^2=n_{\alpha}$, hence $n_{\alpha}$ is a projector
onto the subspace spanned by $\state{\psi^\alpha}, \state{\phi^2}$.
The following projectors can be specified
\<
\ket{\phi^1}\bra{\phi_1}\eq (1-n_{1})(1-n_{2}),\nln
\ket{\phi^2}\bra{\phi_2}\eq n_{1}n_{2},\nln
\ket{\psi^1}\bra{\psi_1}\eq n_{1}(1-n_{2}),\nln
\ket{\psi^2}\bra{\psi_2}\eq n_{2}(1-n_{1}).
\>
From these projectors many useful combinations can be assembled
straightforwardly.
The Cartan generators can be rewritten in terms of number operators as follows
\[
\gen{H}_1=n_{1}+n_{2}-1,\qquad
\gen{H}_3=n_{2}-n_{1}.
\]

For dealing with spin chains we introduce multi-site oscillators by
adding a site index $j,k$. The oscillator algebra
becomes
\[\label{eq:oscalg}
\acomm{c^{}_{\alpha,j}}{c^\dagger_{\beta,k}}=\delta_{jk}\delta_{\alpha\beta},
\qquad
\acomm{c^{}_{\alpha,j}}{c^{}_{\beta,k}} = \acomm{c^\dagger_{\alpha,j}}{c^{\dagger}_{\beta,k}} = 0.
\]
By means of the projectors specified above we can spell out the Hamiltonian
of our model in oscillator notation.
Thus for the operator from \tabref{tab:HamiltonianGeneral}
we have the expression listed in \tabref{tab:SecondQuantHam}.
\begin{table}
\<
\ham_{j,k}\eq
\frac{A-B}{q+q^{-1}}\,(c_{1,j}^\dagger c_{2,j}^\dagger c_{2,k}^{} c_{1,k}^{}
                       +c_{1,k}^\dagger c_{2,k}^\dagger c_{2,j}^{} c_{1,j}^{})
\nl
-\frac{D-E}{q+q^{-1}}\,(c_{1,j}^\dagger c_{2,k}^\dagger c_{2,j}^{} c_{1,k}^{}
                       +c_{1,k}^\dagger c_{2,j}^\dagger c_{2,k}^{} c_{1,j}^{})
\nl
+\frac{1}{q+q^{-1}}\,c_{1,j}^\dagger c_{1,k}^{}\bigbrk{q^{-1} \kappa^{-1} C (1-n_{2,j})n_{2,k}-q\kappa Fn_{2,j}(1-n_{2,k})}
\nl
+\frac{1}{q+q^{-1}}\,c_{2,j}^\dagger c_{2,k}^{}\bigbrk{\kappa^{-1} C (1-n_{1,j})n_{1,k}-\kappa Fn_{1,j}(1-n_{1,k})}
\nl
+\frac{1}{q+q^{-1}}\,c_{1,k}^\dagger c_{1,j}^{}\bigbrk{q^{-1}\kappa F(1-n_{2,j})n_{2,k}-q\kappa^{-1} C n_{2,j}(1-n_{2,k})}
\nl
+\frac{1}{q+q^{-1}}\,c_{2,k}^\dagger c_{2,j}^{}\bigbrk{\kappa F(1-n_{1,j})n_{1,k}-\kappa^{-1} C n_{1,j}(1-n_{1,k})}
\nl
+c_{1,j}^\dagger c_{1,k}^{}\bigbrk{G(1-n_{2,j})(1-n_{2,k})-Ln_{2,j}n_{2,k}}
\nl
+c_{2,j}^\dagger c_{2,k}^{}\bigbrk{G(1-n_{1,j})(1-n_{1,k})-Ln_{1,j}n_{1,k}}
\nl
+c_{1,k}^\dagger c_{1,j}^{}\bigbrk{L(1-n_{2,j})(1-n_{2,k})-Gn_{2,j}n_{2,k}}
\nl
+c_{2,k}^\dagger c_{2,j}^{}\bigbrk{L(1-n_{1,j})(1-n_{1,k})-Gn_{1,j}n_{1,k}}
\nl
+A
+(K-A)(n_{1,j}+n_{2,j})
+(H-A)(n_{1,k}+n_{2,k})
\nl
+(A+D-H-K)(n_{1,j}n_{1,k}+n_{2,j}n_{2,k})
\nl
+\lrbrk{A-2H+\frac{qA+q^{-1}B}{q+q^{-1}}}n_{1,k}n_{2,k}
+\lrbrk{A-2K+\frac{q^{-1}A+qB}{q+q^{-1}}}n_{1,j}n_{2,j}
\nl
+\lrbrk{A-H-K+\frac{qD+q^{-1}E}{q+q^{-1}}}n_{1,j}n_{2,k}
+\lrbrk{A-H-K+\frac{q^{-1}D+qE}{q+q^{-1}}}n_{2,j}n_{1,k}
\nl
+\lrbrk{-A-D+2H+2K-\frac{qA+q^{-1}B}{q+q^{-1}}-\frac{q^{-1}D+qE}{q+q^{-1}}}n_{2,j}n_{1,k}n_{2,k}
\nl
+\lrbrk{-A-D+2H+2K-\frac{qA+q^{-1}B}{q+q^{-1}}-\frac{qD+q^{-1}E}{q+q^{-1}}}n_{1,j}n_{1,k}n_{2,k}
\nl
+\lrbrk{-A-D+2H+2K-\frac{q^{-1}A+qB}{q+q^{-1}}-\frac{qD+q^{-1}E}{q+q^{-1}}}n_{1,j}n_{2,j}n_{2,k}
\nl
+\lrbrk{-A-D+2H+2K-\frac{q^{-1}A+qB}{q+q^{-1}}-\frac{q^{-1}D+qE}{q+q^{-1}}}n_{1,j}n_{2,j}n_{1,k}
\nl
+\lrbrk{3A+B+3D+E-4H-4K}n_{1,j}n_{2,j}n_{1,k}n_{2,k},
\nn
\>
\caption{Oscillator form of the $\Qgrp(\alg{su}(2)\times\alg{su}(2))$
spin chain Hamiltonian.}
\label{tab:SecondQuantHam}
\end{table}
%

\subsection{Connection with the One-Dimensional Hubbard Model}
\label{sec:Nondefcase}

In two special limits the spin chain with undeformed $\alg{h}$
symmetry becomes equivalent to the one-dimensional Hubbard model \cite{Beisert:2006qh}.
This conclusion was drawn first by comparison
of the Bethe equations of both models
and second by comparison of their R-matrices.
Here we show explicitly how the Hubbard model
can be embedded into the $\alg{h}$ spin chain
on the level of Hamiltonians
as a preparation for the more complicated
comparison with the Alcaraz--Bariev models.

The problem in the comparison of Hamiltonians consists in the fact that
the original Hubbard model Hamiltonian
exhibits manifest $\alg{su}(2)$ symmetry
while our Hamiltonian
has manifest $\Qgrp(\alg{su}(2)\times\alg{su}(2))$ symmetry.
We therefore need to recover an additional
manifest $\alg{su}(2)$ symmetry and set $q=1$
before the Hamiltonians can be compared.

\paragraph{Hubbard Hamiltonian.}

The standard Hubbard model Hamiltonian
based on the electronic oscillator algebra \eqref{eq:oscalg}
reads
\[\label{eq:HubbardHamClass}
\mathcal{H}^{\textrm{Hub}}_{j,k} =
\sum\limits_{\alpha=1,2}
\lrbrk{c^{\dagger}_{\alpha,j} c^{}_{\alpha,k} + c^{\dagger}_{\alpha,k} c^{}_{\alpha,j}}
+ U n_{1,j}n_{2,j}.
\]
It has one manifest $\alg{su}(2)$ symmetry
and a twisted $\alg{su}(2)$ symmetry.
In order to match with our Hamiltonian in
\tabref{tab:HamiltonianGeneral,tab:HamCoeff}
we should first make the second $\alg{su}(2)$ symmetry manifest
by applying a transformation \eqref{eq:allowedtransf}
\[
\ham'_{j,k}=
\Twist^{}_{j,k}\ham_{j,k}^{\mathrm{Hub}}\Twist_{j,k}^{-1}
+\quarter U (1-n_{1,j}-n_{1,k}-n_{2,j}-n_{2,k}
-2n_{1,j}n_{2,j}+2n_{1,k}n_{2,k})
\]
with the twist
\[\label{eq:TwistHubbard}
\Twist_{j,k} =
\exp \bigbrk{\ihalf \pi (n_{1,k}+n_{2,k}-1)}
=
\exp \bigbrk{\ihalf \pi \gen{H}_{1,k}}
.
\]
The resulting Hamiltonian reads
\[
\ham'_{j,k}=
\sum\limits_{\alpha=1,2}
\lrbrk{-ic^{\dagger}_{\alpha,j} c^{}_{\alpha,k}+ic^{\dagger}_{\alpha,k} c^{}_{\alpha,j}}
+\quarter U \sum_{\ell=k,j}\bigbrk{n_{1,\ell}n_{2,\ell}+(1-n_{1,\ell})(1-n_{2,\ell})-\half}.
\]
It has manifest $\Qgrp(\alg{su}(2)\times\alg{su}(2))$ invariance
with $q=1$
and takes the form in \tabref{tab:HamiltonianGeneral} with
\[\label{eq:HubbardTwistCoeff}
A'=B'=-D'=-E'=\sfrac{1}{4} U,\quad
G'=-L'=i,\quad
C',F'=-2 (i\kappa)^{\pm 1},\quad
H'=K'=0.
\]
%

\paragraph{Comparison of Hamiltonians.}

It was shown in \cite{Beisert:2006qh}
that the Hubbard model corresponds to the values
$(0,\infty)$ or $(\infty,0)$
of the parameters $(\xp{},\xm{})$.
Moreover we have to set $q=1$ for the undeformed setup.

In fact the Hamiltonian is singular at these
parameter values and there are several ways in
which the singular limit can be taken.
For instance, we can first set $q=1$ and then take the limit
\[
(\xp{},\xm{})\to (0,\infty).
\]
Alternatively one can take the limits
\[
(\xp{},\xm{})\to(0,\antimap(0))\qquad\mbox{or}\qquad
(\xp{},\xm{})\to(\antimap(\infty),\infty)
\]
and only afterwards $q\to 1$.
We rescale the Hamiltonian to match the above $G'$
\[
\ham'_{j,k}=\frac{i}{G}\,\ham_{j,k}.
\]
This leads to,
see \eqref{eq:IdentityHam}
\[
A'=B'=-D'=-E'=\frac{1}{4g}\,,\quad
G'=-L'=i,\quad
C'=F'=-2,\quad
H'=K'=0.
\]
which agrees exactly with the twisted
Hubbard Hamiltonian \eqref{eq:HubbardTwistCoeff}
upon identification of $\kappa=-i$ and
\[
U=\frac{1}{g}\,.
\]
%

\subsection{Connection with the Alcaraz--Bariev Models}
\label{sec:ABCompare}

Here we study the relation between our model
and an integrable spin chain proposed by Alcaraz and Bariev \cite{Alcaraz:1999aa}
(AB model).

\paragraph{The Alcaraz--Bariev Hamiltonian.}

Alcaraz and Bariev \cite{Alcaraz:1999aa}
proposed a spin chain model
based on the electronic states introduced in \secref{eq:Electronic}.
The Hamiltonian is the following general deformation
of the Hubbard Hamiltonian \eqref{eq:HubbardHamClass}
\<
\ham^{\mathrm{AB}}_{j,k}\eq
(c_{1,j}^{\dagger} c^{}_{1,k}+c_{1,k}^{\dagger} c^{}_{1,j})
(1 + t_{11}n_{2,j} + t_{12}n_{2,k}+ t'_{1}n_{2,j}n_{2,k})
\nl
+
(c_{2,j}^{\dagger} c^{}_{2,k}+c_{2,k}^{\dagger} c^{}_{2,j})
(1 + t_{21}n_{1,j} + t_{22}n_{1,k}+ t'_{2}n_{1,j}n_{1,k})
\nl
+J(c_{1,j}^{\dagger}c_{2,k}^{\dagger} c^{}_{2,j}c^{}_{1,k}
  +c_{1,k}^{\dagger}c_{2,j}^{\dagger} c^{}_{2,k}c^{}_{1,j} )
+t\indup{p}(c_{1,j}^{\dagger}c_{2,j}^{\dagger}c^{}_{2,k}c^{}_{1,k}
  + c_{1,j}^{\dagger}c_{2,j}^{\dagger}c^{}_{2,k}c^{}_{1,k})
\nl
+V_{11}n_{1,j}n_{1,k}
+V_{12}n_{1,j}n_{2,k}
+V_{21}n_{2,j}n_{1,k}
+V_{22}n_{2,j}n_{2,k}
+Un_{1,j}n_{2,j}
\nl
+V_3^{(1)}n_{2,j}n_{1,k}n_{2,k}
+V_3^{(2)}n_{1,j}n_{1,k}n_{2,k}
+V_3^{(3)}n_{1,j}n_{2,j}n_{2,k}
+V_3^{(4)}n_{1,j}n_{2,j}n_{1,k}
\nl
+V_4 n_{1,j}n_{2,j}n_{1,k}n_{2,k},
\>
where
\[\begin{array}{rclcrclcrcl}
t_{11}\eq t_4-1,&&
t_{12}\eq t_3-1,&&
t'_1\eq t_5-t_3-t_4+1,
\\[3pt]
t_{21}\eq t_1-1,&&
t_{22}\eq t_2-1,&&
t'_2\eq t_5-t_1-t_2+1.
\end{array}
\]
It was found to be integrable in four
cases which we will denote by A$^\pm$ and B$^\pm$.
The coefficients are related as follows
in the case $A^{\pm}$
\[\begin{array}[b]{c}
t_1 = \epsilon t_2 = t_3 =\epsilon t_4 =\sin{\vartheta},\quad
t_5=\epsilon=\pm 1,
\\[3pt]
J=-\epsilon t\indup{p}=-\half \epsilon U=V_{12}e^{2\eta}=V_{21}e^{-2\eta}=\cos{\vartheta},
\\[3pt]
V_{11}=V_{22}=V_3^{(1)}=V_3^{(2)}=V_3^{(3)}=V_3^{(4)}=V_4=0,
\end{array}
\]
and in the case B$^\pm$
\<\label{eq:caseb}
\begin{array}[b]{c}
t_1 = \epsilon t_2 = \epsilon t_3 e^{2\eta} = t_4 e^{-2\eta} =\sin{\vartheta}, \quad
t_5=\epsilon=\pm 1,
\\[3pt]
J=-\epsilon t\indup{p}=V_{12}e^{2\eta}=V_{21}e^{-2\eta}=\cos{\vartheta},
\quad
U=2t_p +\sin\vartheta\tan\vartheta(e^\eta-\epsilon e^{-\eta})^2,
\\[3pt]
V_{11}=V_{22}=V_3^{(2)}=V_3^{(4)}=V_4=0,\quad V_3^{(1)}=-V_3^{(3)}=V_{12}-V_{21}.
\end{array}
\>
with the free parameters $\vartheta,\eta$.
The parameter $\epsilon=\pm 1$ distinguishes between
the models A$^\pm$, B$^\pm$, respectively.

Let us introduce a replacement of the parameter $\eta$
that will become very useful in the following discussion.
In the A$^\pm$ case we shall set
\[\label{eq:ABreparA}
e^{2\eta}=\frac{\epsilon}{\xi} \frac{1-\xi\cos\vartheta}{\xi-\cos\vartheta}\,,
\]
while for the case B$^\pm$ we use the definition
\[\label{eq:ABreparB}
e^{2\eta}=\epsilon \xi\, \frac{1-\xi\cos\vartheta}{\xi-1\cos\vartheta}\,.
\]
%

\paragraph{$\Qgrp(\alg{su}(2)\times\alg{su}(2))$ Symmetry.}

First of all we note that the AB Hamiltonian
does not immediately match
the expression of the general $\Qgrp(\alg{su}(2)\times\alg{su}(2))$-invariant
Hamiltonian in \tabref{tab:HamiltonianGeneral}.
However, we can apply the transformation \eqref{eq:allowedtransf}
which preserves integrability in order to restore the
$\Qgrp(\alg{su}(2)\times\alg{su}(2))$ invariance.

We find that the AB Hamiltonian can be brought into
the form in \tabref{tab:HamiltonianGeneral} if some of the parameters
are related: Explicitly we find that the $V_k$ parameters must be
related by
\[\label{eq:Vks}
\begin{array}[b]{rclcrcl}
V_{11}\eq V,
&&
V_3^{(1)}\eq -e^{2if_3}q J+e^{2if_1}q^{-1}t\indup{p}-2V,
\\[3pt]
V_{12}\eq e^{2if_3}q^{-1} J+V,
&&
V_3^{(2)}\eq -e^{2if_3}q^{-1} J+e^{2if_1}q^{-1}t\indup{p}-2V,
\\[3pt]
V_{21}\eq e^{2if_3}qJ+V,
&&
V_3^{(3)}\eq -e^{2if_3}q^{-1} J+e^{2if_1}qt\indup{p}-2V,
\\[3pt]
V_{22}\eq V
&&
V_3^{(4)}\eq -e^{2if_3}qJ+e^{2if_1}qt\indup{p}-2V,
\\[3pt]
V_4\mathrel{}&\hspace{-2\arraycolsep}=\hspace{-2\arraycolsep}&
\multicolumn{5}{l}{\mathrel{}e^{2if_3}(q+q^{-1})J-e^{2if_1}(q+q^{-1})t\indup{p}+4V,}
\end{array}\]
with a new parameter $V$. Furthermore the $t_k$ parameters must
take the form
\[\label{eq:tks}
t_1=t_0,
\quad
t_2=-e^{2if_1+2if_3}t_0,
\quad
t_3=-e^{2if_1}q^{-1}t_0,
\quad
t_4=e^{2if_3}qt_0,
\quad
t_5=-e^{2if_1+2if_3},
\]
with a new parameter $t_0$.
The parameters for the transformation take the form
\<
\half a_1=a_2\eq -\quarter e^{2if_1}(q+q^{-1})-\quarter U,
\nln
\half b_1=b_2\eq -\quarter e^{2if_1}(q-q^{-1})-\quarter U,
\nln
a_3=b_3\eq 0,
\>
and the twist parameters are constrained by
\[
e^{4if_1}=
e^{4if_2}=
e^{4if_3}=
e^{2if_3+2if_2}=1.
\]
Note that we have set $a_0=1$.
We have also adjusted $a_2,b_2$ such that
$H'=K'=0$.
The resulting values for the parameters of \tabref{tab:HamiltonianGeneral}
read
\[\label{eq:ABCoeffs}
\begin{array}[b]{rclcrcl}
A'\eq\quarter e^{2if_1}(q+q^{-1})t\indup{p}+\quarter U,
&&
D'\eq-\quarter e^{2if_1}(q+q^{-1})t\indup{p}-\quarter U+V,
\\[3pt]
B'\eq-\sfrac{3}{4} e^{2if_1}(q+q^{-1})t\indup{p}+\quarter U,
&&
E'\eq e^{2if_3}(q+q^{-1})J+\sfrac{3}{4} e^{2if_1}(q+q^{-1})t\indup{p}-\quarter U+V,
\\[3pt]
C'\eq-e^{if_1+if_3}\kappa t_0(q+q^{-1}),
&&
F'\eq-e^{-if_1-if_3}\kappa^{-1} t_0(q+q^{-1}),
\\[3pt]
G'\eq e^{-if_1+if_2-if_3},
&&
L'\eq e^{+if_1+if_2-if_3},
\\[3pt]
H'\eq 0,
&&
K'\eq 0.
\end{array}
\]
The ten parameters $A',\ldots,L'$ are thus given
as functions of six parameters
$J,U,t_0,t\indup{p},V$ and $\kappa$.
Three further parameters are implicitly given by $a_0,a_2$
(overall scaling and identical shift) and $b_2$
(affecting $H$ and $K$ only)
which preserve the form in \tabref{tab:HamiltonianGeneral}
and which we have fixed above.
One of the ten parameters cannot be chosen continuously
due to the relation
$G'/L'=e^{2if_1}=\pm 1$.

\paragraph{Comparison of Hamiltonians.}

We find that in the cases A$^\pm$ the symmetry can be restored
only if $e^{2\eta}=\epsilon$ or $\cos\vartheta=0$.
Moreover, $q=\pm 1$. This means that we cannot
explain the two-parametric model A$^\pm$ in general
using our methods.
We shall therefore disregard this case here.
In order to relate the cases A$^\pm$ to our model
we would have to find a more general transformation than \eqref{eq:allowedtransf}
to recover the symmetry. It would be interesting to find out if this is
possible.

Conversely, the models B$^{\pm}$ can be brought to the
form in \tabref{tab:HamiltonianGeneral},
and in what follows we shall exclusively consider these cases.
We match the $t_k$ in
\eqref{eq:tks,eq:caseb}
by demanding
\[
\epsilon=-e^{2if_1+2if_3},
\qquad
q=e^{2\eta}e^{2if_3}.
\]
The correct expressions for $V_k$ in \eqref{eq:Vks} follow by imposing
the remaining four relations in \eqref{eq:caseb}
\[
V=0,
\qquad
J=-\epsilon t\indup{p},
\qquad
t\indup{p}^2+t_0^2=1,
\qquad
U=
2t\indup{p}+(t\indup{p}-t\indup{p}^{-1})\bigbrk{\epsilon (e^{2\eta}+e^{-2\eta})-2}.
\]
Incidentally, these are precisely the constraints
for our integrable Hamiltonian \eqref{eq:IntConstr}
when using the general AB coefficients \eqref{eq:ABCoeffs}.
Therefore it should be possible to find suitable
parameters $\xpm{},g$ in \tabref{tab:HamCoeff}
to match the case B$^\pm$.

We start the comparison by making the convenient choice
\[
f_1=-\quarter \pi (1+\epsilon),\qquad
f_2=f_3=0,\qquad
\epsilon=-e^{2if_1},
\qquad
q=e^{2\eta}.
\]
To match $G'=e^{-if_1}$
we have rescale our Hamiltonian
in \tabref{tab:HamiltonianGeneral,tab:HamCoeff}
\[
\ham'_{j,k}=
\frac{e^{-if_1}}{G}\,\ham_{j,k}.
\]
Furthermore we have to match our normalization $C'=F'$
by setting
\[
\kappa=e^{-if_1}.
\]
The ratio $L'/G'=L/G=e^{2if_1}=-\epsilon$ is determined through the $\xpm{}$
\[
\frac{L}{G}=q^{-1}\frac{\xp{}-\antimap(\xp{})}{\xm{}-\antimap(\xm{})}
=-\epsilon.
\]
This constraint is satisfied when (our) $U^2=-\epsilon$
or when $q^{2C}=-\epsilon$. Both values are possible, let us discuss them separately:

The condition $q^{2C}=-\epsilon$ holds for the following two pairs $(\xp{},\xm{})$
\[\label{eq:xpmq1}
\xp{}=\frac{iq}{2g}\,\frac{1\pm \sqrt{1-4\epsilon g^2(q^{1/2}-\epsilon q^{-1/2})^2}}{1-\epsilon q}\,,
\quad
\xm{}=\frac{\epsilon i}{2g}\,\frac{1\mp\sqrt{1-4\epsilon g^2(q^{1/2}-\epsilon q^{-1/2})^2}}{1-\epsilon q}\,.
\]
We match $\vartheta$ of B$^\pm$ using the coefficients $A'-B'$ or $C'$
\<
\cos\vartheta \eq -\epsilon t\indup{p}
= \frac{B'-A'}{e^{2\eta}+e^{-2\eta}}
= \frac{e^{-if_1}}{e^{2\eta}+e^{-2\eta}}\,\frac{B-A}{G}
=\pm\frac{g(e^{\eta}-\epsilon e^{-\eta})^2}{\sqrt{1-4\epsilon g^2(e^{\eta}-\epsilon e^{-\eta})^2}}\,,
\nln
\sin\vartheta\eq t_0
=-\frac{C'}{e^{2\eta}+e^{-2\eta}}
=-\frac{e^{-if_1}}{e^{2\eta}+e^{-2\eta}}\,\frac{C}{G}
=\mp\frac{\sqrt{1-g^2(e^{2\eta}-e^{-2\eta})^2}}{\sqrt{1-4\epsilon g^2(e^{\eta}-\epsilon e^{-\eta})^2}}\,.
\>
These two relations are compatible with the identity $\cos^2\vartheta+\sin^2\vartheta=1$ and
they can be solved for $g$
\[
g=\frac{\cos\vartheta}{4\cosh(\eta+if_1)\sqrt{\cosh^2(\eta+if_1)-\cos^2\vartheta}}\,.
\]
In the parametrization \eqref{eq:ABreparB} the above relations simplify significantly
\[
g=\frac{\xi(1-\xi\cos\vartheta)(\xi-\cos\vartheta)}
{\cos\vartheta(\xi-e^{+i\vartheta})(\xi-e^{-i\vartheta})(\xi^2-1)}\,,
\qquad
q=\epsilon\xi\, \frac{1-\xi\cos\vartheta}{\xi-\cos\vartheta}\,,
\qquad
\xpm{}=i\epsilon \xi^{\pm 1}.
\]

Conversely, the condition $U^2=-\epsilon$ holds for the following two pairs $(\xp{},\xm{})$
(note the change of sign w.r.t.\ \eqref{eq:xpmq1})
\[
\xp{}=\frac{iq}{2g}\,\frac{1\pm \sqrt{1+4\epsilon g^2(q^{1/2}+\epsilon q^{-1/2})^2}}{1+\epsilon q}\,,
\quad
\xm{}=\frac{-\epsilon i}{2g}\,\frac{1\pm\sqrt{1+4\epsilon g^2(q^{1/2}+\epsilon q^{-1/2})^2}}{1+\epsilon q}\,.
\]
We match $\vartheta$ of B$^\pm$ using the coefficients $A'-B'$ or $C'$
\<
\cos\vartheta \eq -\epsilon t\indup{p}
= \frac{B'-A'}{e^{2\eta}+e^{-2\eta}}
= \frac{e^{-if_1}}{e^{2\eta}+e^{-2\eta}}\,\frac{B-A}{G}
=\pm\frac{\sqrt{1-g^2(e^{2\eta}-e^{-2\eta})^2}}{\sqrt{1+4\epsilon g^2(e^{\eta}+\epsilon e^{-\eta})^2}}\,,
\nln
\sin\vartheta\eq t_0
=-\frac{C'}{e^{2\eta}+e^{-2\eta}}
=-\frac{e^{-if_1}}{e^{2\eta}+e^{-2\eta}}\,\frac{C}{G}
=\pm\frac{g(e^{\eta}+\epsilon e^{-\eta})^2}{\sqrt{1+4\epsilon g^2(e^{\eta}+\epsilon e^{-\eta})^2}}\,.
\>
These two relations are compatible with the identity $\cos^2\vartheta+\sin^2\vartheta=1$ and
they can be solved for $g$
\[
g=\frac{\sin\vartheta}{4\sinh(\eta+if_1)\sqrt{\sinh^2(\eta+if_1)+\sin^2\vartheta}}\,.
\]
Again the parametrization
\eqref{eq:ABreparB} simplifies the above relations
\<
g\eq \frac{\sin\vartheta (1-\xi\cos\vartheta)(\xi-\cos\vartheta)}
       {\cos\vartheta (\xi-e^{+i\vartheta})(\xi-e^{-i\vartheta})((\xi+\xi^{-1})\cos\vartheta-2)}\,,
\nln
q\eq\epsilon\xi\, \frac{1-\xi\cos\vartheta}{\xi-\cos\vartheta}\,,
\qquad
\xp{}=\frac{i \epsilon\xi\sin\vartheta}{\cos\vartheta-\xi}\,,
\qquad
\xm{}=\frac{i \epsilon\sin\vartheta}{1-\xi\cos\vartheta}\,.
\>

Note that several different points in the
constrained parameter space of $\xpm{},g$
correspond to the same Hamiltonian
given in terms of $\vartheta$.
Furthermore, we have not been very careful about selection of branches
of $U=\sqrt{U^2}$, $q^C=\sqrt{q^{2C}}$ in
\eqref{eq:q2c,eq:U2} as well as for the above square roots.
We expect that these signs ambiguities are equivalent to shifting
$\vartheta$ and $i\eta$ by multiples of $\pi/2$.

\paragraph{Comparison of Bethe Equations.}

Here we will show how the B$^\pm$ models proposed by Alcaraz and Bariev \cite{Alcaraz:1999aa}
can be embedded into our $\Qgrp(\alg{h})$-spin chain model
by comparing Bethe equations.

Consider the Alcaraz--Bariev Bethe equations \cite{Alcaraz:1999aa}
\<
1\eq\lrbrk{z_k}^K
\prod\limits_{j=1}^{M}
\frac{\sinh(\lambda_k-\lambda^{(1)}_j+\eta)}{\sinh(\lambda_k-\lambda^{(1)}_j-\eta)}\,,
\nln
1\eq\prod_{j=1}^{N}
\frac{\sinh(\lambda^{(1)}_k-\lambda_j+\eta)}{\sinh(\lambda^{(1)}_k-\lambda_j-\eta)}
\mathop{\prod_{j=1}}_{j\neq k}^{M}
\frac{\sinh(\lambda^{(1)}_k-\lambda^{(1)}_j-2\eta)}{\sinh(\lambda^{(1)}_k-\lambda^{(1)}_j+2\eta)}\,.
\>
By inserting $\Phi(z)$ into the expression for $\exp(2\lambda_k)$ in
\cite{Alcaraz:1999aa} we can write the relation between $z_k$ and
$\lambda_k$ as follows
\[\label{eq:ABlambda}
\exp(2\lambda_k)=e^{-2\eta} \frac{z^2_k+(\epsilon e^{+2\eta}-1) t\indup{p}^{-1} z_k-\epsilon e^{+2\eta}}
                       {z_k^2 + (\epsilon e^{-2\eta}  -1)t\indup{p}^{-1}z_k-\epsilon e^{-2\eta}}\,.
\]
The energy of an eigenstate is given by
\[\label{eq:ABEnergy}
E=\sum_{j=1}^N (z_j+z_j^{-1}).
\]

It is straightforward to match the auxiliary (second) Bethe equation
with our Bethe equations
\eqref{eq:BetheEquationsHomo}
where $f_2=f_3=0$
by equating $q=e^{2\eta}$ and
\<\label{eq:ParMatchBetheAB}
u(y_k)\eq \exp(2\lambda_k-2\lambda_0)-\frac{ig^{-1}}{q-q^{-1}}\,,
\nln
w_k\eq\exp(2\lambda^{(1)}_k-2\lambda_0)-\ihalf g^{-1}\frac{q+q^{-1}}{q-q^{-1}}\,.
\>
Matching of the first Bethe equation requires
\[\label{eq:zandy}
z_k=e^{-if_1}q^{-C-1/2}U^{-1} \frac{y_k-\xp{}}{y_k-\xm{}}\,,
\qquad
y_k = \frac{\xp{}-e^{if_1}q^{C+1/2}U \xm{} z_k }
{1-e^{if_1}q^{C+1/2}U z_k}.
\]
We can bring the above relations \eqref{eq:ParMatchBetheAB,eq:zandy}
between $\lambda_k$ and $y_k$ in the same form as \eqref{eq:ABlambda}
\[\label{eq:Ourlambda1}
\exp(2\lambda_k)=\exp(2\lambda_0)\lrbrk{u(y_k)+\frac{ig^{-1}}{q-q^{-1}}}
=\exp(2\lambda_0)\frac{ig^{-1}}{q-q^{-1}}\,y_k\antimap(y_k)
\]
with
\[\label{eq:Ourlambda2}
y_k\antimap(y_k)=q^{-2}\xp{}\antimap(\xp{})
\frac{z_k^2-e^{-if_1}U^{-1}q^{C+1/2}\bigbrk{2+(U^2-q^{-2C})/q^{-1}\xp{}\antimap(\xp{})}z_k+e^{-2if_1}q}
{z_k^2-e^{-if_1}U^{-1}q^{C-1/2}(U^2+q^{-2C})z_k+e^{-2if_1}q^{-1}}\,.
\]
We set
$q^{2C}=e^{2if_1}=-\epsilon$ and use \eqref{eq:xpmq1}.
Noting that
$q^{-1}\xp{}\antimap(\xp{})=q\xm{}\antimap(\xm{})=-\epsilon$
the prefactors imply that we have to choose the shift parameter
$\lambda_0$ as follows
\[
\exp(2\lambda_0)=\epsilon ig(q-q^{-1})\,.
\]
Some further manipulations then show that our
relation
\eqref{eq:Ourlambda1,eq:Ourlambda2}
is precisely the same as
the one from the Alcaraz--Bariev equations
\eqref{eq:ABlambda}.
Finally, the eigenstate energies $E$ \eqref{eq:ABEnergy}
match with our result $E'$ \eqref{eq:engABC,eq:momspec,eq:EnergyTrans}
using the transformation parameters
\[
a_0=e^{if_1}\,\frac{1}{G}\,,
\qquad
a_1=2e^{if_1}\,\frac{A}{G}\,,
\qquad
a_2=e^{if_1}\,\frac{A}{G}\,,
\qquad
a_3=0.
\]

\paragraph{The Spectrum of the A$^\pm$ Cases.}

The Bethe equations derived by Alcaraz and Bariev
are almost the same for the A$^\pm$ and B$^\pm$ cases.
Actually, the only difference is the definition
of a global parameter $r$ as a function of $\eta$, namely
\[
\cosh 2r=\epsilon \sin^2\vartheta+\cos^2\vartheta \cosh 2\eta
\qquad\mbox{vs.}\qquad
\cosh 2r=\cosh 2\eta
\]
for cases A and B, respectively.
This implies that set of solutions to the Bethe equations
for model A with $\eta=\eta\indup{A}$
is the same as the set of solutions for model B with $\eta=\eta\indup{B}$
where $\eta\indup{A}$ and $\eta\indup{B}$ are related by
\[
\epsilon \sin^2\vartheta+\cos^2\vartheta \cosh 2\eta\indup{A}=\cosh 2\eta\indup{B}
\]
Moreover the dispersion relation is the same for both cases, and
therefore the spectra of the models agree.
Note that our reparametrization
\eqref{eq:ABreparA,eq:ABreparB} of $\eta$ has been chosen carefully
so that the above relation holds for equal $\xi=\xi\indup{A}=\xi\indup{B}$.

In fact this leads to a puzzle concerning our above results:
We have not been able to find a transformation of the Hamiltonian
for model A that restores manifest $\Qgrp(\alg{su}(2)\times\alg{su}(2))$ symmetry.
Therefore we have not been able to relate this Hamiltonian to ours.
Nevertheless the matching of spectra implies that the
two Hamiltonians are indeed related by a similarity transformation
for a suitable choice of parameters.
Consequently, there must exist a more elaborate transformation to
make the $\Qgrp(\alg{su}(2)\times\alg{su}(2))$ symmetry
of the Hamiltonian of model A manifest and match it with our
Hamiltonian.

\paragraph{Comparison of Scattering Matrices.}

To shed more light onto the equivalence of the A$^\pm$ and B$^\pm$ cases
we shall compare the corresponding scattering matrices.
These were derived in \cite{Alcaraz:1999aa}
for excitations $c^\dagger_1$ and $c^\dagger_2$
with momentum $p=-i\log z$
above the (ferromagnetic) vacuum
and they can be compared to the results of \secref{sec:scatstate}.
We will use
the parametrization \eqref{eq:ABreparA,eq:ABreparB}
which allows us to compare the expressions directly without the further need to
transform parameters.
It turns out that the diagonal scattering matrix elements
precisely match
\[
(S\indup{A})^{12}_{12}(z_1,z_2)=(S\indup{A})^{21}_{21}(z_1,z_2)=
(S\indup{B})^{12}_{12}(z_1,z_2)=(S\indup{B})^{21}_{21}(z_1,z_2).
\]
The off-diagonal elements also agree up to a simple factor
\[
(S\indup{A})^{12}_{21}(z_1,z_2)=(S\indup{B})^{12}_{21}(z_1,z_2)\,\frac{f(z_1)}{f(z_2)}\,,\qquad
(S\indup{A})^{21}_{12}(z_1,z_2)=(S\indup{B})^{21}_{12}(z_1,z_2)\,\frac{f(z_2)}{f(z_1)}\,.
\]
with
\[
f(z)=\frac{1+\epsilon\xi z}{z+\epsilon\xi}\,.
\]

The fact that the off-diagonal elements differ by reciprocal
factors of $f(z_1)/f(z_2)$ alone implies that the spectra are
the same.
Therefore there must exist a similarity transformation between
the two Hamiltonians. In fact the transformation is simple for
scattering states: Each excitation $c^\dagger_2$ with momentum $p$
is multiplied by a factor of $f(e^{ip})$ while the
other excitations $c^\dagger_1$ are left alone. It is not hard
to convince oneself that this leads to the above relative
factors between the S-matrix elements.
However, it is not as straightforward to express the similarity
transformation for spin chain states because the
the transformation must be non-local
due to the momentum dependence of the factors $f(e^{ip})$.
Astonishingly the Hamiltonian remains local after the similarity
transformation which implies that the transformation must be of
a very special kind.

\paragraph{An Alternative Coalgebra?}

The above discussion has made it clear that the
A$^\pm$ and B$^\pm$ Hamiltonians are equivalent.
Consequently the A$^\pm$ Hamiltonian
must obey a $\Qgrp(\alg{su}(2)\times\alg{su}(2))$ symmetry
albeit in a non-standard form.
It is conceivable that
this symmetry is generated by an alternative coproduct
$\Qgrp(\alg{h})$.
This is in fact a promising idea because it
would explain the locality of the A$^\pm$ Hamiltonian:
Assuming that there exist an alternative
(fundamental) R-matrix, we could right away derive
a nearest neighbor Hamiltonian which is likely to be
the one of the A$^\pm$ cases.
Thus it would be exciting to find an alternative
coalgebra structure for $\Qgrp(\alg{h})$
and also find a suitable similarity transformation
to the canonical one derived in this paper.
Unfortunately this issue is outside the scope of the present paper.

\subsection{Quantum-Deformed EKS Model}

At the $(\xp{},\xm{})$ values $(0,0)$, $(\infty,\infty)$ as well as
$(\antimap(0),\antimap(0))$, $(\antimap(\infty),\antimap(\infty))$
one obtains the quantum-deformation of the EKS model
\cite{Essler:1992py}.
For instance at $(\xp{},\xm{})=(0,0)$ the parameters
of the rescaled Hamiltonian
$
\ham'_{j,k}=(q^{1/2}/G)\ham_{j,k}
$
read
\[A'=-B'=-D'=E'=\half (q+q^{-1}),\quad
G=L^{-1}=q^{1/2},\quad
C'=F'=H'=K'=0.
\]
It is not too hard to see that after a twist this leads to
the quantum-deformed EKS model,
i.e.\ an standard integrable nearest-neighbor spin chain
with $\Qgrp(\alg{u}(2|2))$ symmetry and spins
in the four-dimensional fundamental representation.

Note that this model is somewhat singular.
For instance the Bethe equations \eqref{eq:BetheEquationsHomo}
do not apply directly, but the limit
$(\xp{},\xm{})\to (0,0)$ has to be taken carefully.
In particular, there are two allowed limits of the $y_k$,
either $y_k\to 0$ or $y_k\to\antimap(0)$.
Let $N_1$ of them be of the first kind and $N_3=N-N_1$ of the second.
They should scale like
\[
y_k\sim\exp(2\lambda^{(1)}_k) q\xm{},
\qquad
y_{k+N_1}\sim\antimap\bigbrk{\exp(2\lambda^{(3)}_k) q\xm{}}.
\]
Furthermore, the $w_k$ must scale like
\[
w_k\sim -\frac{i}{2g}\,\frac{q+q^{-1}}{q-q^{-1}}+\frac{\exp(2\lambda^{(2)}_k)q\xm{}}{g^2(q-q^{-1})^2}\,.
\]
%
The limit of the Bethe equations \eqref{eq:BetheEquationsHomo}
agrees with the standard form of $\Qgrp(\alg{u}(2|2))$
Bethe equations, cf.\ \cite{Reshetikhin:1987bz,Ribeiro:2005aa},
up to a twist
\<
1\eq e^{+K\hbar}\lrbrk{\frac{\sinh(\lambda^{(1)}_k-\hbar)}{\sinh(\lambda^{(1)}_k+\hbar)}}^K
\prod\limits_{j=1}^{M}
\frac{\sinh(\lambda^{(1)}_k-\lambda^{(2)}_j+\hbar)}
     {\sinh(\lambda^{(1)}_k-\lambda^{(2)}_j-\hbar)}\,,
\nln
1\eq
\prod_{j=1}^{N_1}
\frac{\sinh(\lambda^{(2)}_k-\lambda^{(1)}_j+\hbar)}
     {\sinh(\lambda^{(2)}_k-\lambda^{(1)}_j-\hbar)}
\mathop{\prod_{j=1}}_{j\neq k}^{M}
\frac{\sinh(\lambda^{(2)}_k-\lambda^{(2)}_j-2\hbar)}
     {\sinh(\lambda^{(2)}_k-\lambda^{(2)}_j+2\hbar)}
\prod_{j=1}^{N_3}
\frac{\sinh(\lambda^{(2)}_k-\lambda^{(3)}_j+\hbar)}
     {\sinh(\lambda^{(2)}_k-\lambda^{(3)}_j-\hbar)}\,,
\nln
1\eq e^{-K\hbar}
\prod\limits_{j=1}^{M}
\frac{\sinh(\lambda^{(3)}_k-\lambda^{(2)}_j+\hbar)}
     {\sinh(\lambda^{(3)}_k-\lambda^{(2)}_j-\hbar)}
\>
with $q=e^{2\hbar}$. Here the two different kinds of
$y_k$ have led to the additional Bethe equation
needed for standard $\Qgrp(\alg{u}(2|2))$ chains.

Finally, the limit of the total momentum
\eqref{eq:momspec}
and the total energy
\eqref{eq:engABC}
yields
\<
e^{iP}\eq e^{\hbar (N_3-N_1)}
\prod_{j=1}^{N_1}
\frac{\sinh(\lambda^{(1)}_k+\hbar)}{\sinh(\lambda^{(1)}_k-\hbar)}\,,
\nln
E\eq \cosh(2\hbar)K+\sum_{j=1}^{N_1}
\frac{\sinh^2(2\hbar)}{\sinh(\lambda^{(1)}_k+\hbar)\sinh(\lambda^{(1)}_k-\hbar)}
\,.
\>
We observe that only the $\lambda^{(1)}_k$ carry momentum and energy,
the $\lambda^{(3)}_k$ do not contribute.

\subsection{Another Quantum-Deformation of the Hubbard Model}\label{sec:AnotherMod}

Some curious values for the parameters $(\xp{},\xm{})$
appear to be $(0,\antimap(0))$, $(\antimap(\infty),\infty)$
and $(\antimap(0),0)$ $(\infty,\antimap(\infty))$.
In the limit $q\to 1$ they obviously approach the
values $(0,\infty)$ and $(\infty,0)$ corresponding
to the Hubbard model.
One has to be careful in using these values because several
common combinations of $\xpm{}$ turn out to be singular
and have to be regularized properly, e.g.\ by taking
a limit.

For definiteness we shall use the pair
\[
(\xp{},\xm{})=(0,\antimap(0));
\]
the other combinations yield similar results.
For convenience we rescale the Hamiltonian
\[
\ham'_{j,k}=\frac{iq^{1/2}}{G}\,\,\ham_{j,k}.
\]
The regularized coefficients of this Hamiltonian read
\[
\begin{array}[b]{c}
\displaystyle
A'=B'=-D'=-E'=\frac{1-2(q-q^{-1})^2g^2}{4g\sqrt{1-(q-q^{-1})^2g^2}}\,,
\\[15pt]
C'=F'=q+q^{-1},
\quad
G'=+iq^{+1/2},
\quad
L'=-iq^{-1/2},
\quad
H'=K'=0.
\end{array}
\]
In particular, we find $A'=B'$ and $D'=E'$.
This is interesting because
the pair-hopping terms of the oscillator Hamiltonian
in \tabref{tab:SecondQuantHam} vanish
just as for the Hubbard model
(we set $\kappa=-i$ for convenience)
\<
\ham'_{j,k}\eq
A'\sum_{\ell=j,k}\bigbrk{(1-n_{1,\ell})(1-n_{2,\ell})+n_{1,\ell}n_{2,\ell}-\half}
\nl
+iq^{+1/2}c_{1,j}^\dagger c_{1,k}^{}
\bigbrk{1-(1-q^{+1/2})n_{2,j}}
\bigbrk{1-(1-q^{-3/2})n_{2,k}}
\nl
+iq^{+1/2}c_{2,j}^\dagger c_{2,k}^{}
\bigbrk{1-(1-q^{-1/2}) n_{1,j}}
\bigbrk{1-(1-q^{-1/2}) n_{1,k}}
\nl
-iq^{-1/2}c_{1,k}^\dagger c_{1,j}^{}
\bigbrk{1-(1-q^{+3/2}) n_{2,j}}
\bigbrk{1-(1-q^{-1/2}) n_{2,k}}
\nl
-iq^{-1/2}c_{2,k}^\dagger c_{2,j}^{}
\bigbrk{1-(1-q^{+1/2}) n_{1,j}}
\bigbrk{1-(1-q^{+1/2})n_{1,k}}.
\>
Note that this Hamiltonian is not obviously hermitian.
It would be interesting to find out
if its spectrum is nevertheless real (for some non-hermitian twist).

The closed chain with this Hamiltonian can be diagonalized
by the Bethe equations \eqref{eq:BetheEquationsHomo}.
The regularized momentum relation \eqref{eq:momspec} reads
\[
e^{ip_k}=iq^{1/2}g\frac{q-q^{-1}+ig^{-1}/y_k}{\sqrt{1-(q-q^{-1})^2g^2}}
\,,
\]
and the energy of an eigenstate \eqref{eq:engABC} takes the form
\[
E'=\frac{1-2(q-q^{-1})^2g^2}{4g\sqrt{1-(q-q^{-1})^2g^2}}\, (K-2N)
+\sum_{j=1}^{N} \bigbrk{iq^{+1/2} e^{ip_j}-iq^{-1/2} e^{-ip_j}}.
\]
Note that this dispersion relation is not isotropic.
This fact is related to our choice of $\xpm{}$.
E.g.~choosing $\xpm{}$ such that the dispersion
relation is isotropic leads
to the Alcaraz--Bariev model discussed in \secref{sec:ABCompare}.
It may be desirable to investigate this model further
to see if it has interesting (physical) properties.

\section{Conclusions and Outlook}\label{sec:conclusions}

In this paper we have considered quantum deformations
of the threefold central extension $\alg{h}=\alg{psu}(2|2)\ltimes \Reals^3$
of the Lie superalgebra $\alg{psu}(2|2)$.
We have set up the Hopf algebra $\Qgrp(\alg{h})$
together with its fundamental R-matrix
and applied it to derive quantum deformations of the
one-dimensional Hubbard model.

A first important result is that the quantum deformation $\Qgrp(\alg{h})$
of the Lie superalgebra $\alg{h}$ is possible
despite its non-standard structure (central extension).
Moreover, we have constructed an invariant fundamental R-matrix
which obeys the Yang-Baxter equation,
crossing symmetry and fusion (with a suitable choice of overall phase factor).
This suggests that the Hopf algebra can be made quasi-cocommutative
and quasi-triangular.
Curiously, the matrix structure of the fundamental R-matrix is
determined by $\Qgrp(\alg{h})$ invariance alone.
This feature is related to the
representation theory $\Qgrp(\alg{h})$
which we have outlined and which is largely analogous to
the undeformed superalgebras $\alg{h}$ and $\alg{su}(2|2)$.
We have also discussed hermiticity conditions and
we were forced to choose $q$ to be real. Usually
in quantum groups $\Qgrp(\alg{g})$
one can also choose $q$ from the unit circle
(such that $q+q^{-1}$ is real), and it
would be important to understand if and how hermiticity
can be achieved in that case for our algebra.

We have applied the nested Bethe ansatz to the R-matrix in order
to diagonalize it and to write down the Bethe equations
for a closed spin chain. The Bethe equations are structurally
similar to the Lieb--Wu equations for the one-dimensional Hubbard model:
One should view the former as the quantum deformation
of the latter; the Lieb-Wu equations are of the
algebraic (XXX-like) type, while ours are trigonometric (XXZ-like).
A question for future work is whether an elliptic (XYZ-like)
deformation of our fundamental R-matrix and the Lieb--Wu equations exists.

In the final section we have derived an integrable Hamiltonian
from the R-matrix and compared directly it to the one-dimensional
Hubbard model and some of its generalizations.
Our Hamiltonian has manifest $\Qgrp(\alg{su}(2)\times\alg{su}(2))$ symmetry,
three independent parameters and it can be
further deformed in several canonical ways
(rescaling, shifts, twists) while preserving integrability.
By adjusting the parameters we were able
to match the one-parametric Hubbard chain as well
the cases B$^\pm$ of a two-parametric
Hamiltonian proposed by Alcaraz and Bariev \cite{Alcaraz:1999aa}.
We have also identified a potentially interesting
combination of parameters for which
our Hamiltonian has no pair-hopping terms.
We have not managed to explain the cases A$^\pm$ of the Alcaraz--Bariev
Hamiltonian in general, but we have provided an argument based on
the Bethe equations why this should be possible nevertheless.
It would therefore be important to show explicitly how to
relate the cases A$^\pm$, perhaps by finding an alternative coalgebra
for $\Qgrp(\alg{h}(2))$.
Furthermore an investigation of the possible condensed matter theory applications
of quantum-deformations of the one-dimensional Hubbard model is worth performing.
Do these models also display signs of superconductivity like the undeformed model?
It is also worth finding out if any of the other electronic models
discussed in the introduction can be obtained from our Hamiltonian.

Here we have investigated the Hopf algebra $\Qgrp(\alg{h})$,
but the analysis is far from complete.
In fact, the symmetry algebra in the undeformed setup is
much larger, it appears to be a Yangian (double) \cite{Beisert:2007ds}.
The corresponding quantum deformation would be a
quantum affine algebra.
One will need this algebra in order to formulate the universal R-matrix
of which the R-matrix derived in this paper is the fundamental representation.
A classical and undeformed analysis \cite{Beisert:2007ty}
has shown that the affine algebra is not just $\alg{h}[u,u^{-1}]$.
It is rather a deformation of $\alg{u}(2|2)[u,u^{-1}]$
which includes only \emph{one} tower of central charges
and one tower of inner \emph{automorphisms}.
Therefore it is quite clear that this quantum affine algebra is not Kac--Moody,
and one cannot directly apply the general framework associated with such algebras.

Finally, we should ask how our results can be applied to the AdS/CFT correspondence.
Quantum-deformed (XXZ-like) spin chains with
$\Qgrp(\alg{su}(2))$ and $\Qgrp(\alg{su}(3))$ symmetry
have indeed appeared in this context,
in particular for so-called beta-deformed $\superN=4$ gauge theory \cite{Leigh:1995ep,Lunin:2005jy}
with complex deformation parameter
\cite{Roiban:2003dw,Berenstein:2004ys,Frolov:2005ty,Mansson:2007sh}.
However, the argument used in \cite{Berenstein:2004ys}
tells us that the symmetry considered in this paper
cannot apply to conformal gauge theories such as \cite{Leigh:1995ep,Lunin:2005jy}:
The two bosonic subalgebras $\alg{su}(2)$ of $\alg{psu}(2|2)$
play different roles; one originates from the internal symmetry
which is deformed, and the other one originates from conformal symmetry
which remains undeformed. In contrast, quantum deformations typically apply to the
whole of a Hopf algebra, as is the case for our algebra.
Nevertheless, it seems possible to deform the internal $\alg{su}(4)$ symmetry
and the associated $S^5$, so why should it not be possible to deform
$\alg{su}(2,2)$ and the associated $AdS_5$?
Let us speculate about such a complete quantum deformation
of the AdS/CFT correspondence:
an ``AdS$_q$/CFT$_q$'' duality.
The field theory would have to have
quantum-deformed conformal and internal symmetries, and
it would most probably be formulated on some sort of non-commutative spacetime.
The string theory would be formulated
on a quantum-deformed $(AdS_5\times S^5)_q$ background,
which would be both curved and non-commutative. Perhaps
the most convenient definition would be in terms of
coset spaces like $\grp{SO}_q(6)/\grp{SO}_q(5)$
with non-commutative coordinates.
Despite the expected dire technical complications,
the planar limit of these dual models
would stand a good chance of being integrable,
and the world sheet S-matrix would then be given through our
quantum-deformed R-matrix.
It would be exciting to find out if this proposed picture
is more than daydreaming.

\subsection*{Acknowledgements}

The authors are grateful to thank Fabian Spill
for collaboration at earlier stages of the project
and for many discussions.
The authors would also like to thank
Sergey Frolov,
Frank G\"ohmann,
Alexander Gorsky
for fruitful discussions.
P.K.\ thanks MPI f\"ur Gravitationsphysik in Potsdam for hospitality
where main part of this work was performed.
The work of P.K.\ was partially supported by a grant of the
President of Russian Federation (NS-7293.2006.2)
and a grant of the Dynasty Foundation awarded by the Scientific Board of ICFPM.

\appendix

\section{Some Useful Relations}\label{sec:usefulQ}

The fundamental representation
is defined in terms of the variables $\xpm{}$
constrained by the quadratic relation \eqref{eq:xpmrel}
\[\label{eq:xpmrel3}
\frac{\xp{}}{q}+\frac{q}{\xp{}}-q\xm{}-\frac{1}{q\xm{}}
+ig(q-q^{-1})\lrbrk{\frac{\xp{}}{q\xm{}}-\frac{q\xm{}}{\xp{}}}=\frac{i}{g}\,.
\]
An expression which depends
on $\xpm{}$ can therefore be written in many ways
which are equivalent upon \eqref{eq:xpmrel3}.
For example one may choose to eliminate all
$(\xm{})^k$ with $k\neq 0,1$
(because the constraint is quadratic in $\xm{}$)
in order to get a unique representative within the equivalence
class of the expression.
Unfortunately this particular expression
is typically not the most economical one.
However such a representative can be used for the purpose to show
that a certain expression is identically zero
(or that two expressions are equivalent).

In this appendix we would like to present various identities
involving the $\xpm{}$
which were used in the paper or may prove useful otherwise.


\paragraph{One Set of $\xpm{}$.}

\[
u= q^{-1}\specmap(\xp{})-\frac{i}{2g}=q\specmap(\xm{})+\frac{i}{2g}
=\frac{i}{g}\,\frac{q^{\mp 1}\xpm{}\antimap(\xpm{})-\half(q+q^{-1})}{q-q^{-1}}
\]
\<
q^{-1}\xp{}\antimap(\xp{})=q\xm{}\antimap(\xm{})
\eq -ig(q-q^{-1})u+\half(q+q^{-1})
\nln\eq -igq^{-1}(q-q^{-1})\specmap(\xp{})+q^{-1}
\nln\eq -igq(q-q^{-1})\specmap(\xm{})+q
\nln\eq\frac{q^{C}U-q^{-C}U^{-1}}{q^{-C}U-q^CU^{-1}}
\>
\[
\frac{\xp{}-\antimap(\xm{})}{\xm{}-\antimap(\xp{})}=
\frac{(\xp{})^2+1}{(\xm{})^2+1}\,
\frac{\xm{}+\antimap(\xm{})}{\xp{}+\antimap(\xp{})}
=
q^{2C}U^{2}
\]
\[
q/\xp{}-q^{-1}/\xm{}-ig^{-1}
=-\frac{\xp{}-\xm{}}{q^{-1}\xp{}\antimap(\xp{})}
=-\frac{\xp{}-\xm{}}{q\xm{}\antimap(\xm{})}
\]
\[
q^{C}U\frac{(\xm{})^2+1}{\xm{}+\antimap(\xm{})}
=q^{-C}U^{-1}\frac{(\xp{})^2+1}{\xp{}+\antimap(\xp{})}
=\frac{\xp{}-\xm{}}{q^C U-q^{-C}U^{-1}}
\]
%

\paragraph{One $y$.}

\[
\frac{y^2+1}{y+\antimap(y)}\,
\frac{\antimap(y)^2+1}{y+\antimap(y)}
=1-(q-q^{-1})^2g^2
\]

\[
\specmap(y)=\frac{i}{g}\,\frac{y\antimap(y)-1}{q-q^{-1}}
\]

\paragraph{One Set of $\xpm{}$ and a $y$.}

\[
\frac{\xp{}-\antimap(y)}{\xm{}-\antimap(y)}\,
\frac{\antimap(\xm{})-y}{\antimap(\xp{})-y}
=q^{2C}U^2
\]
\[
\frac{\xp{}-y}{\xm{}-y}\,\frac{\antimap(\xp{})-y}{\antimap(\xm{})-y}
=
\frac{\specmap(\xp{})-\specmap(y)}{\specmap(\xm{})-\specmap(y)}
\]
%

\paragraph{Two Sets of $\xpm{}$.}

\[
\frac{\xp{1}-\xp{2}}{\xm{1}-\xm{2}}\,
\frac{\antimap(\xm{1})-\antimap(\xm{2})}{\antimap(\xp{1})-\antimap(\xp{2})}
=
\frac{\xp{1}-\antimap(\xm{2})}{\xm{1}-\antimap(\xp{2})}\,
\frac{\antimap(\xm{1})-\xp{2}}{\antimap(\xp{1})-\xm{2}}
=q^{2C_1+2C_2}U_1^2U_2^2
\]
\[
\frac{\xp{1}-\xp{2}}{\xm{1}-\xm{2}}\,
\frac{\xp{1}-\antimap(\xp{2})}{\xm{1}-\antimap(\xm{2})}
=q^{2C_1+2}U_1^2
\]
\[
\frac{\xp{1}-\xm{2}}{\xm{1}-\xp{2}}\,
\frac{\xp{1}-\antimap(\xm{2})}{\xm{1}-\antimap(\xp{2})}
=q^{2C_1}U_1^2
\frac{\specmap(\xp{1})-\specmap(\xm{2})}{\specmap(\xm{1})-\specmap(\xp{2})}
\]

\section{Braiding from Scattering Problem}
\label{sec:scatterbraid}

An equivalent approach to derive and incorporate the
braiding factors for the coproduct \eqref{eq:bradedcopr}
is based on using an S-matrix which acts as a permutation operator
and commutes with the coproduct $\comm{\copro(\gen{J})}{\smat}=0$
for each generator $\gen{J}\in\Qgrp(\alg{h})$.
This relation is equivalent to cocommutativity \eqref{eq:cocorel}
upon identifying the S-matrix with the R-matrix
as usual $\smat=\perm\rmat$.

What we want to stress here is the fact that for the quantum deformed case the
introduction of the braiding in the coproduct is necessary to
allow for a quasi cocommutative Hopf algebra.
If one would only take the quantum-deformed algebra with the coproduct
\eqref{eq:qcoproduct} there cannot be any R-matrix which
transforms $\copro$ to $\coproop$ for the central elements.
We will identify the additional braiding element with the central
charges in such way that the coproduct will indeed become cocommutative
on the center.

In particular, let us consider two short
modules with central charges $\srep{C_1,P_1,K_1}$ and
$\srep{C_2,P_2,K_2}$.
For the coproduct \eqref{eq:qcoproduct} we have
\[
\copro(\gen{P})= \gen{P}\otimes 1+q^{2\gen{C}}\otimes\gen{P},
\]
but
\[
\coproop(\gen{P})= 1\otimes\gen{P}+\gen{P}\otimes q^{2\gen{C}}.
\]
We need to change the coproduct by introducing an additional braiding factor
to make it quasi-cocommutative.
The importance of having a quasi-cocommutative Hopf algebra lies in the fact
that they have an R-matrix which not only intertwines the modules,
but satisfies the quasi-triangularity condition from which the Yang--Baxter equation follows.

\paragraph{Pairwise Scattering.}

\begin{figure}\centering
\includegraphics{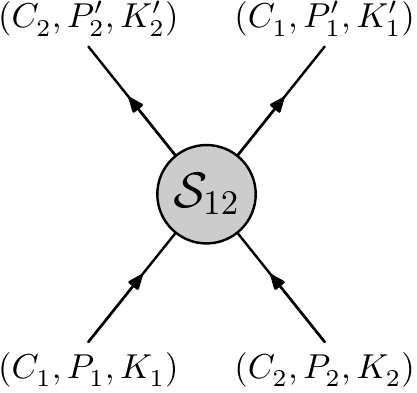}
\caption{Scattering process and transformation of central charges.}
\label{fig:scatter}
\end{figure}

Now we will consider the representation structure when the
S-matrix acts on chains of multiplets.
First we consider the scattering matrix of two short multiplets
(see \figref{fig:scatter})
\[
\smat_{12}: \srep{\vec{C}_1}\otimes\srep{\vec{C}_2}
\longrightarrow\srep{\vec{C}'_2}\otimes\srep{\vec{C}'_1}.
\]
For each generator $\gen{J}\in\Qgrp(\alg{h})$ we want the S-matrix
to be invariant
\[
[\copro(\gen{J}),\smat]=0.
\]
In particular, this relation must hold for the central charges.

An obvious way to conserve the total charge $C_1+C_2$
is to demand that the individual charges
are merely interchanged
\[\label{eq:EnergyConserv}
C'_1 = C_1,\quad C'_2 = C_2 .
\]
Taking the coproduct \eqref{eq:qcoproduct}
of the other central charges $\gen{P},\gen{K}$
and evaluating them before and after scattering we have
\<
P_1 +P_2 q^{2C_1} \eq P'_2 + P'_1 q^{2C_2},
\nln
K_1 q^{-2C_2} +K_2  \eq K'_2 q^{-2C_1} + K'_1 .
\>
Combining the short multiplet constraint \eqref{eq:short} with \eqref{eq:EnergyConserv}
we have the following equations to be solved
\[
P_1 K_1 = P'_1 K'_1, \quad P_2 K_2 = P'_2 K'_2.
\]
The substitution
\[\label{eq:substitution}
\begin{array}[b]{rclcrcl}
P_1 \eq 1-q^{2C_1}U_1,&&
K_1\eq q^{-2C_1}-U_1^{-1},
\\[3pt]
P_2\eq(1-q^{2C_2}U_2)U_1,&&
K_2\eq(q^{-2C_2}-U_2^{-1})U_1^{-1}
\end{array}
\]
simplifies the calculations dramatically and we get%
\footnote{There are two roots for the second order equation
-- one of them corresponds to trivial scattering and we did not mention it here.}
\[\label{eq:p1pr}
\begin{array}[b]{rclcrcl}
P'_1\eq(1-q^{2C_1}U_1)U_2,&&
K'_1\eq(q^{-2C_1}-U_1^{-1})U_2^{-1},
\\[3pt]
P'_2\eq1-q^{2C_2}U_2,&&
K'_2\eq q^{-2C_2}-U_2^{-1}.
\end{array}
\]
%

\paragraph{Factorized Scattering.}

A basic requirement
for a factorized $K$-particle S-matrix $\smat_{\pi}$ for any
permutation $\pi\in S_K$ is that it forms a representation of the
permutation group $S_K$, i.e. $\smat_{\pi}\smat_{\pi'}=\smat_{\pi\pi'}$. It imposes certain
relations on $P_i$ and $K_i$. We consider the permutation that
interchanges three modules as follows
\[
\smat_\pi: \srep{\vec C_1} \otimes \srep{\vec
C_2}\otimes\srep{\vec C_3} \mapsto \srep{\vec
C'_3}\otimes\srep{\vec C'_1} \otimes \srep{\vec C'_2}.
\]
This process can be represented in two different ways. The
factorized scatterings ($23\to 3'2'$ and $1'3'\to 3''1''$)
lead to the following relations
\[
P'_3= P_2+P_3 q^{2C_2}-P'_2 q^{2C_3},\quad
P'_1=P_1
\]
for the process $23\to 3'2'$ and
\<\label{eq:p3pr}
P''_3\eq P'_1+P'_3 q^{2C_1}-P''_1 q^{2C_3}\nln \eq(P_1+P_2
q^{2C_1})-(P'_1+P'_2 q^{2C_1})q^{2C_3}+P_3q^{2C_1 +2C_2},\nln
P_2''\eq P_2'
\>
for the process $1'3'\to 3''1''$.

\paragraph{Fusion.}

Furthermore, the overall process can be represented as a pairwise scattering of
composite multiplet
$\srep{\vec{C}_{12}}$ with
$\srep{\vec{C}_3}$. The former multiplet is long but for particular values of
$\vec{C}_1$ and $\vec{C}_2$ it
splits into two short multiplets. The central charges $P'_{1,2,3}$ and $K'_{1,2,3}$
become related by the following constraints
\<\label{eq:compositescatt}
P_1+P_2q^{2C_1}+P_3q^{2C_1+2C_2} \eq P'_3+P'_1q^{2C_3}+P'_2q^{2C_1+2C_3}\,,\nln
K_1q^{-2C_2-2C_3}+K_2q^{-2C_3}+K_3 \eq K'_3q^{-2C_1-2C_2}+K'_1q^{-2C_2}+K'_2.
\>
The results \eqref{eq:p3pr} and solution of \eqref{eq:compositescatt} should
correspond and it constrains the form of central charges.
An educated guess consists in choosing them in the following way
\<
P_k \eq g\alpha(1-q^{2C_k}U^2_k)\prod\limits_{j=1}^{k-1}U^2_j,
\nln
K_k \eq
\frac{g}{\alpha}(q^{-2C_k}-U^{-2}_k)\prod\limits_{j=1}^{k-1}U^{-2}_j.
\>
which completely agrees with \eqref{eq:PKident}.

The following remark is noteworthy here. We have derived
the dependence of the central charges $P$ and $K$ on the braiding factors $U$
by demanding the condition that
scattering with a pair of particles can be treated as a successive
scattering with one particle and then with another one. Alternatively we
can treat $U$ as a function of $P$ and $K$ and put $UU^{-1}=1$ as a
constraint (which is for sure satisfied in \eqref{eq:substitution}
automatically). If this constraint is not satisfied then the Hopf
algebra for $\alg{h}=\alg{su}(2|2)\ltimes\mathbb{R}^2$ cannot
be quasi-triangular.
We can drop one central charge
(say $K$ and thus have $\alg{su}(2|2)\ltimes\mathbb{R}^1$ algebra)
then this relation does
not constrain us any longer
and the Hopf algebra based on
$\alg{su}(2|2)\ltimes\mathbb{R}^1$
is quasi-triangular.

\paragraph{Fundamental S-Matrix.}

We write down the explicit form of the fundamental S-matrix
$\smat=\perm\rmat$ in \tabref{tab:Smatrix}.
Its coefficients are the same as for the fundamental R-matrix
given in \tabref{tab:Scoeffs}.
\begin{table}
\<
\smat\state{\phi_1^1\phi_2^1}\eq
A_{12}\state{\phi_2^1\phi_1^1}
\nln
\smat\state{\phi^1_1\phi^2_2}\eq
\frac{q A_{12}+q^{-1} B_{12}}{q+q^{-1}}\state{\phi_2^1\phi_1^2}
+\frac{A_{12}-B_{12}}{q+q^{-1}}\state{\phi_2^2\phi_1^1}
+\frac{q^{-1}C_{12}}{q+q^{-1}}\state{\psi_2^1\psi_1^2}
-\frac{C_{12}}{q+q^{-1}}\state{\psi_2^2\psi_1^1} \nln
\smat\state{\phi_1^2\phi_2^1}\eq \frac{A_{12}-
B_{12}}{q+q^{-1}}\state{\phi_2^1\phi_1^2} +\frac{q^{-1}A_{12}+q
B_{12}}{q+q^{-1}}\state{\phi_2^2\phi_1^1}
-\frac{C_{12}}{q+q^{-1}}\state{\psi_2^1\psi_1^2}
+\frac{q C_{12}}{q+q^{-1}}\state{\psi_2^2\psi_1^1}
\nln
\smat\state{\phi_1^2\phi_2^2}\eq
A_{12}\state{\phi_2^2\phi_1^2}
\nn\\[10pt]
\smat\state{\psi_1^1\psi_2^1}\eq
D_{12}\state{\psi_2^1\psi_1^1}
\nln
\smat\state{\psi_1^1\psi_2^2}\eq
\frac{qD_{12}+q^{-1} E_{12}}{q+q^{-1}}\state{\psi_2^1\psi_1^2}
+\frac{D_{12}-E_{12}}{q+q^{-1}}\state{\psi_2^2\psi_1^1}
+\frac{q^{-1}F_{12}}{q+q^{-1}}\state{\phi_2^1\phi_1^2}
-\frac{F_{12}}{q+q^{-1}}\state{\phi_2^2\phi_1^1}
\nln
\smat\state{\psi_1^2\psi_2^1}\eq
\frac{D_{12}-E_{12}}{q+q^{-1}}\state{\psi_2^1\psi_1^2}
+\frac{q^{-1}D_{12}+q E_{12}}{q+q^{-1}}\state{\psi_2^2\psi_1^1}
-\frac{F_{12}}{q+q^{-1}}\state{\phi_2^1\phi_1^2}
+\frac{q F_{12}}{q+q^{-1}}\state{\phi_2^2\phi_1^1}
\nln
\smat\state{\psi_1^2\psi_2^2}\eq
D_{12}\state{\psi_2^2\psi_1^2}
\nn\\[10pt]
\smat\state{\phi_1^a\psi_2^\beta}\eq
G_{12}\state{\psi_2^\beta\phi_1^a}
+H_{12}\state{\phi_2^a\psi_1^\beta}
\nln
\smat\state{\psi_1^\alpha\phi_2^b}\eq
K_{12}\state{\psi_2^\alpha\phi_1^b}
+L_{12}\state{\phi_2^b\psi_1^\alpha}\nonumber
\>
\caption{The fundamental S-matrix of $\Qgrp(\alg{h})$. }
\label{tab:Smatrix}
\end{table}

\bibliography{su22q}
\bibliographystyle{nb}

\end{document}